\documentclass{aa}  
\usepackage{graphicx}
\usepackage{txfonts}
\usepackage{hyperref}
\usepackage{subfigure}
\usepackage{float}
\usepackage{comment}
\usepackage{layouts}
\usepackage{multirow}
\usepackage{xcolor}

\newcommand*\diff{\mathop{}\!\mathrm{d}}

\newcommand{\zerodel}{.\kern-\nulldelimiterspace}

\newcommand{\NC}[1]{\textcolor{blue}{#1}}

\newcommand{\SD}[1]{\textcolor{purple}{#1}}

\newcommand{\XMM}{XMM-\textit{Newton}~}

\begin{document}

   \title{Investigating the turbulent hot gas in X-COP galaxy clusters} 

   \author{S. Dupourqué \inst{1}
          \and 
          N. Clerc \inst{1}
          \and
          E. Pointecouteau \inst{1}
          \and 
          D. Eckert \inst{2}
          \and
          S. Ettori \inst{3}
          \and
          F. Vazza \inst{4,5,6}
          }

   \institute{IRAP, Université de Toulouse, CNRS, CNES, UT3-PS, Av. du Colonel Roche 9, 31400, Toulouse, France
         \and
             Department of Astronomy, University of Geneva, Ch. d’Ecogia 16, CH-1290 Versoix, Switzerland
        \and INAF, Osservatorio di Astrofisica e Scienza dello Spazio, via Piero Gobetti 93/3, 40129 Bologna, Italy
        \and Dipartimento di Fisica e Astronomia, Universit\'{a} di Bologna, Via Gobetti 93/2, 40122, Bologna, Italy
\and  Hamburger Sternwarte, University of Hamburg, Gojenbergsweg 112, 21029 Hamburg, Germany
\and Istituto di Radioastronomia, INAF, Via Gobetti 101, 40122, Bologna, Italy}

   \date{Received ???; accepted ???}

 
  \abstract
   {Turbulent processes at work in the intracluster medium perturb this environment, impacting its properties, displacing gas, and creating local density fluctuations that can be quantified via X-ray surface brightness fluctuation analyses. Improved knowledge of these phenomena would allow for a more accurate determination of the mass of galaxy clusters, as well as a better understanding of their dynamic assembly.}
   {In this work, we aim to set constraints on the structure of turbulence using X-ray surface brightness fluctuations. We seek to consider the stochastic nature of this observable and to constrain the structure of the underlying power spectrum.}
   {We propose a new Bayesian approach, relying on simulation-based inference to account for the whole error budget. We used the X-COP cluster sample to individually constrain the power spectrum in four regions and within $R_{500}$. We spread the analysis on the entire set of 12 systems to alleviate the sample variance. We then interpreted the density fluctuations  as the result of either gas clumping or  turbulence.}
   {For each cluster considered individually, the normalisation of density fluctuations correlates positively with the Zernike moment and centroid shift, but negatively with the concentration and the Gini coefficient. The spectral index within $R_{500}$ and evaluated over all clusters is consistent with a Kolmogorov cascade. The normalisation of density fluctuations, when interpreted in terms of clumping, is consistent within $0.5 R_{500}$ with the literature results and numerical simulations; however, it is higher between 0.5 and $1 R_{500}$. Conversely, when interpreted on the basis of turbulence, we deduce a non-thermal pressure profile that is lower than the predictions of the simulations within 0.5 $R_{500}$, but still in agreement in the outer regions. We explain these results by the presence of central structural residues that are remnants of the dynamic assembly of the clusters.}
   {}


   \maketitle
%

\section{Introduction}

The intracluster medium (ICM) is the primary baryonic component of galaxy clusters. The hot gas ($T\sim 10^7-10^8$ K) is governed by many physical processes that introduce perturbations on various scales. In the inner parts of galaxy clusters, the strong cooling and accreting gas trigger feedback from the active galactic nucleus (AGN) hosted by the central galaxy \citep{mcnamara_mechanical_2012, voit_global_2017}. In the outer parts, merger events and accretion from the cosmic web generate shocks, adiabatic compression, and turbulent cascades \citep{nelson_evolution_2012}. Throughout the history of the dynamic assembly of galaxy clusters, these perturbations will play a predominant role in the non-thermal heating of the ICM \citep{bennett_disturbing_2021}, generating non-thermal pressure, which drives the hydrostatic mass bias when it is not accounted for in the mass estimation of massive halos when assuming their hydrostatic equilibrium \citep{piffaretti_total_2008, lau_residual_2009, meneghetti_weighing_2010, nelson_weighing_2014, biffi_nature_2016, pratt_galaxy_2019}.

Turbulence is a stochastic process that occurs in high Reynolds number flows. It can be interpreted qualitatively as the transport of kinetic energy from a large injection scale to a small viscous scale, which then dissipates as heat in the medium. Putting constraints on the turbulence that occurs in the intracluster medium is of great interest when it comes to characterising non-thermal heating, since turbulent motions can be responsible for more than $80\%$  of these effects -- and potentially all of it \citep{vazza_turbulent_2018, angelinelli_turbulent_2020}. 
The level of baseline turbulence implied by most of the modern cosmological simulations also appears to be in line with the implications of the latest discoveries of very extended radio emission, namely, out to at
least $R_{500}$  in clusters of galaxies classified as 'mega-halos' \citep{cuciti_galaxy_2022}. The detection of  such unprecedentedly large volumes of cluster environments filled with relativistic particles and magnetic fields is currently best explained by the ubiquitous re-acceleration of electrons by turbulence, which is expected to be present at a similar level in all clusters on these radii, at a level compatible with simulations \citep[see also][]{botteon_magnetic_2022}.

Direct observations of this phenomenon is possible via studies of the centroid shifts and broadening of the ICM emission lines to derive  the bulk and turbulent motions, respectively. Upper limits on the fraction of energy in the form of turbulent motions have been derived using the \XMM spectrometer for samples of clusters \citep{sanders_constraints_2011, pinto_chemical_2015}. A novel approach using the EPIC-pn detector has allowed for the direct measurement of bulk flows in the Coma and Perseus clusters \citep{sanders_measuring_2020}, showing consistent results with \textit{Hitomi} and was also applied to the Virgo and Centaurus clusters \citep{gatuzz_velocity_2022, gatuzz_measuring_2022}. How these measurements translate into a correction for the hydrostatic bias is illustrated in, for instance, \cite{ota_constraining_2018, ettori_tracing_2022}. The first credible spatially resolved measurements of this type in the X-rays were achieved with the \textit{Hitomi} observatory \citep{the_hitomi_collaboration_quiescent_2016} at the centre of the Perseus cluster. Future direct measurements  will be obtained with the coming of the new generation of X-ray integral fields unit, that is  \textit{XRISM}/Resolve  in the coming years \citep{xrism_science_team_science_2020} and \textit{Athena}/X-IFU in the long term \citep{barret_athena_2020}. 

It is also possible to characterise these phenomena using indirect observables. The X-ray surface brightness (XSB) fluctuations are mostly due to density fluctuations in the ICM, and have been studied for the first time in the Coma cluster by \cite{schuecker_probing_2004}, and later by \cite{churazov_x-ray_2012}, deriving constraints on the density fluctuation power spectrum on scales ranging between 30 kpc and 500 kpc. Various theoretical works have suggested a strong link between turbulent velocities and density fluctuations \citep[e.g.][]{zhuravleva_relation_2014, gaspari_relation_2014, mohapatra_turbulence_2020, mohapatra_turbulent_2021, simonte_exploring_2022}, indicating that the study of brightness fluctuations could constrain the turbulent processes that occur in the ICM. This methodology was applied to Perseus \citep{zhuravleva_gas_2015}, and in the cool cores of ten galaxy clusters \citep{zhuravleva_gas_2018}. These studies constrained turbulent velocities to $\lesssim 150 \text{ km s}^{-1}$ at scales smaller than $ 50 \text{ kpc}$, which is consistent with direct measurements from \textit{Hitomi}.

Turbulent processes, and thereby the resulting density fluctuations, originate from chaotic processes that can be assimilated to random fields observed in spatially finite regions. The stochastic nature of this observable coupled with the finite size of the observations means it intrinsically carries an additional variance, which is later referred to as the sample variance. This effect has been studied for the structure function of turbulent velocities for XRISM \citep{zuhone_mapping_2016} and Athena X-IFU mock observations \citep{clerc_towards_2019, cucchetti_towards_2019}, and it has been dominant in the error budget at spatial scales $\gtrsim 50 \text{ kpc}$. It is expected to play a significant role in fluctuation analyses, leading to an underestimation of the error budget when it is not accounted for. 
In this work, we propose a novel approach based on the forward modelling of observables related to surface brightness fluctuations, which allows, for the first time, the full error budget associated with their stochastic nature to be considered. In this work, we apply this methodology to the \XMM ~Cluster Outskirts Project \citep[X-COP,][]{eckert_xmm_2017} sample and derive individual constraints on the statistical properties and spatial distribution of the density fluctuations via the characterisation of the 3D power spectrum of the associated random field. We combine these constraints under the assumption of comparable dynamics within the clusters to obtain stronger constraints on the whole sample, which can be related to the gas clumping and the turbulent processes occurring in the ICM.

In Sect.~\ref{sec:data_method}, we describe our methodology as well as the validation process. In Sect.~\ref{sec:results}, we present the individual results and the joint constraints on density fluctuations. In Sect.~\ref{sec:discussion}, we interpret these constraints as being gas clumping or coming from turbulent processes and contributing to non-thermal heating. We also inspect the correlations between the obtained parameters and the dynamic state of the clusters and discuss the various limitations of our approach. Throughout this paper, we assume a flat $\Lambda$CDM cosmology with $H_0 = 70 \text{ km s}^{-1}$ and $\Omega_m = 1 - \Omega_\Lambda = 0.3$. Scale radii are defined according to the critical density of the Universe at the corresponding redshift. The Fourier transform conventions are highlighted in Appendix \ref{app:fourier_convention}. 
\section{Data and method}
\label{sec:data_method}

\subsection{X-COP Sample}

The \textit{XMM-Newton} Cluster Outskirts Project \citep[X-COP,][]{eckert_xmm_2017} is an XMM Very Large Program dedicated to the study of the X-ray emission of cluster outskirts. This sample comprises 12 massive ($\sim 3 - 10 \times10^{14} M_{\odot}$) and local ($0.04<z<0.1$) galaxy clusters, for a total exposure time of $\sim 2$ Ms. The targets were selected from the first $Planck$ catalogue \citep{the_planck_collaboration_planck_2014} as (i) resolved sources regarding {\it Planck}'s spatial resolution (i.e. all clusters have $R_{500} > 10'$), (ii) with high signal-to-noise ratio ($S/N>12$), and (iii) in directions that ensure low hydrogen column densities to prevent soft-X ray absorption ($N_H < 10^{21} \text{ cm}^{-2}$). The thermodynamical properties of the X-COP sample have been characterised \cite{ghirardini_universal_2019}. The sample was further used to derive the first observational constraints on the amount of non-thermal pressure in the outskirts of clusters \cite{eckert_non-thermal_2019}.

\subsection{Data preparation}

This project is based on the publicly available X-COP data products \citep{ghirardini_universal_2019,ettori_hydrostatic_2019}\footnote{\url{https://dominiqueeckert.wixsite.com/xcop}}. Here, we recall the main steps of the data reduction, image extraction, and point source subtraction. We refer to \citet{ghirardini_universal_2019, eckert_non-thermal_2019} for further details on the data preparation procedure.
    
    First, the data were reduced using the XMMSAS v13.5 software package and the Extended Source Analysis Software (ESAS) procedure; namely, the raw data were screened using the \texttt{emchain} and \texttt{epchain} executables to extract raw event files. We then extracted the light curves of the entire field of view to exclude time periods affected by soft proton flares.

From the cleaned event files, we extracted count maps from each observation in the [0.7-1.2] keV from the three detectors of the EPIC instrument (MOS1, MOS2, and pn). We then computed the corresponding exposure maps including the telescope vignetting using the \texttt{eexpmap} task. Finally, we created non X-ray background maps from the filter-wheel-closed event files, appropriately rescaling the filter-wheel-closed data to match the count rates observed in the unexposed corners of the individual CCDs. The contribution of residual soft protons was estimated by measuring the ratio between the high-energy count rates inside and outside field of view (IN-OUT) and its relation to the residual soft proton component, calibrated over $\sim500$ blank-sky pointings \citep{ghirardini_xmm_2018}. 

For each cluster, the individual images, exposure maps, and background maps were stacked and concatenated into joint EPIC mosaic images. Point sources were extracted by joining the [0.5-2] keV and [2-7] keV energy bands using the XMMSAS tool \texttt{ewavelet} with a $S/N = 5$ threshold and the detected point sources were masked when their count rate exceeded a threshold set by the peak of the logN-logS distribution, such that the source exclusion threshold is homogeneous over the entire field.

\subsection{Modelling surface brightness}
\label{sec:mean-profile}

To characterise the surface brightness fluctuations of X-COP clusters, we choose first to subtract the bulk of the main cluster emission. We determined an average emissivity profile, assuming radial symmetry, and biaxiality in directions orthogonal to the line of sight. This was achieved by fitting an average surface brightness profile on the X-ray images in the [0.7-1.2] keV energy band. In this section, we note our use of the $\Vec{r} = (x,y,\ell)$ in the 3D parametrisation and $\Vec{\rho} = (x,y)$ in the 2D parametrisation, with $\ell$ as the coordinate along the line of sight and $(x,y)$ as the coordinates mapping the dimensions on the image. The true surface brightness $\Sigma_{X}$ is given by:  

\begin{equation}
\Sigma_{X}(\Vec{\rho}) =  \frac{1}{4\pi(1+z)^4}\int^{+\infty}_{-\infty}\Lambda \left(T, Z\right) \: n_e^2(\vec{r}) \: \diff \ell
    \label{eq:true_surface_brightness}
,\end{equation}

where $\Lambda$ is the intra-cluster gas cooling function in the [0.7-1.2] keV band, $T$ is the 3D-temperature, $Z$ is the metallicity, and  $n_e$ is the 3D electron density. The surface brightness we observe is convolved with the \XMM responses functions, and partially absorbed by the Galactic hydrogen column density. We then define the observed surface brightness $S_X$ as :

\begin{equation}
S_{X}(\Vec{\rho}) = \int^{+\infty}_{-\infty}\Psi(\vec{r}) \: n_e^2(\vec{r}) \: \diff \ell + B
    \label{eq:surface_brightness_definition}
,\end{equation}

where $\Psi(\vec{r})$ encompasses the cooling function, the cosmological dimming, the Galactic absorption, and the convolution with \XMM response functions, while $B$ is a constant surface brightness background left as a free parameter. The 3D density profile is modelled using a modified version of the Vikhlinin functional form \citep{vikhlinin_chandra_2006}, defined as follows:

\begin{equation}
    n_e^2(\Vec{r}) = n_{e,0}^2 \left(\frac{r}{r_c}\right)^{-\alpha}\frac{(1+r^2/r_c^2)^{-3\beta+\alpha/2}}{(1+r^\gamma/r_s^\gamma)^{\epsilon / \gamma}} 
    \label{eq:vikhlinin_density_profile}
,\end{equation}

where $n_{e,0}$ is the central density, $r_c$ and $r_s$ are two scale radii, and $\alpha, \beta, \gamma, \epsilon$ are  parameters that define the slopes and smoothness of transition between the power laws in this equation. We set $\alpha =0,$ as done by \cite{shi_analytical_2016}, to remove the central singularity, which causes problems due to low values of $r$ on the image. We fix $\gamma=3$ to prevent degeneracies with the $\epsilon$ parameter, as the data do not allow for a simultaneous determination of $\epsilon$ and $\gamma$.

\begin{figure*}
\centering
\subfigure{
\includegraphics[width=0.47\textwidth]{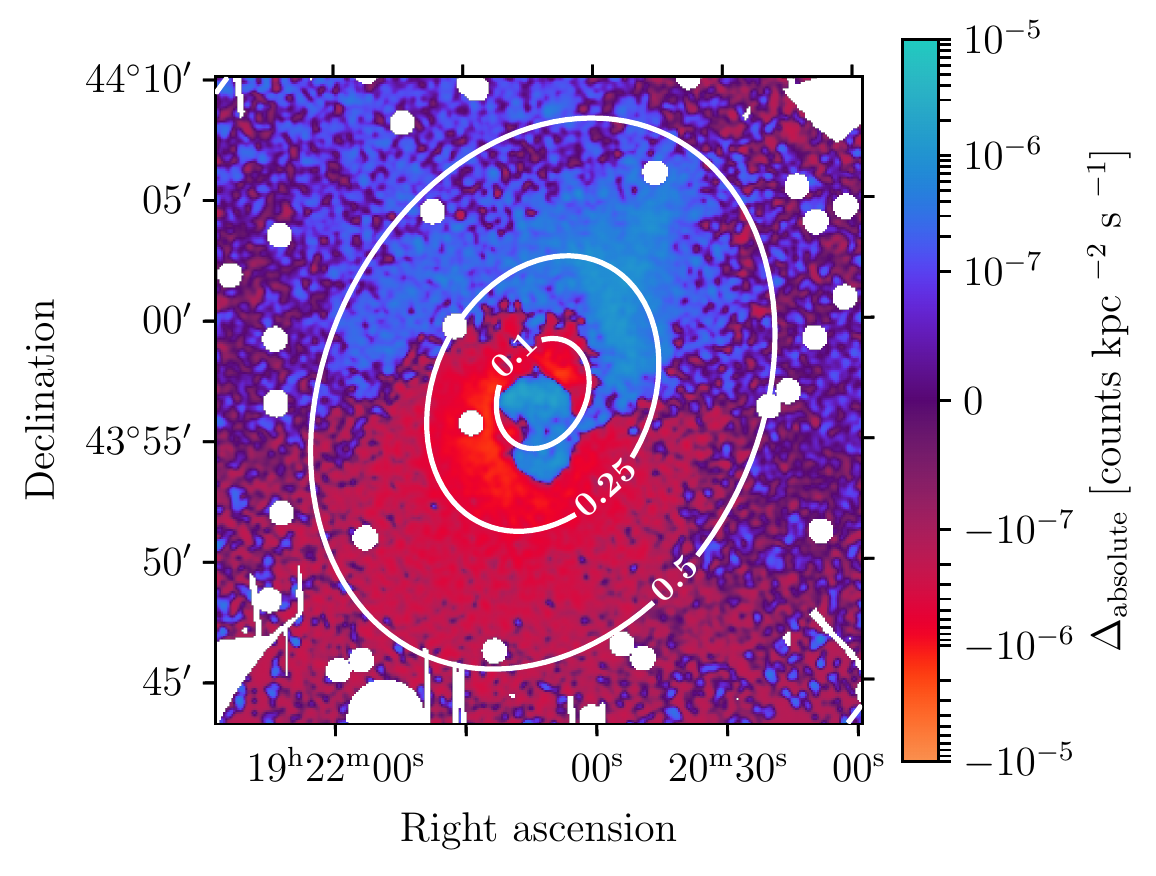}}
\label{fig:subfig:absolute-fluctuations} 
\hspace{0.2in}
\subfigure{
\includegraphics[width=0.47\textwidth]{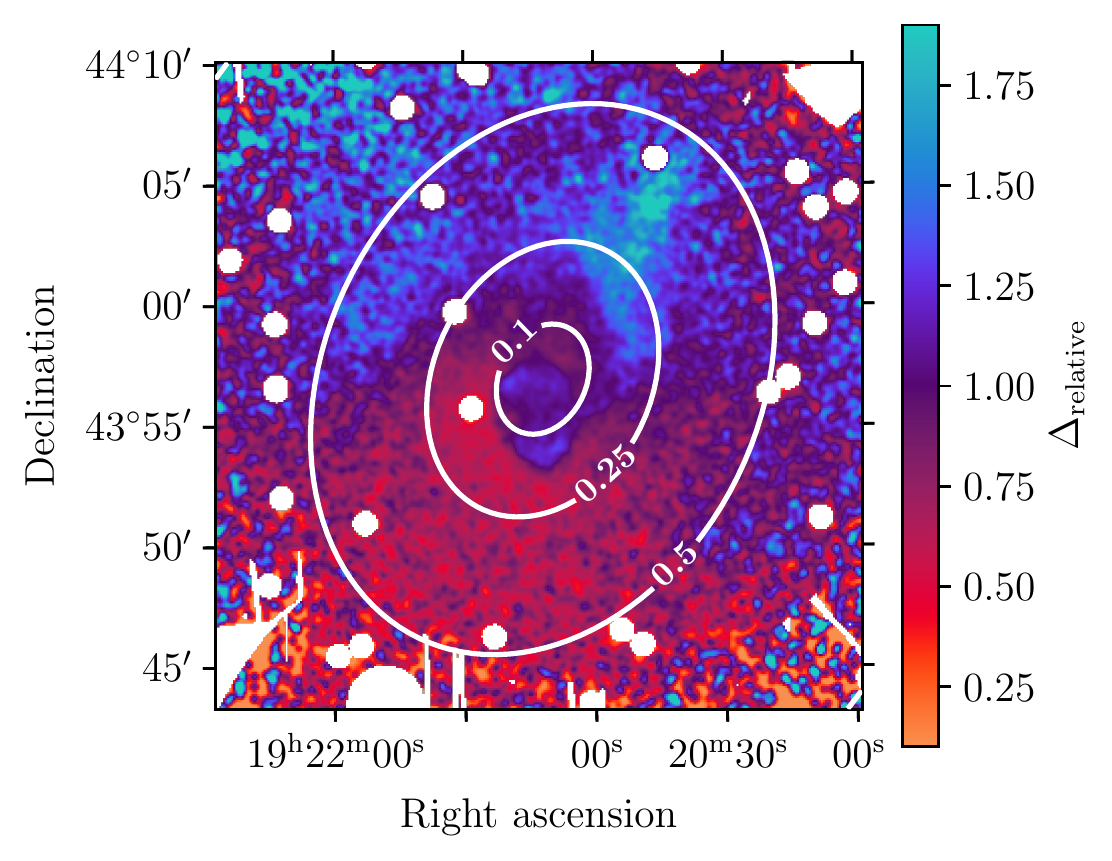}}
\label{fig:subfig:relative-fluctuations} 
\caption{Comparison between the absolute fluctuations $\Delta_{\text{ absolute}}$ from Eq. (\ref{eq:sb-fluc-diff}) (\textit{left}) and the relative fluctuations $\Delta_{\text{ relative}}$ from Eq. (\ref{eq:sb-fluc-ratio}) (\textit{right}) using the surface brightness data and best fit model image of A2319. For display  purposes, the images are filtered by a Gaussian kernel of 7.5". The successive contours represent the distance to the centre in units of $R_{500}$}.
\label{fig:comparison-relative-absolute}
\end{figure*}

We computed the cooling rate $\Lambda$ and the Galactic absorption using a functional approximation fitted to the count rate derived from a \texttt{"PhAbs*APEC"} model under \texttt{XSPEC 12.11.1,} as detailed in Appendix \ref{app:cooling_function}. Galactic absorption admits non-negligible fluctuations ($\sim10\%$) at the scale of the large X-COP mosaics. To take them into account, we used the $N_H$ data from the HI4PI survey \citep{hi4pi_collaboration_hi4pi_2016}, which provide a mapping of the Galactic column density at the spatial scale of 12-16 arcmin. We performed the projection along the line of sight with a double exponential quadrature \citep{takahasi_double_1973, mori_double-exponential_2001}. The centre position on the model image is left free and fitted as a Gaussian perturbation of the cluster centre, as provided in the Planck catalogue \citep{the_planck_collaboration_planck_2014}. The effective centre is then parametrised as:

\begin{equation}
\begin{pmatrix}
x_c\\
y_c
\end{pmatrix}
=
\begin{pmatrix}
\mathcal{N}(1, 0.5) & 0\\
0 & \mathcal{N}(1, 0.5)
\end{pmatrix}
\begin{pmatrix}
x_{\text{Planck}}\\
y_{\text{Planck}}
\end{pmatrix}
.\end{equation}

To take non-sphericity into account, we fit for an elliptical brightness distribution rather than a spherical one. The positions $(x,y)$ are distorted in a surface-conserving way into a new configuration $(\tilde{x},\tilde{y})$, introducing two additional free parameters to describe the ellipse angle $\theta$ and eccentricity $e$:

\begin{equation}
\begin{pmatrix}
\tilde{x} \\
\tilde{y}
\end{pmatrix}
=
\begin{pmatrix}
\frac{1}{\sqrt{1-e^2}} & 0\\
0 & {\sqrt{1-e^2}}
\end{pmatrix}
\begin{pmatrix}
\cos{\theta} & -\sin{\theta}\\
\sin{\theta} & \cos{\theta}
\end{pmatrix}
\begin{pmatrix}
x-x_c \\
y-y_c
\end{pmatrix}
\label{eq:ellipse}
.\end{equation}

This is formally equivalent to assuming a biaxiality of the 3D distribution of the cluster density over the image directions.
The distributions of the free parameter posteriors are obtained using Bayesian inference. 
We define a Poisson likelihood on images rebinned within $R_{500}$ with a Voronoi tessellation for a target count of 100 counts per bin \citep{cappellari_adaptive_2003}. As an example, for A3266, the median characteristic bin size and its 14\textsuperscript{th}-86\textsuperscript{th} percentile is $10^{+10}_{-4}$ arcsec. The model count for each bin is approximated by the product of the surface brightness $S_{X}$ estimated in the centre times the area and exposure of the bin. 
This is a reasonable approximation, since the size of the bins is small over the regions where the  surface brightness varies a lot (see App. \ref{app:binning_approximation} for further discussions). The total number of photon counts is obtained by adding the estimated background to the model counts, as defined in Eq. \ref{eq:model_counts}. We mask the regions associated with point sources and also remove identified substructures and groups such as in A1644, A2142, A644, A85, and RXC1825. The posterior distributions are sampled with the No U-Turn Sampler \citep{hoffman_no-u-turn_2014}, as implemented in the \texttt{numpyro} library \citep{bingham_pyro_2019, phan_composable_2019}. The prior distributions are displayed in App.~\ref{app:mean_model_parameters}.

\subsection{Fluctuations and power spectrum}
\label{sec:definition-flucutuation}

We want to quantify the surface brightness fluctuations under the assumption that they come exclusively from intrinsic density fluctuations. In this framework, the density profile can be decomposed into a rest component $n_{e}^2$ and relative fluctuations $\delta$ whose amplitude is smaller than unity: 

\begin{equation}
    S_{X}(\Vec{\rho}) =  \int^{+\infty}_{-\infty} \Psi(\vec{r})~ n_{e}^2(\Vec{r})~(1+\delta(\Vec{r}))^2 \diff \ell
    \label{eq:sb_def}
.\end{equation}

By linearising the equation (\ref{eq:sb_def}) in $\delta$, it follows that, at first order, the raw image corresponds to the sum of an unperturbed image $S_{X,0}$ and a surface brightness fluctuation map: 

\begin{equation}
    S_{X}(\Vec{\rho}) \simeq S_{X,0}(\Vec{\rho}) + 2 \int^{+\infty}_{-\infty} \epsilon_0(\Vec{r})~\delta(\Vec{r}) \diff \ell
    \label{eq:sx_ref_and_perturbed}
,\end{equation}

where $\epsilon_0(\Vec{r}) = \Psi(\vec{r})~ n_{e,0}^2(\Vec{r})$ is the unperturbed emissivity of the cluster. The fluctuations can be defined in a theoretically equivalent way by taking the difference or ratio of the perturbed to the unperturbed surface brightness. However, from an observational perspective, both approaches have their advantages and disadvantages, which we quantitively discuss in Sect.~\ref{sec:definition_aperture}. We define the absolute and relative fluctuations, $\Delta_{\text{abs}/\text{rel}}$, as follows:

\begin{equation}
    \Delta_{\text{ abs}}(\Vec{\rho}) \overset{\operatorname{def}}{=} \frac{S_{X}(\Vec{\rho}) - S_{X,0}(\Vec{\rho})}{2} \simeq\int^{+\infty}_{-\infty} \epsilon_0(\Vec{r})\delta(\Vec{r}) \diff \ell
\label{eq:sb-fluc-diff}
,\end{equation}

\begin{equation}
    \Delta_{\text{ rel}}(\Vec{\rho}) \overset{\operatorname{def}}{=} \frac{S_{X}(\Vec{\rho})}{S_{X,0}(\Vec{\rho})} \simeq 1 + 2\frac{\int^{+\infty}_{-\infty} \epsilon_0(\Vec{r})\delta(\Vec{r}) \diff \ell}{\int^{+\infty}_{-\infty} \epsilon_0(\Vec{r})\diff \ell} 
\label{eq:sb-fluc-ratio}
.\end{equation}

In practice, we use, we used the definition of \cite{zhang_planck_2022} as a variant of Eq.~\ref{eq:sb-fluc-ratio} for relative fluctuation maps, which performs better in low statistic regions while having a similar qualitative behaviour. The difference \citep[as presented in e.g.][]{khatri_thermal_2016} generates maps of fluctuations whose amplitude decreases with distance from the centre. This method defines fluctuations everywhere on the image, but biases their amplitude by over-representing fluctuations in the brightest areas of the image. The ratio \citep[e.g.][]{churazov_x-ray_2012} instead defines well-scaled fluctuations, at the expense of intensifying the noise in the low-brightness regions. In Fig. (\ref{fig:comparison-relative-absolute}), we compare the fluctuation maps obtained using the absolute and relative definition for A2319. It should be noted that both methods suffer from the lack of signal over regions of the order of $R_{500}$, here the photon statistics is insufficient to perform an efficient fluctuation analysis. To study the contribution of each spatial scale to surface brightness fluctuations, it is natural to perform the analysis in Fourier space. We define the power spectrum of the fluctuations, $\mathcal{P}_{2D,\Delta}(\Vec{k_\rho}),$ as follows: 

\begin{equation}
    \mathcal{P}_{2D, \Delta}(k_\rho) = \frac{1}{2\pi} \int |\Hat{\Delta}(\Vec{k_\rho}) |^2 \diff \varphi
,\end{equation}

with $\Hat{\Delta}$ the Fourier transform of the two-dimensional (2D) map $\Delta$ and $\varphi$ the azimuthal angle on the image. The numerical evaluation of $P_{2D,\Delta}$ is performed using the method described by \cite{arevalo_mexican_2012}, which computes the variance of images filtered by Mexican hats on a characteristic scale to estimate the azimuthal average of the power spectrum. A detailed explanation of the method is provided in Appendix \ref{app:mexican_hat_analytic}. Other approaches found in the literature compute the structure function, $\mathcal{SF}_{2D}(\rho),$ of the image instead of the $\mathcal{P}_{2D}(k)$ \citep[e.g.][]{roncarelli_measuring_2018, clerc_towards_2019, cucchetti_towards_2019}. We emphasise that these two approaches are fully equivalent, since both $\mathcal{SF}_{2D}$ and $\mathcal{P}_{2D}$ are measures of the second order correlation of the fluctuation field and are related by the following bijection, which can be derived using the Wiener–Khinchin theorem: 

\begin{equation}
    \mathcal{SF}_{2D}(\Vec{\rho}) = 2\int \left(1-e^{2i\pi\Vec{k_\rho}.\Vec{\rho}}\right)\mathcal{P}_{2D}(\Vec{k_\rho})\diff{\Vec{k_\rho}}
.\end{equation}

We stuck to the formulation in power spectrum as it is more practical in our chosen framework, in particular, to generate the density fluctuation random fields (see Sect.~\ref{sec:density-fluctuations-as-grf}).

\begin{table}
    \centering
    \begin{tabular}{c|c|c|c}
    \hline
        Region & Radius & Min scale & Max scale \\
        \hline
        \hline
        (I) & $0 < r < R_{500}/10$ & $0.02 \, R_{500}$ & $0.10 \, R_{500}$ \\
        (II) & $R_{500}/10 < r < R_{500}/4$ & $0.02 \, R_{500}$ & $0.23 \, R_{500}$ \\
        (III) & $R_{500}/4 < r < R_{500}/2$ & $0.02 \, R_{500}$ & $0.43 \, R_{500}$\\
        (IV) & $R_{500}/2 < r < R_{500}$ & $0.02 \, R_{500}$ & $0.86 \, R_{500}$ \\
        \hline
    \end{tabular}
    \caption{Regions used in the analysis, minimum and maximum scale achievable in Fourier space according to Nyquist–Shannon theorem}
    \label{tab:region_definition}
\end{table}

\subsection{Density fluctuations as a Gaussian random field}
\label{sec:density-fluctuations-as-grf}
The next step in the analysis is to properly relate $\mathcal{P}_{2D, \Delta}$ to the 3D power spectrum of density fluctuations. Previous work based on this method uses a proportional relationship between the 2D and 3D spectra, involving the Fourier transform of the assumed emissivity \citep[see Eq.~11 in ][]{churazov_x-ray_2012}. This approximation is valid as long as the surface brightness can be considered constant, but is no longer valid when considering large areas with strong gradients and important variations in the surface brightness. Moreover, this approach does not consider the stochastic nature of this observable and the associated sample variance. To address this, we chose to consider density fluctuations as a random field. Since many works suggest deep connections between density fluctuations and turbulent processes \citep{zhuravleva_relation_2014, gaspari_relation_2014, simonte_exploring_2022}, we  modelled the density fluctuation as a Gaussian random field (GRF) with a 3D power spectrum, $\Bar{\mathcal{P}}_{3D, \delta}$, noted with a bar to distinguish it from the 3D power spectrum measured from a single realisation of the random field. By noting $\left<.\right>$ the averaging operator over realisations,
it is defined as: 

\begin{equation}
    \Bar{\mathcal{P}}_{3D, \delta}(\Vec{k_r}) \delta^D(\vec{k_r}+ \Vec{k_r'})  = \left<\tilde{\delta} (\vec{k_r}) \tilde{\delta} (-\vec{k_r'})\right> = \left<|\tilde{\delta} (\vec{k_r}) |^2\right>
.\end{equation}

The link between turbulence and density fluctuations suggests both their power spectra take a similar form. As such, we adopted the simplest possible turbulent model as a Kolmogorov cascade. We chose the following functional form as proposed by \cite{zuhone_mapping_2016}, which has no direction dependence under the isotropic hypothesis: 

\begin{equation}
\Bar{\mathcal{P}}_{3D, \delta}(k)= \sigma^2 \frac{e^{-\left(k/k_{\text{dis}}\right)^2} e^{-\left(k_{\text{inj}}/k\right)^2} k^{-\alpha} }{\int 4\pi k^2 \diff k \, e^{-\left(k/k_{\text{dis}}\right)^2} e^{-\left(k_{\text{inj}}/k\right)^2} k^{-\alpha} }
\label{eq:p3dmodel}
,\end{equation}

where $k_{\text{inj}}$ and $k_{\text{disp}}$ are, respectively, the injection and dissipation scales, $\alpha$ is the inertial range spectral index, and $\sigma_\delta^2$ is the variance of fluctuations. For ease of understanding, we used spatial scales of injection rather than frequency scales, defined as $\ell_{\text{inj},\text{dis}} = 1/k_{\text{inj},\text{dis}}$ and expressed in units of $R_{500}$. Since the dissipation scale \citep[of order $\gtrsim 10^{-3} ~R_{500}$,][]{lazarian_turbulence_2015} is expected to be much lower than the spatial resolution of our images ($\sim 10^{-2} ~R_{500}$ at the average redshift of the X-COP clusters), we set it to $10^{-3} ~R_{500}$ for all X-COP clusters. This choice has little effect on the other parameters of the spectrum, since the exact value of the dissipation length will very marginally affect the normalisation of the spectrum and, therefore, $\sigma_\delta$. The GRF hypothesis is valid as long as the fluctuations within the clusters can be represented by an isotropic homogeneous field. However, as clusters are dynamical objects, they are subject to, for instance, merging, accretion, or sloshing events at the centre, which will induce residuals that cannot be described by a GRF. This assumption is discussed further in Sect.~\ref{sec:discussion-grf}.
\subsection{Optimal definition and observable}
\label{sec:definition_aperture}
As seen in Sect.~\ref{sec:mean-profile}, the definition of fluctuation and aperture size (and, more generally, the shape of the mask we choose) has various implications with regard to the signal we measure. For example, the fluctuation map, $\Delta_{\rm rel}$, obtained through a ratio is expected to over-represent the Poisson noise for radii $\sim R_{500}$, and, conversely, for the maps, $\Delta_{\rm abs}$, obtained with a subtraction, the very low significance of the signal at this distance to the centre does not provide additional information. To quantify the comparison between absolute and relative approaches, we used the numerical simulations (further explained in Sect.~\ref{sec:sbi}) to study the signal-to-noise ratio (S/N) that can be obtained in each case. 

We simulated 100 mock surface brightness maps for density fluctuations with $\sigma_\delta = 0.32$, $ \ell_{\text{inj}}= 0.05 R_{500}$, $\alpha = 11/3$ (as defined in Eq.~\ref{eq:p3dmodel}) and computed the associated $\mathcal{P}_{2D}$ in the regions defined in Table \ref{tab:region_definition}. The maximum scale was defined using Nyquist-Shannon criterion, which is half of the highest scale accessible in the mask. For a ring of inner radius, $R_{\min}$, and outer radius, $R_{\max}$, it is given by $(R_{\max}^2-R_{\min}^2)^{1/2}$. To quantify the valuable information in the power spectrum, we compared it to the power spectrum that is obtained by switching off density fluctuations, $\mathcal{N}_{\text{2D}}$. This quantity represents the fluctuations which are expected when observing a perfect cluster at rest, without any density fluctuations. It can be obtained by forward modelling the power spectrum, but with a zero-normalisation density fluctuation field,  accounting for the dispersion due to the mean model fit and the Poisson noise by computing it for models drawn from the posterior distribution and making various realisation of the count image. We define the S/N of the power spectrum as follows for each realisation of the mock density fluctuations:

\begin{equation}
\label{eq:snr-definition}
    \text{S/N}_{\Delta} \overset{\operatorname{def}}{=} \frac{\mathcal{P}_{\text{2D, }\Delta}}{\mathcal{N}_{\text{2D, }\Delta}} ;\quad \Delta\in \{\text{abs, rel}\}
.\end{equation}

\begin{figure}
    \centering
    \includegraphics[width=\hsize]{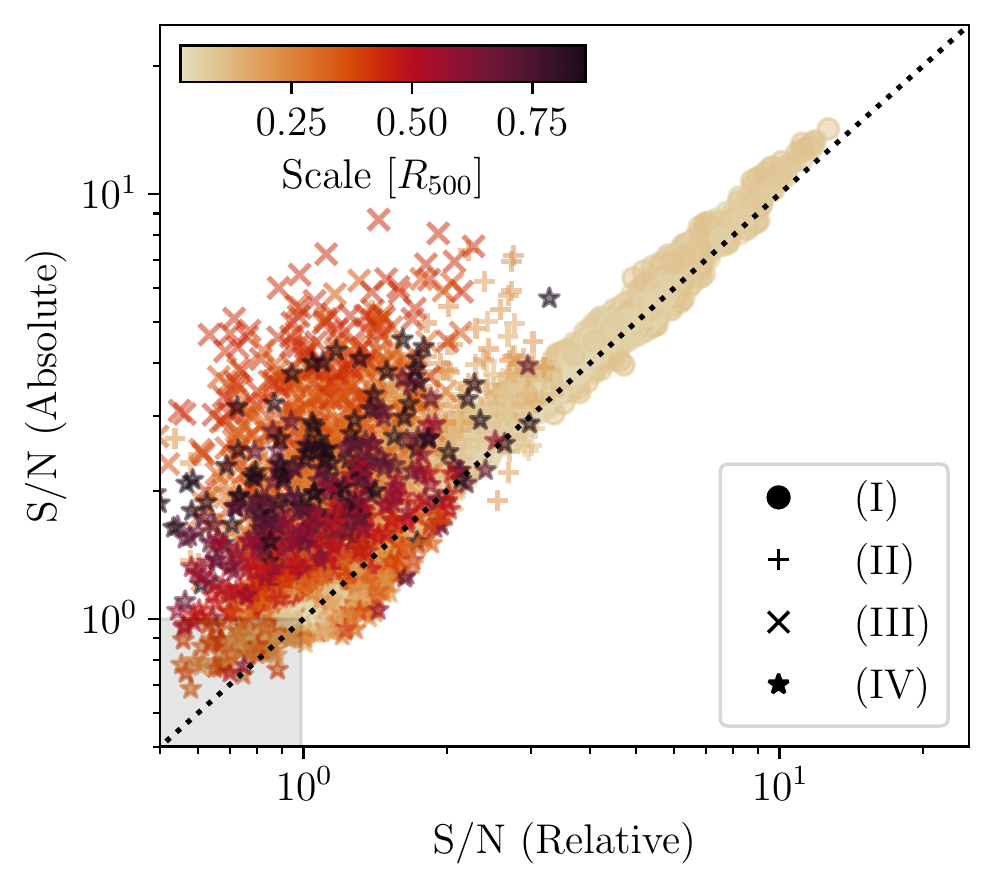}
    \caption{Comparison of the S/N values obtained using the absolute and relative definition for surface brightness fluctuations, for different spatial scales and in the four regions defined in Table~(\ref{tab:region_definition}). The comparison is made for 100 mock spectra using the best-fit model and the exposure map of A3266, with density fluctuation parameters fixed at $\sigma_{\delta} = 0.32$, $ \ell_{\text{inj}} =0.05 R_{500}$ and $\alpha = 11/3$. The grey shaded area represents points with signal below noise level.}
    \label{fig:snr_comparison}
\end{figure}

This quantity can trace the excess of surface brightness fluctuation when compared to what is expected with the cluster emissivity and Poisson noise. In Fig. \ref{fig:snr_comparison}, we show the comparison between the S/N obtained with the absolute method and the relative method, for the four regions of analysis and for several spatial scales. For the most central regions, the S/N metric values are high and both methods perform equivalently well. For the outermost regions, the absolute method performs better than the relative method for large spatial scales, and both perform equally well on smaller scales. With this in mind, we favour the absolute method in the following. 

\subsection{Error budget}
\label{sec:error_budget}

Constraining density fluctuations requires controlling the error budget associated with surface brightness fluctuations. The Bayesian characterisation of the cluster emission (see Sect.~\ref{sec:mean-profile}) allows us to estimate the uncertainties related to the mean surface brightness best-fit model, in the form of posterior distribution of the model parameters. The other major source of error is the Poissonian nature of the image counts, which introduces constant power at all spatial scales.

Observables related to the velocity (or density fluctuation) field are subject to an additional variance intrinsically linked to their stochastic nature. The measured statistics of the density fluctuation field $\delta$ will vary from one realisation to another, due to finite size effects introduced by the cluster itself and the limited field of view. Any quantity that depends on $\delta$ will therefore be affected by this 'sample variance', including the surface brightness fluctuation power spectrum, $\mathcal{P}_{\text{2D}, \Delta}$. This variance  increases at higher spatial scales due to the lack of large-scale modes and becomes predominant in the error budget at spatial scales of $\gtrsim 50$ kpc in the structure functions of \cite{cucchetti_towards_2019}. In the following, we implement a forward modelling approach which, for the first time, accounts for the sample variance induced by the finite sizes of galaxy clusters and limited fields of view.

\subsection{Simulation-based inference}
\label{sec:sbi}

The Bayesian framework is well suited to the integration of the error budget, which should be reflected in the likelihood of our problem. In a classical Bayesian inference, we seek to obtain the posterior distribution $p(\Vec{\theta}|\Vec{x})$ of the parameters $\Vec{\theta}$ of our model given an observation $\Vec{x}$, by inverting the likelihood distribution $p(\Vec{x}|\Vec{\theta})$, namely, the probability distribution of an observation given the parameters. In practice, the posterior distribution is sampled by estimating the likelihood for many parameters with Markov chain Monte Carlo (MCMC) methods. This is the process we describe in Sect. \ref{sec:mean-profile}, assuming a Poisson distribution in each pixel, which allows us to define a corresponding likelihood. However, in the case of surface brightness fluctuations, there is no simple way to define an analytical and closed form for the likelihood function of a spectrum $\mathcal{P}_{\text{2D}, \Delta}$ given the density fluctuation parameters, as we cannot grasp the underlying distribution. Therefore, the Bayesian inference of the $\Bar{\mathcal{P}}_{3D, \delta}$ parameter distributions falls under the scope of approximate Bayesian computation methods, which only require the ability to model observables $\Vec{x}$ (including all the sources of variance one wishes to consider) for any set of parameters $\Vec{\theta}$ instead of an analytical likelihood function.

In this work, we chose to use a neural network (NN) that learns the likelihood of our problem, based on simulated observations. First, we mocked many couple of observables $\vec{x}_i$ using parameters $\vec{\theta}_i$ drawn from the prior distribution $p(\Vec{\theta})$ selected for the inference. We then trained the neural network to learn an estimator $q(\Vec{x}|\Vec{\theta})$ of the likelihood $p(\Vec{x}|\Vec{\theta})$ using the couples $(\vec{x}_i, \vec{\theta}_i)$. This approximation is accomplished by adjusting a normalising flow, which is formally an invertible bijection between two probability distributions. The normalising flow itself is built using several layers of density estimators, implemented by  masked autoencoders \citep{germain_made_2015}, so we ended up training a masked autoregressive flow \citep{papamakarios_masked_2017, papamakarios_sequential_2019}. Once the training had converged, the posterior distribution was sampled by evaluating the likelihood with the previously trained flow, sampling for $p(\Vec{\theta}|\Vec{x}) \propto q(\Vec{x}|\Vec{\theta})p(\Vec{\theta})$ with a classical MCMC approach.

We  used the \texttt{sbi} library \citep{tejero-cantero_sbi_2020} implementation to perform this simulation-based inference. The prior distribution used for the parameters of Eq. \ref{eq:p3dmodel} are shown in Table~\ref{tab:all_parameters}. We generated mock X-ray images with surface brightness fluctuations by projecting an emissivity field with an additional density fluctuation field, as highlighted in Eq. \ref{eq:sb_def}. To achieve this, we defined emissivity cubes dimensioned as the X-COP data in the $(x, y)$ directions and with the same spatial resolution, but expanded to $\pm 5 R_{500}$ along the line of sight. We used the same 3D models as in  Sect.~\ref{sec:mean-profile}, that is:\ density, temperature, cooling, and ellipticity, along with their best-fit parameters, as well as the individual properties of each observation, that is, the exposure maps, background maps, and $N_H$ maps, to mock the expected rest surface brightness of each cluster in the X-COP sample. The density fluctuations $\delta$ are generated by drawing a single realisation of a Gaussian random field, assuming a Kolmogorov-like power spectrum, as defined in Eq. \ref{eq:p3dmodel}. These fluctuations are co-added to the modelled emissivity and projected along the line of sight in the same standard as Eq. \ref{eq:sx_ref_and_perturbed}. We then choose to use the $\text{S/N}_{\Delta}$, as defined in Eq.~\ref{eq:snr-definition} as an observable, since this quantity reduces the proper contribution of each cluster by dividing by the spectrum of the expected emission without density fluctuations and best represents the excess power due to their presence. It is evaluated on 20 logarithmically spaced scales between the minimum and maximum scale accessible in each region, as defined in Table~\ref{tab:region_definition}. Generating 300,000 realisations of the fluctuation field allows us to properly learn the likelihood function for each region of analysis. The posterior parameters are then sampled using the NUTS sampler, as implemented in the \texttt{pyro} library \citep{bingham_pyro_2019}. We checked that the number of simulations is sufficient by testing the neural network asserting proper convergence of the recovered parameters for a NN trained with an arbitrary number of simulations. We do not observe any significant improvement in training for more than 100,000 simulations.

\begin{figure*}[ht]
\subfigure{
\includegraphics[width=0.47\textwidth]{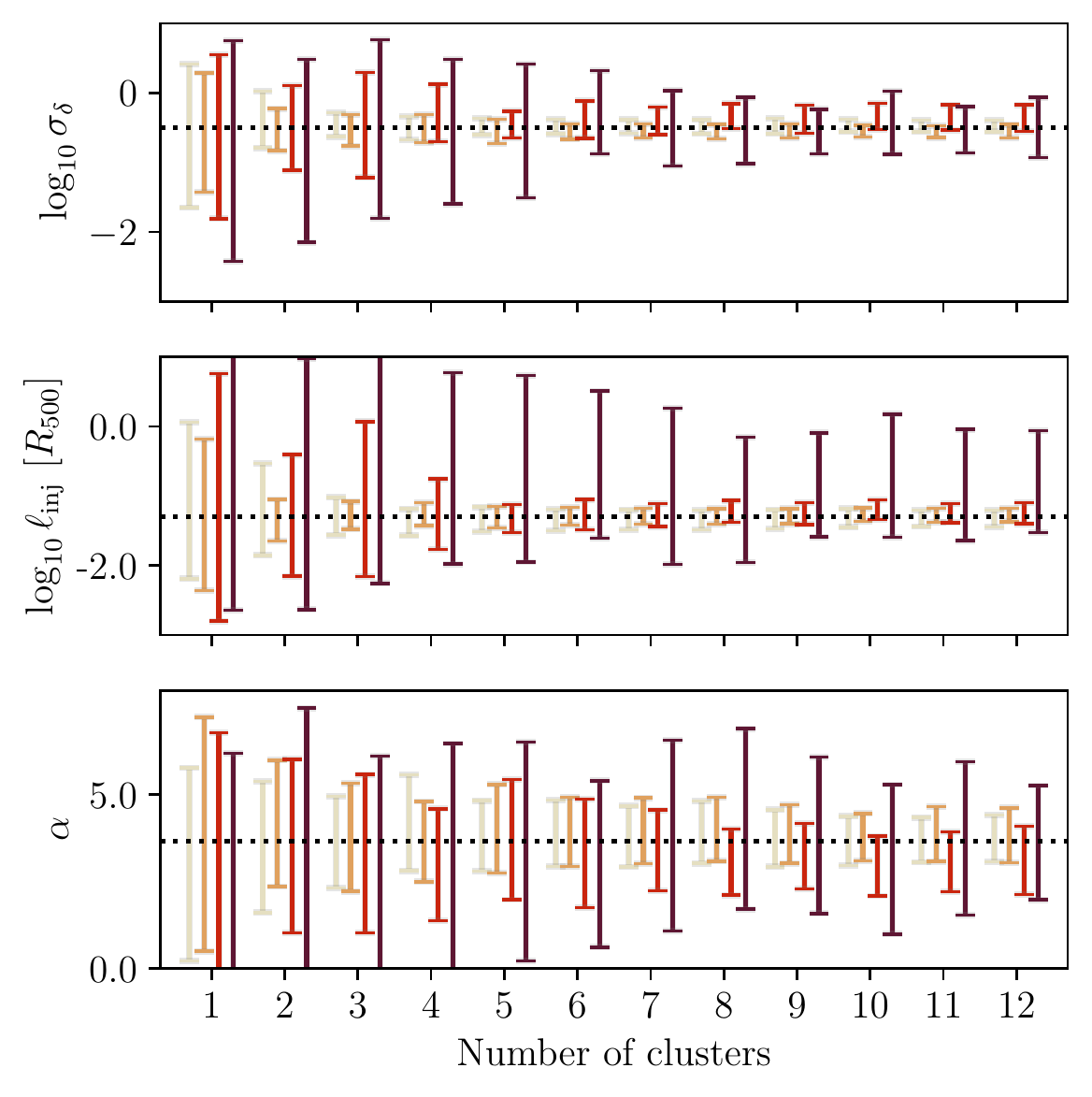}}
\label{fig:subfig:joint_convergence} 
\hspace{0.2in}
\subfigure{\includegraphics[width=0.47\textwidth]{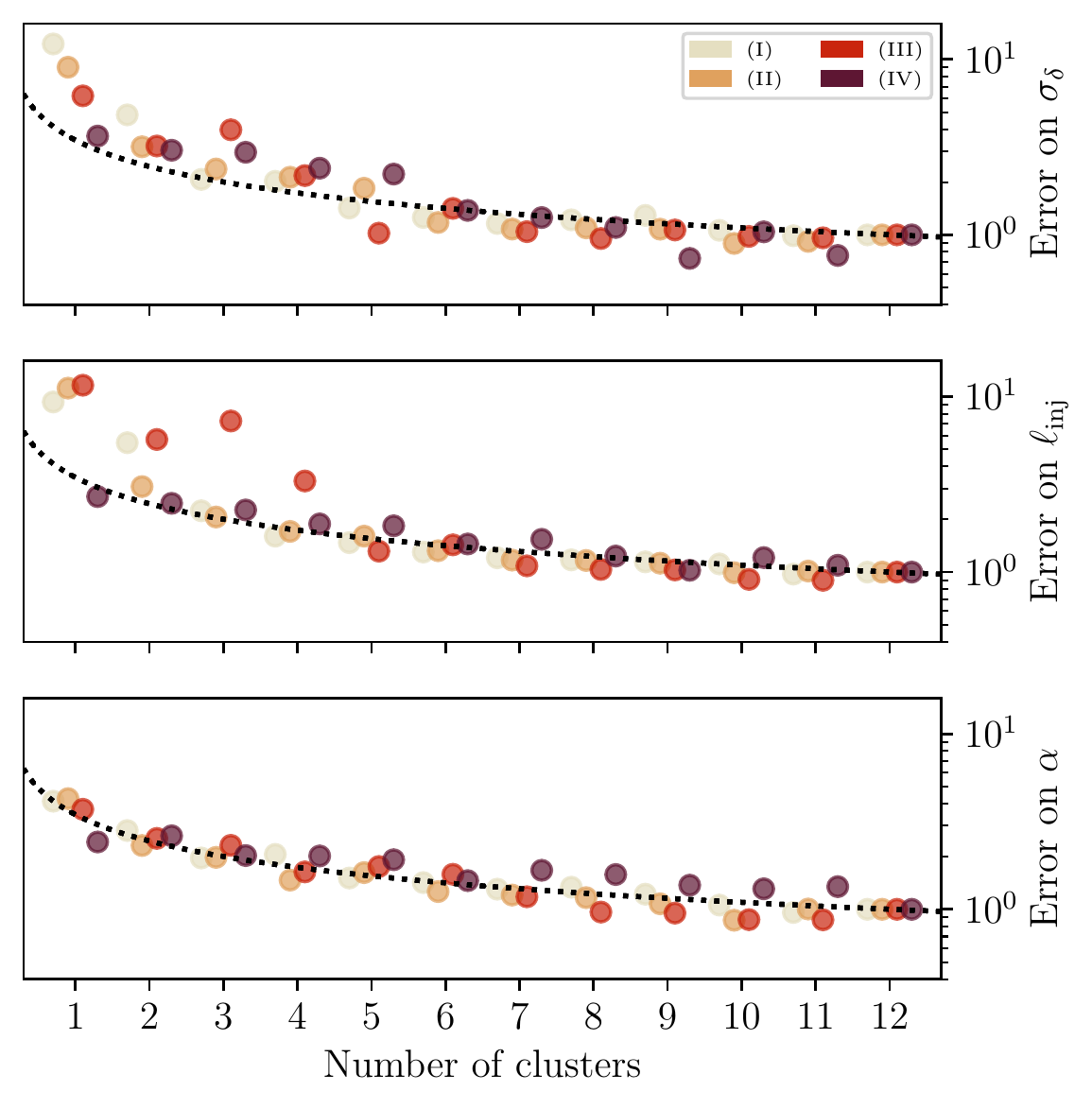}}
\label{fig:subfig:joint_error}
  \caption{
  Estimation of the posterior parameters and qualitative behaviour of the associated errors for mock observations with known parameters with an increasing number of clusters in the joint likelihood.  For each point, we randomly draw $N$ clusters ten times, compute the posterior parameter distributions and average the mean and standard deviation to alleviate the selection effect. The parameters are in each of the four regions as defined in Table~(\ref{tab:region_definition}). Left panel : mean and standard deviation of the parameters estimated for an increasing number cluster in the joint likelihood. The black line represents the true parameters. Right panel : error on each parameter compared to the number of cluster used to find the parameters, rescaled so that it equal to 1 for $N=12$. The black line shows the expected $N^{-1/2}$ behaviour.}
     \label{fig:sbi_sample_convergence}
\end{figure*}

\subsection{Validation}

To validate our methodology, we applied it to mock clusters with known power spectrum and statistical properties similar to the actual X-COP clusters with known density fluctuation parameters and compared them with the parameters inferred via the method outlined in Sect.~\ref{sec:sbi}, in the multiple regions defined in Table \ref{tab:region_definition}. We performed a validation by producing, for each mock cluster, a realisation of the density fluctuations with $\sigma_{\delta} = 0.32$, $ \ell_{\text{inj}} =0.05~R_{500}$ and $\alpha = 11/3$ as the parameters to recover, along with 300000 simulations for parameters drawn from the priors in Table~\ref{tab:all_parameters}. We tackled the effect of sample variance by defining a joint likelihood for all our clusters, which is equivalent to stating that each cluster is an individual observation of the same fluctuation process. Since all spatial coordinates are scaled to $R_{500}$, we can define a joint likelihood as follows:

\begin{equation}
\label{eq:joint_likelihood}
    \log \mathcal{L}_{\text{joint}} = \sum_{\text{cluster}} \log \mathcal{L}_{\text{cluster}}
,\end{equation}

where $\mathcal{L}_{\text{cluster}}$ is the likelihood estimated with the previously trained NN for each cluster in the X-COP sample. The effectiveness of this approach is illustrated in Figure (\ref{fig:sbi_sample_convergence}). In this case, we randomly drew N clusters ten times to alleviate cluster selection effects on the reconstructed parameters. We computed the joint parameters for each draw and average them. The error on each parameter decreases with the number of clusters used to define the likelihood, behaving as a $1/\sqrt{N}$ dependence is  expected to, when adding independent observations.

\section{Results}
\label{sec:results}

The data accompanying this analysis is available online on the repository\footnote{\url{https://github.com/renecotyfanboy/turbulence_xcop}} associated with this article.
\subsection{Mean profile and fluctuation maps}

\begin{figure*}
\includegraphics[width=\hsize]{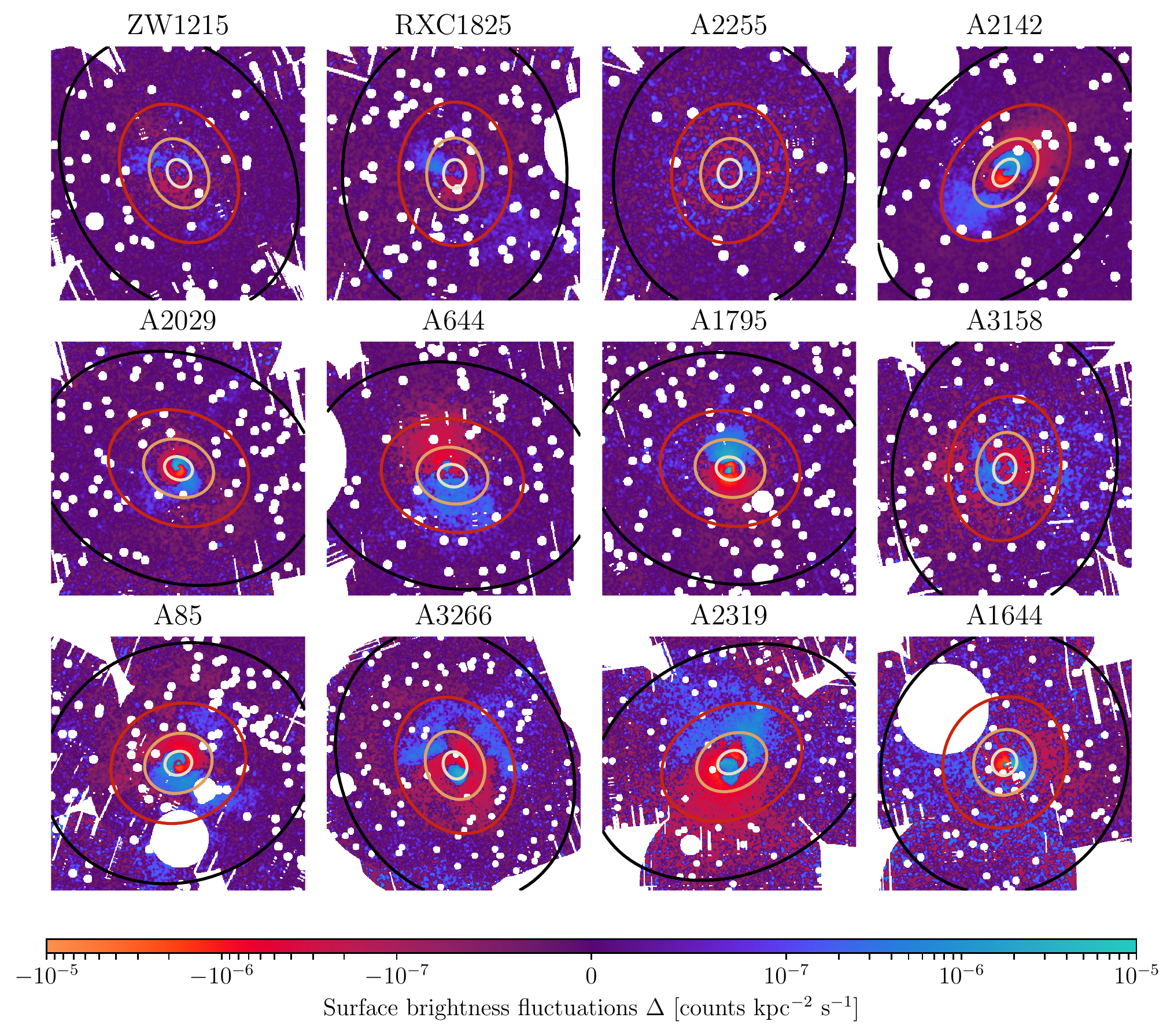}
\caption{Surface brightness fluctuation maps, $\Delta_{\rm abs}$, from the absolute method (see Sect.~\ref{sec:definition_aperture}) for the X-COP cluster sample. The successive contours represent the annular regions I, II, III, and IV with outer radii of 0.1, 0.25, 0.5 and 1 $R_{500}$, respectively. For display  purposes, the images are filtered by a Gaussian kernel of 7.5". Each image is $\sim 2.5$ Mpc on a side. The images are ranked in ascending order of disturbance for the cluster as measured by the $C_Z$ coefficient (see Sect.~\ref{sec:morpho_indicators}).}
\label{fig:fluctuation-maps}
\end{figure*}

The marginalised parameter mean and 1$\sigma$ dispersion are displayed in Table~\ref{tab:mean-model-density} and Table~\ref{tab:mean-model-morpho}. We ensure proper convergence of the Markov chains by computing the $\hat{R}$ statistic, as proposed in \citep{vehtari_rank-normalization_2021} and checking that $\hat{R} < 1.01$. The parameters so found are compatible with those obtained by \cite{ghirardini_universal_2019}, which are  representative of X-COP clusters. The corresponding fluctuation maps computed with Eq. (\ref{eq:sb-fluc-diff}) are shown in Fig. (\ref{fig:fluctuation-maps}). Most of these fluctuation maps show non-Gaussian features in their residual. Sub-mergers or post-mergers can be identified in A2319 and A3266. Spiral-shaped structures in the inner regions of A85, A2029, or A2142 indicate the presence of gas sloshing. 

\subsection{Density fluctuation power spectrum parameters}

\begin{figure*}
\includegraphics[width=\hsize]{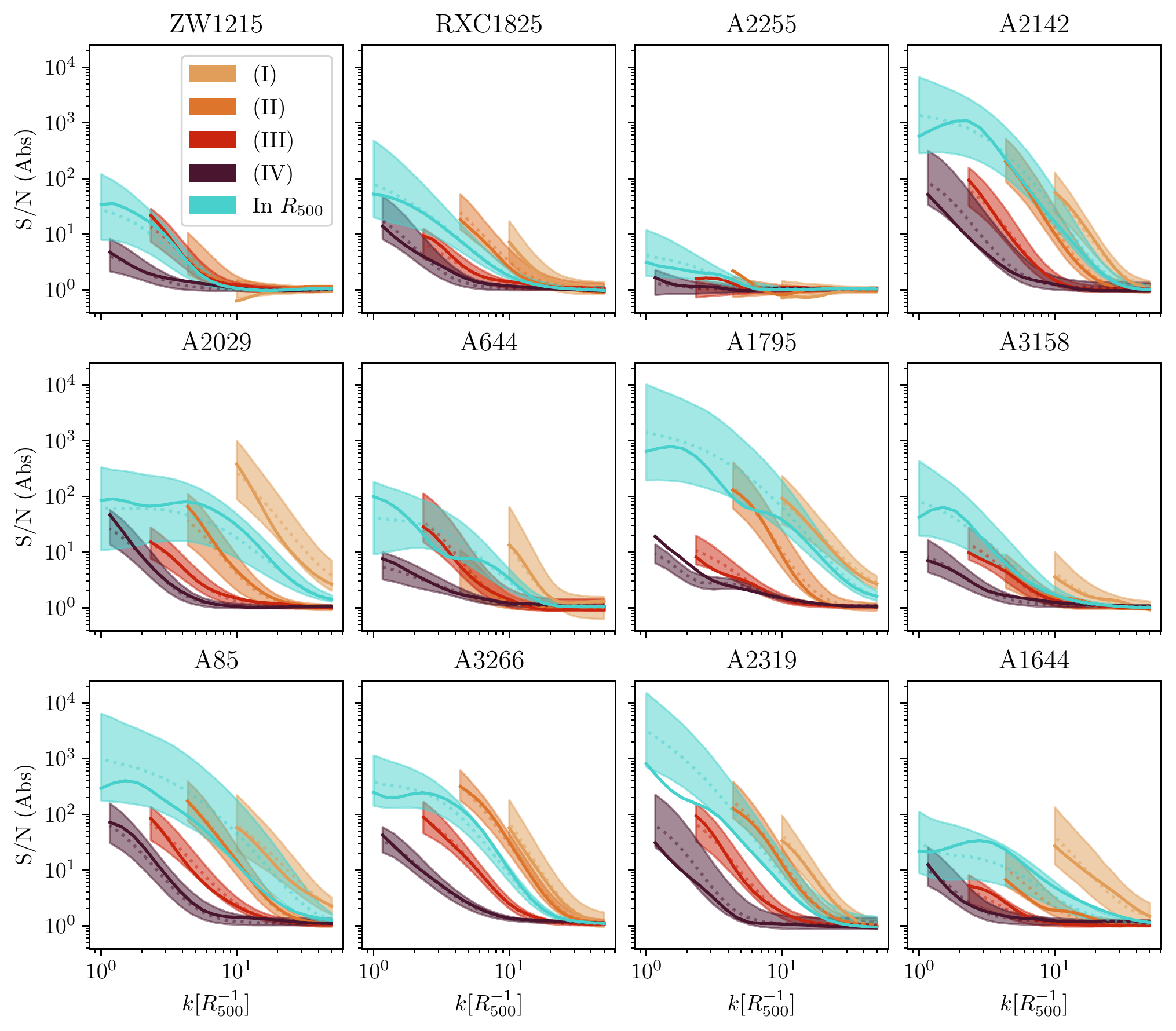}
\caption{Power spectrum S/N as defined in Eq.~\ref{eq:snr-definition} and measured on the surface brightness fluctuation maps with the absolute method (see Sect.~\ref{sec:definition_aperture}) in the four regions as defined in Table~(\ref{tab:region_definition}) and inside $R_{500}$ for each cluster in the X-COP sample. Dotted line and shaded envelopes show the posterior distribution median and 16\textsuperscript{th}-84\textsuperscript{th} percentiles, accounting for the complete error budget for each cluster. The plain line represents the measured observable for each cluster in X-COP. The plots are ranked in ascending order of disturbance for the cluster, as measured by the $C_Z$ coefficient (see Sect.~\ref{sec:morpho_indicators}).}
\label{fig:2d-power-spectra}
\end{figure*}

We calculated, for each fluctuation map, the S/N between the surface brightness fluctuations and the fluctuations expected when only Poisson noise is present (see Sect~\ref{sec:definition_aperture}) for the four different aforementioned regions, which is shown as a plain line in Fig.~\ref{fig:2d-power-spectra}. Separation of the fluctuation maps into several regions provides us a way to discriminate between the different processes at work in the ICM. The central region is expected to be dominated by the presence or absence of a cool core, as well as by AGN-feedback. The second ring should be rather sensitive to the presence or not of sloshing in the core of the cluster. It is worth noting that a sloshing extending at least out to $R_{500}$ has been observed in various clusters, in particular A2142 \citep{rossetti_abell_2013}. The two outer rings should be related to the fluctuations induced by the larger dynamical assembly of the cluster, that is, by the accretion from the cosmic web and infalling halos. We also computed the S/N for the whole region inside $R_{500}$ to get an average statistic on the clusters. Using the methodology presented in Sect.~\ref{sec:sbi}, we constrained the parameters of the density fluctuation power spectrum. 

\begin{figure*}
\centering
\subfigure{
\includegraphics[width=0.47\textwidth]{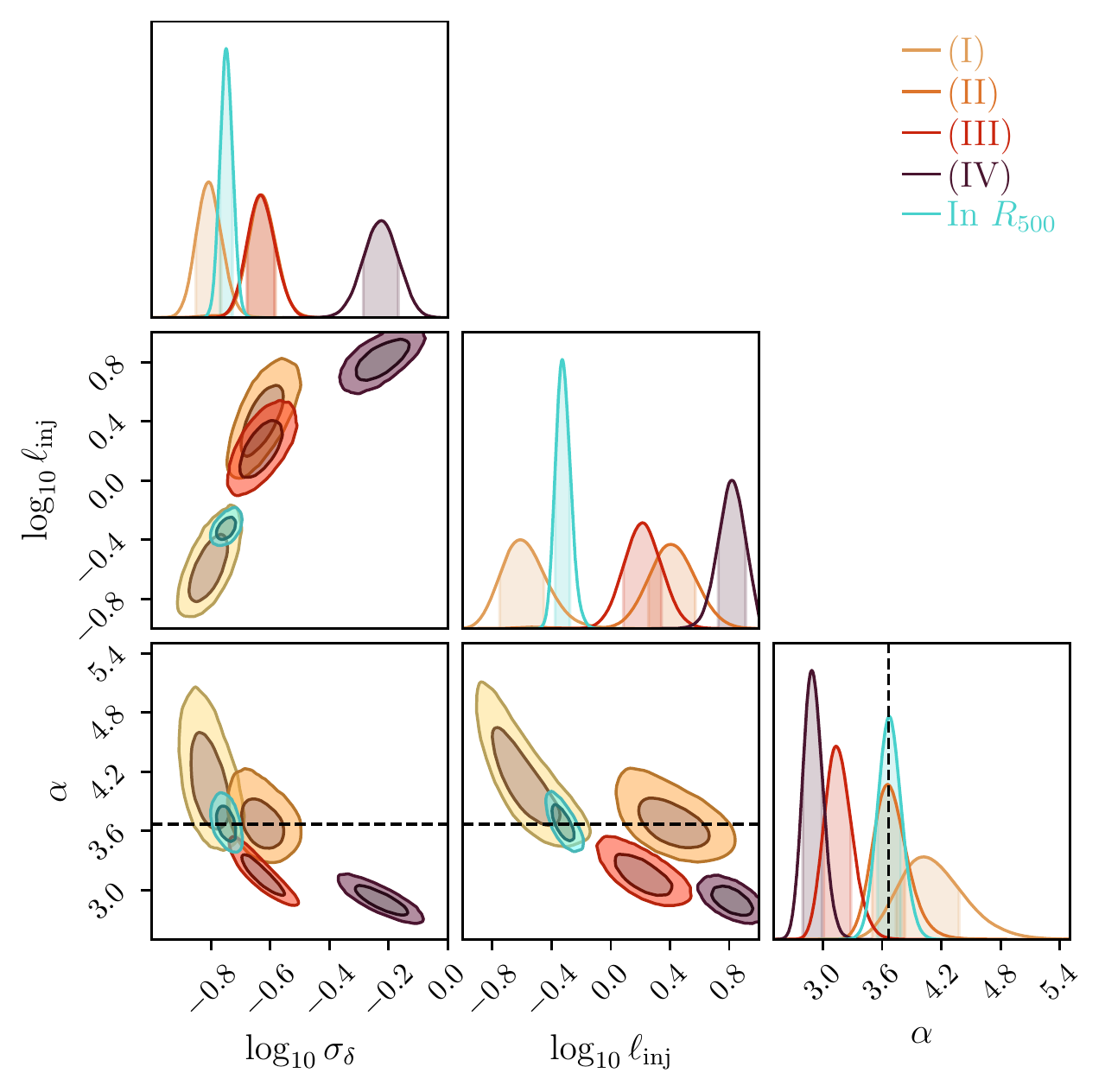}}
\label{fig:subfig:corner} 
\hspace{0.2in}
\centering
\subfigure{
\includegraphics[width=0.47\textwidth]{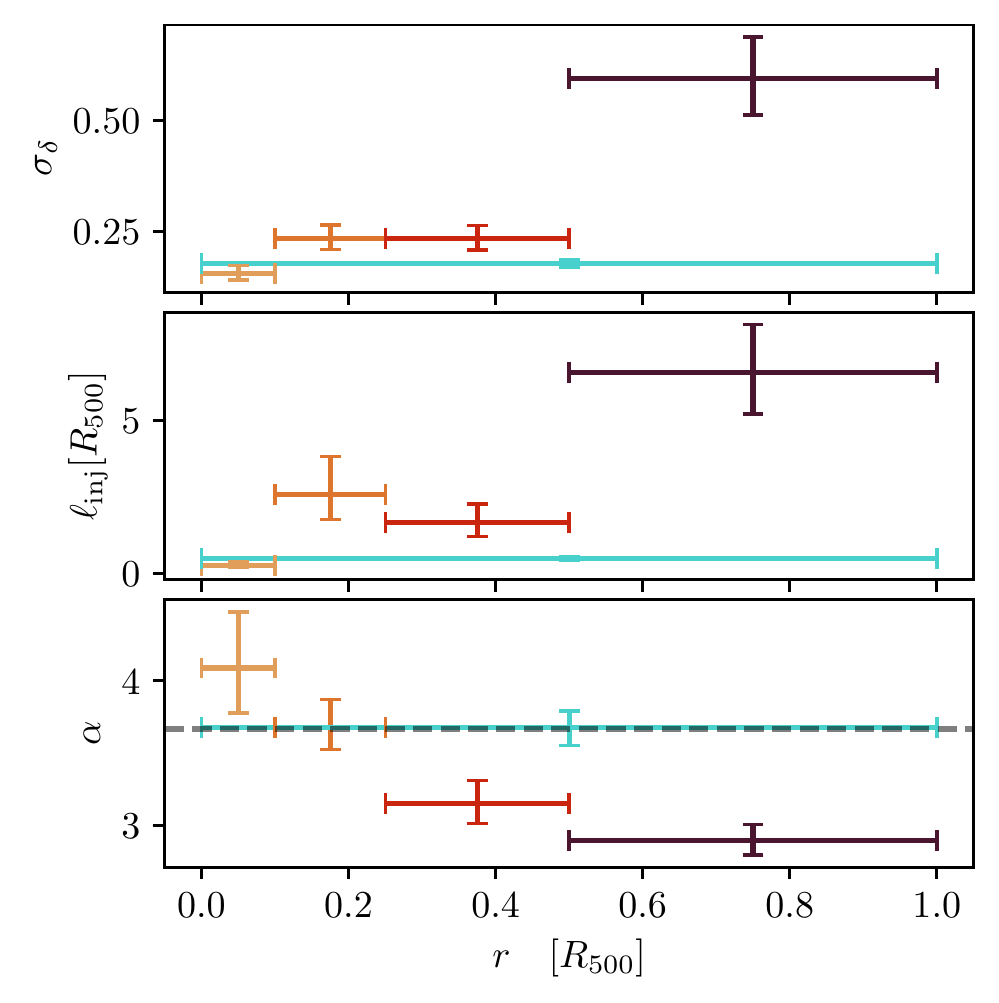}}
\label{fig:subfig:joint} 
\caption{Posterior distributions (\textit{left}) and radial evolution (\textit{right}) of the standard deviation, $\sigma_\delta$, the injection scale, $\ell_\text{inj}$, and the spectral index $\alpha$ of the density fluctuation power spectrum parameters, jointly evaluated on the whole X-COP cluster sample in the four regions as defined in Table~(\ref{tab:region_definition}) and within  $R_{500}$. The colour scale matches the four regions of interest, and the turquoise distribution represents the entire $R_{500}$ region. The black dashed line represents the expected 11/3 index from Kolmogorov-Oboukhov theory.}
\label{fig:corner_plot}
\end{figure*}

The prior distribution and posterior median and 16\textsuperscript{th}-84\textsuperscript{th} percentiles are highlighted in Table~\ref{tab:all_parameters}. The posterior probabilities for the free parameters of the 3D power spectrum derived from the joint fit over the whole X-COP sample are shown in the 'corner plot' of Fig.~\ref{fig:corner_plot}. We also checked the proper convergence by assessing that $\hat{R} < 1.01$ for each parameter. By folding the posterior parameter distributions into our model, we obtain the resulting 2D observables accounting for the complete error budget (including the sample variance), which are shown in Fig.~\ref{fig:2d-power-spectra}. The inner regions are marked by a high slope and low injection scale, indicating the preponderance of finite-sized structures in these regions, which corroborates with the presence of sloshing and assemblage artefacts in our fluctuation maps in Fig.~\ref{fig:fluctuation-maps}. The estimation for all fluctuations inside $R_{500}$ should be more resilient to the central region artefacts, since these structures can be described as marginal realisations of the random field when compared to the fluctuations in the external regions. The parameters in $R_{500}$ jointly estimated on the whole sample converge to a normalisation of $\sigma_\delta \sim 0.18$, an injection of the order of $\ell_{\text{inj}}\sim  0.4 R_{500}$, with a slope of $\alpha \sim 11/3$, compatible with a pure hydrodynamical Kolmogorov cascade. This is comparable to what was determined for the Coma cluster by \cite{schuecker_probing_2004}, studying the pressure fluctuation at spatial scales between 40 and 90 kpc, as well as in \cite{zhuravleva_gas_2015} for the Perseus cluster.

\section{Discussions}
\label{sec:discussion}

\subsection{Correlation with morphological indicators}
\label{sec:morpho_indicators}
The processes that lead to density fluctuations are expected to be related to the dynamical state of the clusters. Indeed, merger events or sloshing in the centre are intrinsically linked to the dynamic assembly of clusters and can be characterised by morphological and dynamical indicators. We consider various indicators that are well correlated with the dynamic state of the cluster \citep{lovisari_x-ray_2017, campitiello_chex-mate_2022} and we seek correlations with the density fluctuations parameters. We chose to use the following indicators:1) the concentration parameter $c_{\text{SB}}$ compares the central emission of the cluster to its total emission. We used the definition from \cite{campitiello_chex-mate_2022}, namely,  the ratio between the surface brightness inside 0.15 $R_{500}$ and $R_{500}$; 2) the centroid shift $w$ relates the variation of the distance of the peak of luminosity and the centroid of the emission for a varying aperture. We used the definition and determination of this parameter from \cite{eckert_constraints_2022}. 3) the Gini coefficient $G$ measures the disparity of surface brightness in the pixels of the image. This coefficient ranges from 1 to 0 for a perfectly homogeneous and perfectly inhomogeneous distribution, respectively, and is computed using the following formula:
    $$
    G = \frac{\sum_{i}^{n} \sum_{j}^{n} \left| s_{i} - s_{j} \right|}{2n^2\Bar{s}},
    $$
     where $s_{i,j}$ is the surface brightness in the $i,j^{\text{th}}$ pixel, $\Bar{s}$ is the mean surface brightness and $n$ the total number of pixels used to compute $G$; 4) finally, the Zernike moments, $Z_i$, relate the decomposition of the image on a basis of orthogonal polynomials that allow us to efficiently describe the morphology of the cluster. Here, we specifically use the Zernike coefficient, $C_Z$, that reflects the asymmetry  of the clusters:  
     $$C_Z = \sum_{n, m \neq 0} \sqrt{| Z_n^m |,}$$ 
     where $Z^m_n$ are coefficients obtained from scalar products of the image and a given Zernike polynomial, as described in \cite{capalbo_three_2021}. The gallery of clusters in Fig.~\ref{fig:fluctuation-maps} is sorted by increasing $C_Z$.

\begin{table}[]
    \centering
    \resizebox{\linewidth}{!}{\begin{tabular}{r|rrrr}
\hline
\hline
Name & $c_{\text{SB}}$ & $w (\times 10^3)$ & $G$ & $C_Z$\\[2mm]
\hline
A1644 & $0.114^{+0.002}_{-0.001}$ & $13.761^{+0.085}_{-0.079}$ & $0.699^{+0.001}_{-0.001}$ & $1.0^{+0.034}_{-0.026}$\\[2mm]
A1795 & $0.52^{+0.001}_{-0.001}$ & $2.36^{+0.019}_{-0.019}$ & $0.823^{+0.001}_{-0.001}$ & $0.477^{+0.007}_{-0.008}$\\[2mm]
A2029 & $0.496^{+0.001}_{-0.001}$ & $0.85^{+0.009}_{-0.01}$ & $0.792^{+0.001}_{-0.001}$ & $0.437^{+0.011}_{-0.01}$\\[2mm]
A2142 & $0.408^{+0.001}_{-0.001}$ & $4.51^{+0.03}_{-0.028}$ & $0.771^{+0.001}_{-0.001}$ & $0.391^{+0.004}_{-0.004}$\\[2mm]
A2255 & $0.078^{+0.002}_{-0.002}$ & $31.391^{+0.269}_{-0.261}$ & $0.694^{+0.002}_{-0.002}$ & $0.389^{+0.011}_{-0.011}$\\[2mm]
A2319 & $0.215^{+0.001}_{-0.001}$ & $33.069^{+0.179}_{-0.176}$ & $0.68^{+0.001}_{-0.001}$ & $0.946^{+0.012}_{-0.013}$\\[2mm]
A3158 & $0.212^{+0.001}_{-0.001}$ & $5.87^{+0.039}_{-0.04}$ & $0.708^{+0.001}_{-0.001}$ & $0.523^{+0.01}_{-0.01}$\\[2mm]
A3266 & $0.17^{+0.001}_{-0.001}$ & $30.837^{+0.166}_{-0.165}$ & $0.665^{+0.001}_{-0.001}$ & $0.79^{+0.016}_{-0.015}$\\[2mm]
A644 & $0.348^{+0.002}_{-0.002}$ & $20.971^{+0.151}_{-0.147}$ & $0.772^{+0.001}_{-0.001}$ & $0.457^{+0.007}_{-0.008}$\\[2mm]
A85 & $0.417^{+0.001}_{-0.001}$ & $3.85^{+0.02}_{-0.019}$ & $0.767^{+0.001}_{-0.001}$ & $0.711^{+0.008}_{-0.009}$\\[2mm]
RXC1825 & $0.153^{+0.001}_{-0.001}$ & $8.031^{+0.061}_{-0.062}$ & $0.625^{+0.001}_{-0.001}$ & $0.333^{+0.007}_{-0.007}$\\[2mm]
ZW1215 & $0.205^{+0.001}_{-0.002}$ & $3.731^{+0.03}_{-0.03}$ & $0.703^{+0.001}_{-0.001}$ & $0.328^{+0.009}_{-0.009}$\\[2mm]
\hline
\end{tabular}}
    \caption{Concentration, $c_{\text{SB}}$, centroid shift, $w$, Gini coefficient, $G,$ and Zernkie moment, $C_Z$, computed for the X-COP cluster sample, with variance from the best-fit elliptical radii and Poisson noise.}
    \label{tab:morpho_indicators}
\end{table}

\begin{figure}[ht]
        \includegraphics[width=\linewidth]{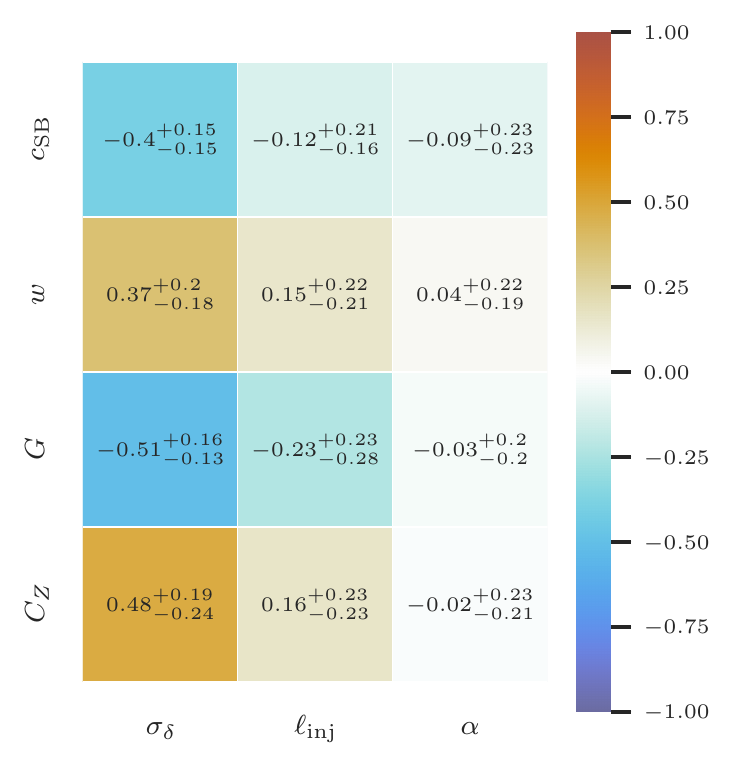}
  \caption{Correlation matrix as measured by the Spearman coefficient between the morphological indicators and the density fluctuation power spectrum parameters. We display the median and difference with the 16\textsuperscript{th}-84\textsuperscript{th} percentiles. The morphological parameters are defined in Sect.~\ref{sec:morpho_indicators} and the density fluctuation parameters are defined in Eq.~\ref{eq:p3dmodel} and evaluated inside $R_{500}$ for each cluster in the X-COP sample.}
     \label{fig:correlation_matrix_sigma}
\end{figure}

We computed these morphological indicators using all the pixels in $R_{500}$, excluding the point sources and assuming spherical symmetry. We accounted for the shot noise by drawing several Poisson realisations of the measured image, with centre position drawn from our best-fit posterior distribution. The resultant morphological indicators are displayed in Table~\ref{tab:morpho_indicators}. We quantified the correlations by drawing 2000 values of each morphological indicators and computing the Spearman correlation coefficient with the same number of density fluctuation parameters drawn from the posterior distributions. In Fig. \ref{fig:correlation_matrix_sigma}, we show the correlation matrix between the density fluctuation parameters measured inside $R_{500}$ and the morphological indicators, as measured by the Spearman correlation coefficient. We observe that the best correlations appear with $\sigma_\delta$ and our 4 morphological indicators. The $\sigma_\delta$ drives the standard deviation and thus the broadness of the density fluctuation distribution. The concentration and Gini are correlated negatively with $\sigma_\delta$, which can be understood as the fact that less concentrated clusters are generally more disturbed and therefore admit a less even distribution of its surface brightness, which goes hand in hand with the emergence of surface brightness fluctuations and thus density fluctuations. Conversely, the positive correlation with the centroid shift and the Zernike moment (which both trace the deviation of the surface brightness from a spherically symmetrical distribution) can be understood as the emergence of density fluctuation and surface brightness will disturb the symmetry of the system and therefore increase the value of these two indicators. We plot the correlation between $\sigma_\delta$ and these parameters along with the best fit as a power-law scaling in Appendix~\ref{app:correlation_morpho}. The range of morphological indicators we use here cannot represent the structure of the fluctuations, so it was to be expected that there would be no significant correlation with the injection scale and the spectral index. Nevertheless, they do reflect the disturbance of the surface brightness, and, in this sense, the correlation with the normalisation of the spectrum, related to the standard deviation of the fluctuations, is self-consistent.

\subsection{Interpretation as gas clumping}
\label{sec:clumping}

\begin{figure}[h]
        \includegraphics[width=\linewidth]{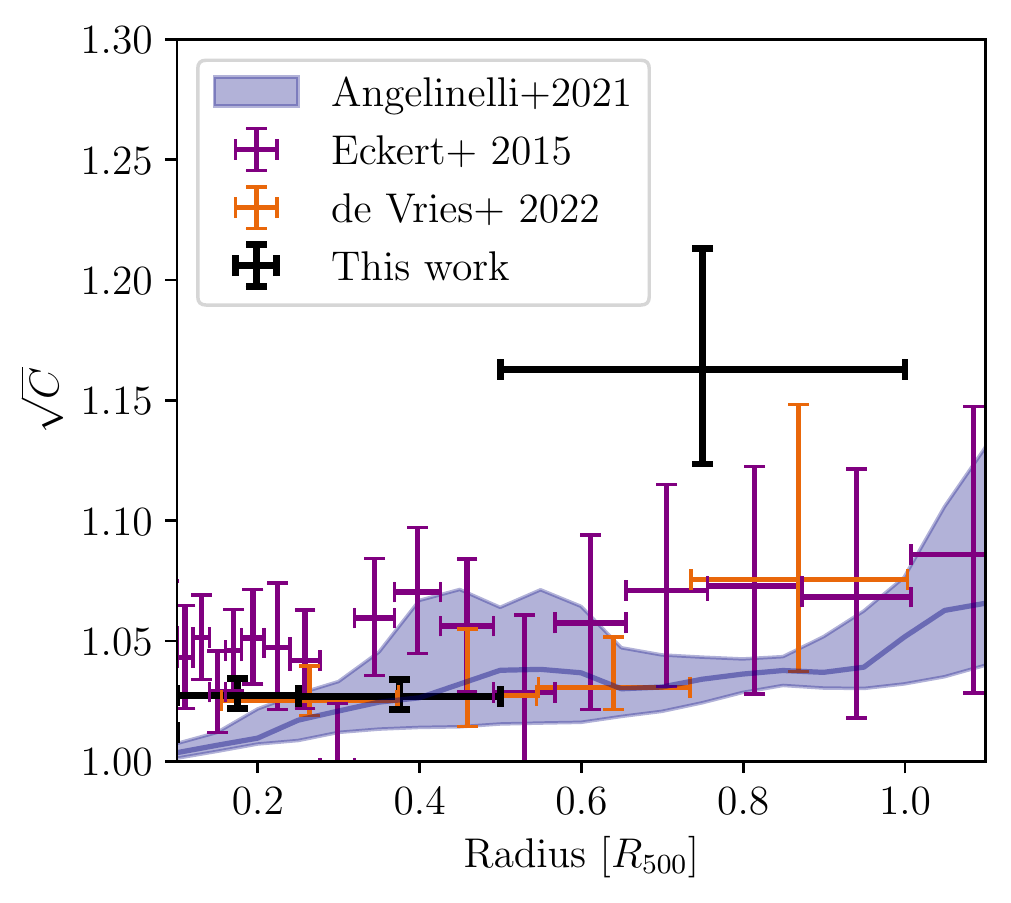}
  \caption{Comparison of the clumping factor obtained with Eq. (\ref{eq:clumping}) for the joint fit across the X-COP sample in each annular region compared to the clumping factor obtained by \cite{eckert_gas_2015} for a sample of 35 clusters, by \cite{devries_chandra_2023} for the Perseus cluster and by \cite{angelinelli_proprieties_2021} for 9 clusters in the ITASCA simulated clusters. The plain line and envelop represent the median and 16\textsuperscript{th}-84\textsuperscript{th} percentiles of the clumping factor profile.}\label{fig:comparaison_clumping}
\end{figure}

Gas clumping refers to the local deviation of the gas density from the average density expected at that location. Throughout this paper, we prefer this definition to the more specific definition of 'accreted substructures', so clumping defines any form of density deviation from the smooth profile. One way of interpreting our density fluctuation measure is in the form of clumping. From the 3D power spectrum of density fluctuations, we can estimate the clumping factor in our analysis regions. Similarly to \cite{zhuravleva_gas_2015}, we define the clumping factor $C$ as follows:

\begin{equation}
\label{eq:clumping}
    C = \frac{\left< n_e^2 \right>}{\left< n_e \right>^2} = 1 + \sigma_\delta^2
,\end{equation}

where $\sigma_{\delta}$ is the normalisation of the density fluctuation power spectrum as defined in Eq.~\ref{eq:p3dmodel}. In Fig.~\ref{fig:comparaison_clumping}, we show the clumping factor we obtained for the regions (II), (III), and (IV) along with a comparison with the approximate clumping factor derived for 35 clusters in \cite{eckert_gas_2015}, the clumping factor estimated in the Perseus cluster with a model-free approach by \cite{devries_chandra_2023}, and the clumping factor estimated by \cite{angelinelli_proprieties_2021} for the Itasca simulated cluster sample. We see a good agreement between the three results in the inner regions. We attribute our lower values to the circularly symmetric profiles used in \cite{eckert_gas_2015}. Furthermore, we see a good agreement with the factor from \cite{devries_chandra_2023}, which was computed from surface brightness fluctuations determined without a spatial surface brightness model (in contrast to what is done in this paper). This may point to the fact that accounting for the ellipticity of the surface brightness minimises the biases introduced by the arbitrary choice of model (discussed further in Sect.~\ref{sec:mean_model_dependency}). However, we observe a tension in the outer region (radii ranging from 0.5 to 1 $R_{500}$), which is presumably due to a change in the nature of the measured fluctuations. This is further discussed in Sect.~\ref{sec:discussion-grf} and in Sect.~\ref{sec:discussion_correlation_mach_density}.

\subsection{Interpretation as turbulent motions}
\label{sec:mach}

Assuming that turbulence in the ICM is the main cause of measured density fluctuations, \cite{zhuravleva_relation_2014} and \cite{gaspari_relation_2014} have shown that the characteristic amplitude of the one-component velocity, which can be related to the 1D Mach number ,$\mathcal{M}_{1D}$, is proportional to the characteristic amplitude of density fluctuations. This relationship was also studied in the case of turbulence in a box and for stratified atmospheres \citep{mohapatra_turbulence_2019, mohapatra_turbulence_2020, mohapatra_turbulent_2021} and is discussed further in the case of stratified turbulence in Sect.~\ref{sec:non-ideal-turbulence}. \cite{simonte_exploring_2022} derived a similar relationship based on cosmological simulations. This relation is meant to link density fluctuations entirely to the random gas motions, which is more or less arguable depending on the relaxed or perturbed nature of the clusters. This is further discussed in Sect.~\ref{sec:discussion_correlation_mach_density}. In this case, the density fluctuation dispersion $\sigma_\delta$ can be linearly related to the velocity dispersion, $\sigma_v$, which we refer to in terms of the 3D Mach number $\mathcal{M}_{3D} = \sigma_v/c_s$, where $c_s$ is the speed of sound in the ICM:

\begin{equation}
    \mathcal{M}_{3D} \simeq (0.63 \pm 0.04)\times \sigma_\delta
    \label{eq:simonte}
.\end{equation}

\begin{figure*}[ht]
\includegraphics[width=\hsize]{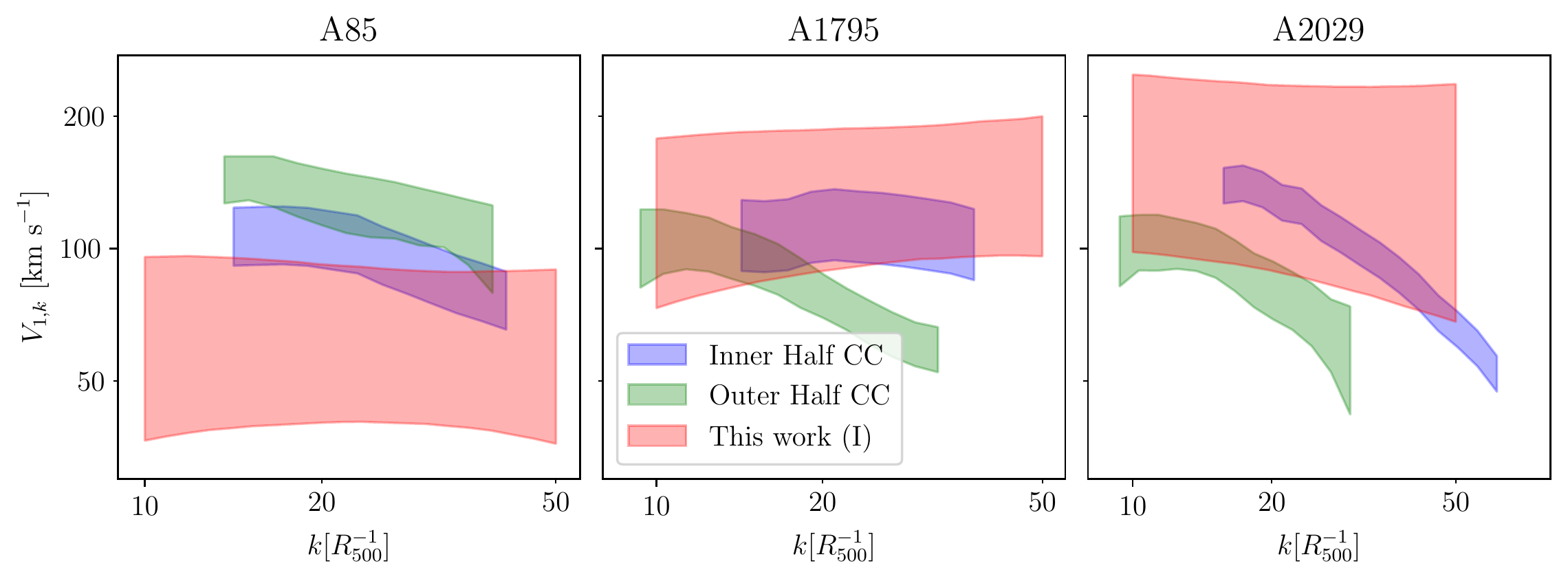}
\caption{Comparison of the one component velocity $V_{1,k}$ from \cite{zhuravleva_gas_2018} in the inner and outer part of the cool-cores of each cluster, and our determination in the central region (I) as defined in Table~\ref{tab:region_definition}. The envelope represents the 16\textsuperscript{th}-84\textsuperscript{th} percentiles of $V_{1,k}$.}
\label{fig:comparaison_v1k}
\end{figure*}

We use the relation derived for the whole sample of \cite{simonte_exploring_2022} (including relaxed and disturbed clusters). This correlation was initially derived in terms of $\sigma_v$ instead of the Mach number, which is equivalent in the limit of an isothermal ICM, but not entirely true in our study involving large scales. The turbulent Mach numbers determined for the X-COP sample are provided in Table~\ref{tab:all_parameters} for the four regions of analysis and in $R_{500}$. The level of turbulence we measure is clearly subsonic, which agrees with the non-thermal pressure support estimated by Eckert+19 using the assumption of a universal gas fraction \citep{eckert_non-thermal_2019}, and also with direct measurements of spectral lines broadening \citep{sanders_constraints_2011, the_hitomi_collaboration_quiescent_2016}. It is also in agreement with previous results from other works based on the statistics of fluctuation in density \citep{zhuravleva_gas_2015, zhuravleva_gas_2018} and in thermodynamical quantities \citep{hofmann_thermodynamic_2016} in various galaxy clusters. We compute the one-component velocity, $V_{1,k}$, as defined by \cite{zhuravleva_gas_2018} in Eq.~4 and use the same scaling relation as proposed in \cite{zhuravleva_relation_2014} to convert the density fluctuation to velocity: 
\begin{equation}
    V_{1,k} = \frac{c_s}{(1\pm 0.3)} \times \sqrt{4\pi k^3 \Bar{\mathcal{P}}_{3D, \delta}(k)}
.\end{equation}
We compare it to the results from \cite{zhuravleva_gas_2018} for the clusters in common with X-COP in Fig~\ref{fig:comparaison_v1k}, where we see a reasonable agreement. The source of differences between our measurements may originate from the different approach used to calculate the fluctuations, as \cite{zhuravleva_gas_2018} used wider \emph{Chandra} energy bands that are combined, whereas we use a single narrow \XMM band that minimises absorption.

\subsection{Non-thermal pressure support}
\label{sec:nth-pressure}

\begin{figure}[ht]
        \includegraphics[width=\linewidth]{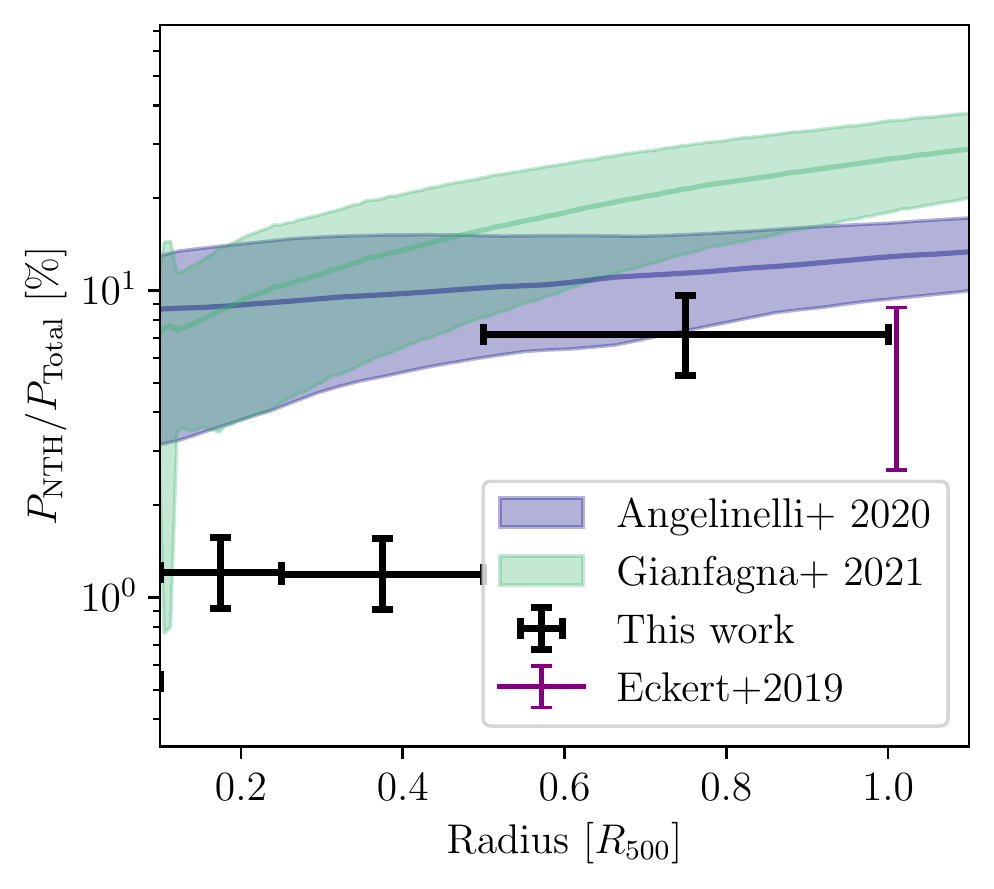}
  \caption{Comparison of the fraction of turbulent  to the hydrostatic pressure in the ICM as measured for the X-COP sample (black points), compared to the previous determination at $R_{500}$ by  \citet[][green point]{eckert_non-thermal_2019}. The blue and green line and associated shade envelope shows the predictions for non-thermal pressure support from the numerical simulations by \cite{gianfagna_exploring_2021} with the MUSIC clusters and \cite{angelinelli_turbulent_2020} with the ITASCA clusters.}   \label{fig:comparaison_turbulent_support}
\end{figure}

Numerical simulations predict that turbulent motions should be the dominant non-thermal pressure component in galaxy clusters \citep{vazza_turbulent_2018, angelinelli_turbulent_2020}. From this perspective, the aim is to characterise the ratio between the non-thermal pressure of turbulent origin and the total pressure $P_{\text{NTH}}/P_{\text{Total}}$. This ratio can be expressed as a function of the 3D turbulent Mach number, $\mathcal{M}_{3D}$ \citep{eckert_non-thermal_2019}, and is expressed as:

\begin{equation}
    \frac{P_{\text{NTH}}}{P_{\text{TOT}}} = \frac{\mathcal{M}_{3D}^2\gamma}{\mathcal{M}_{3D}^2\gamma + 3}
,\end{equation}

where $\gamma = 5/3$ is the polytropic index. The $P_{\text{NTH}}/P_{\text{Total}}$ ratio determined for the X-COP sample are provided in Table~\ref{tab:all_parameters} for the four regions of analysis and inside $R_{500}$. The value of $P_{\text{NTH}}/P_{\text{Total}}$ derived for the  joint $\mathcal{M}_{3D}$ is shown in Fig.~\ref{fig:comparaison_turbulent_support} as a function of radius. It is compared to the previous estimations from \cite{eckert_non-thermal_2019}, and theoretical predictions from numerical simulation by \cite{gianfagna_exploring_2021} and \cite{angelinelli_turbulent_2020}. The level of non-thermal support we obtain is consistent with previous measurements on X-COP clusters at $R_{500}$ \cite{eckert_non-thermal_2019}. Similarly to what these authors found,  we determined  a
value that is a fraction of between 0.5 and 1 $R_{500}$  lower than what is predicted by numerical simulations. This discrepancy is enhanced towards the centre of clusters up to a factor of 10. In our study, the centre is filled by residual structures such as sloshing spirals and cool-cores, which dictate our measurements in these regions. This is further discussed in Sects.~\ref{sec:discussion-grf} and~\ref{sec:discussion_correlation_mach_density}. This also could point to the fact that real clusters may be  more thermalised than those predicted in numerical simulations (e.g. due to an incomplete implementation of the gas physics). In addition, other physical processes, such as cosmic rays, or magnetic fields could contribute to the non-thermal pressure support \citep{ruszkowski_cosmic-ray_2017}. However, the combination of the available radio data for clusters with radio emission and/or enough sources to study Faraday rotation allows for constraints to be set on the magnetic field pressure to $\leq 1-2 \%$, while upper limits drawn from the non-detection of hadronic $\gamma$-rays allows putting a strong upper limit to the level of $\leq 1\%$ within the virial radius of clusters. Combined, these two non-thermal pressure sources should account for $\leq 2-3\%$ of the non-thermal pressure support on the ICM within the virial radius. Numerical estimations on the radial profile of these two components can be found in, for instance, \cite{vazza_non-thermal_2016}. A new paper by \cite{botteon_magnetic_2022} on A2255 may link turbulence at $R_{500}$, or beyond, to the non-thermal pressure. With a first-order estimate, and limited to this perturbed cluster, it is argued that $\sim 10\%$ of turbulent energy (compared to thermal energy) at $R_{500}$ should be enough to explain the diffuse emission detected by LOFAR (assuming a Fermi II acceleration process).

\subsection{Density model dependency}
\label{sec:mean_model_dependency}
The  density model  chosen arbitrarily  necessarily induces fluctuations of its own, thus  introducing a bias in the statistics of brightness fluctuations. We assessed the impact of the choice of model on the statistics of surface brightness fluctuations across the  X-COP clusters.  We compared three models corresponding to increasing level of fidelity: a circular $\beta$-model \citep{cavaliere_reprint_1976}, an elliptical $\beta$-model, and our mean-model (defined in Sect. \ref{sec:mean-profile}). Assuming that the surface brightness fluctuations are distributed according to a log-normal distribution \citep{kawahara_radial_2007}, we fit a dispersion $\delta_{S_{X}}$ inside $R_{500}$: 

$$\log S_X \sim \mathcal{N} \left(\log S_{X,0}, \delta_{S_{X}}\right).$$

\begin{figure}[h]
        \includegraphics[width=\linewidth]{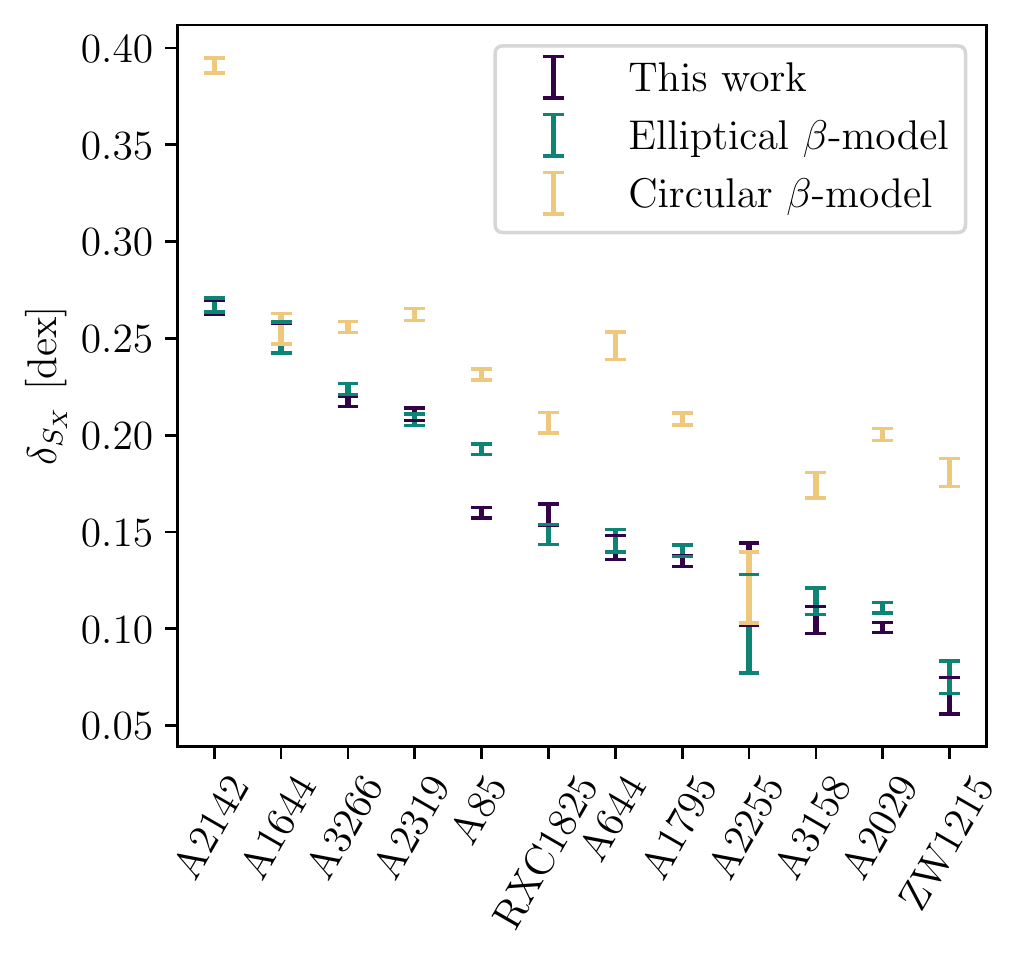}
  \caption{Surface brightness deviation for a circular $\beta$-model, an elliptic $\beta$-model, and the complete model from Sect. (\ref{sec:mean-profile}) for each cluster in the X-COP sample}
     \label{fig:residual_various_models}
\end{figure}

We assume that the mean surface brightness, $S_{X,0}$, in each pixel is  given, respectively, by the best-fit circular and elliptic $\beta$-model and our best-fit model (see Sect.~\ref{sec:results} or Table~\ref{tab:all_parameters}). The results are shown in Fig.~\ref{fig:residual_various_models}. There is a systematic reduction in dispersion when a more accurate model is used, showing that the simplest way to reduce arbitrary fluctuations is to account for the elliptical shape of the surface brightness. The best improvement is the transition between the circular and elliptical $\beta$-model. This is consistent with the results by \citet{zhuravleva_indirect_2022} on numerical simulations, showing that accounting for the ellipticity of clusters can reduce the measured density fluctuations by a factor of up to 2. Moreover, we expect the underlying potential to be intrinsically triaxial \citep[e.g.][]{lau_correlations_2021} which motivates the smooth, elliptical surface brightness model. Including triaxiality could further improve the density modelling, but it cannot be constrained by simply using X-ray images in the [0.7-1.2] keV band. There are very few studies constraining the triaxiality using a combination of lensing, X-ray and/or Sunyaev-Zel'dovich distortion \citep[e.g.][]{filippis_measuring_2005, sereno_measuring_2006, sereno_clump-3d_2017, sayers_clump-3d_2021}. Applying this methodology to X-COP clusters would require a multi-wavelength modelling efforts that are beyond the scope of this paper. Generally speaking, structural residuals induced by modelling flaws affect the parameters of the power spectrum from Eq.~\ref{eq:p3dmodel}. They tend to increase the normalisation as more fluctuations means more variance, change the slope due to the presence of sharp structures and affect the scale of injection, depending on the size of the residuals.

\subsection{Limitations of the Gaussian random field hypothesis}
\label{sec:discussion-grf}

Our methodology is sensitive to density fluctuations, assumed to be homogeneous and isotropic. As such, they can be described simply by a GRF. This assumption is valid as long as the density fluctuations arise from isotropic and homogeneous turbulence, and the density fluctuations are linearly related to the turbulent velocity. In the case of the X-COP clusters, the presence of cool-cores, submergers, and sloshing causes the emergence of structures in the surface brightness fluctuations that are poorly described by a GRF. Even if these structures (e.g. sloshing) cause turbulence in the ICM \citep{zuhone_turbulence_2012}, we are likely to be more sensitive to fluctuations in surface brightness generated by the spiral structure itself rather than to turbulence emerging at the interface of the two phases. The same applies to the presence of hot plasma bubbles from the central AGN feedback, which causes local gas under densities in the central regions with a characteristic size $\lesssim 10 ~\text{kpc}$ \citep{zhang_bubble-driven_2022}. Such spatial scales are out of the reach of \XMM. These rising bubbles could induce density fluctuations, exciting g-modes and sound waves. Although simulations predict that they could contribute up to 20\% of the heat dissipated in the ICM, they would remain elusive in surface brightness fluctuations \citep{choudhury_acoustic_2022}. In contrast, in the outer regions, turbulent processes can be expected to emerge from the dynamic assembly of structures. These processes could be sustained over long periods due to the large relaxation time frames that characterise the cluster outskirts and the propensity for the emergence of magnetohydrodynamic instabilities \citep{perrone_magneto-thermal_2022}. 

We therefore expect the centre of the images to represent the clumping of the gas, that is, in our case, the overdensities present due to the dynamic assembly. Conversely, we can expect the outer regions to represent fluctuations due to turbulent phenomena, since they are close enough to the centre of the cluster for substructure accretion not to bias our results and the clearly identified substructures are masked as part of our analysis. To get rid of the presence of the structures in the centre would require either a fine modelling of the sloshing spiral, as well as the modelling of cool-cores and mergers in the surface brightness residues, or by considering a departure from Gaussianity and informed phase distribution for the random field model we fit, which is beyond the scope of this paper.

\subsection{Implications of non-ideal turbulence}
\label{sec:non-ideal-turbulence}
In galaxy clusters, the gravity influence is expected to stratify the turbulent flows. This effect will tend to suppress parallel motion and stretch the eddies in the direction orthogonal to the gravity field. Depending on the strength of the stratification, the assumption of an isotropic Gaussian field for the velocity can be rejected, since significant differences will appear between the parallel and orthogonal component \citep{mohapatra_turbulence_2020}.

Furthermore, stratification will introduce a new mode of energy cascade. In addition to cascading from high to low spatial scales, kinetic energy can be converted in either direction into gravitational potential energy via buoyancy \citep[e.g.][]{bolgiano_jr_structure_1962}. By moving hot gas to regions of lower density, an additional density contrast appears and increases the total density and surface brightness fluctuations. If stratification is not negligible, not taking this effect into account may imply that the Mach number we calculate may be overestimated. The $\mathcal{M}_{3D}/\sigma_{\delta}$ relationship (see Eq.~\ref{eq:simonte}) proposed by \cite{simonte_exploring_2022} considers these effects in cosmological simulations and adds another degree of anisotropy by including the effect of radial accretion.

\subsection{Limitations of $\sigma_\delta - \mathcal{M}_{3D}$ equivalency}
\label{sec:discussion_correlation_mach_density}

As in the previous section, it is expected that the density fluctuations observed in the centre of clusters are not necessarily representative of the turbulent phenomena taking place there. For instance, in Fig.~\ref{fig:corner_plot}, we observe both a low injection scale and high spectral index, which can be interpreted in a straightforward way as the presence of small-scale, sharp features in the fluctuation maps. The scale of injection then increases and the slope tends to reduce with the radius, reflecting the fact that these structures are less prevalent away from the centre. Previous works on turbulence based on numerical simulations shows that this radial evolution should be much lower for the velocity power spectrum \citep{vazza_turbulence_2012}. \cite{simonte_exploring_2022} showed that the spectrum of density fluctuations was steeper in the centre of clusters than in the outer regions, and that its normalisation increases with the radius, while the velocity spectrum remained fairly invariant. At the same time, they showed a better correlation between density fluctuations and turbulent velocity fluctuations when using only relaxed clusters, similarly to \cite{zhuravleva_indirect_2022}, while the correlation worsens when disturbed clusters, associated with a larger number of clumps in the central regions, are concerned. 
All together, this suggests that the density fluctuations below $R_{500}/2$, are probably more sensitive to the presence of dense substructures and clumps and relatively less to the compression by turbulent motions. On the other hand, our measurement suggests that the region between $R_{500}/2$ and $R_{500}$ gives  an overall better correlation between density and velocity fluctuations.


\section{Conclusions}

We performed an analysis of the spatial statistics of  surface brightness fluctuations across the twelve clusters of the  X-COP sample. We derived the underlying  power spectrum of density fluctuations. For the first time, we have accounted for the stochastic nature of this observable by estimating the associated  sample variance, thus providing a complete error budget.
\begin{itemize}
    \item Using a simulation-based inference approach and modelling the density fluctuations as a Kolmogorov cascade, we can constrain the normalisation, slope, and scale of injection from the surface brightness fluctuations. The increasing trend in normalisation and injection scale with distance from the cluster core is  consistent, respectively, with the general idea that cluster virialisation decreases and that fluctuation-generating processes have dynamic scales that increase further away from the core. 
    \item We correlate the parameters of the density fluctuations with morphological indicators for X-COP clusters to find connections between dynamical state and apparent fluctuations. By properly propagating the uncertainties on each of our parameters, we found a clear correlation between the normalisation of the density fluctuation field and the four parameters of interest, suggesting a link between the measured normalisation of density fluctuations and the dynamical state of  clusters. The normalisation of density fluctuations within $R_{500}$ is positively correlated with the Zernike coefficient and the centroid shift, and negatively correlated with the Gini coefficient and concentration parameter. All these correlations can be interpreted as being due to the fact that the amount of fluctuation increases when the surface brightness of the cluster is disturbed, which is self-consistent in our analysis.
    \item We further interpreted these density fluctuations as gas clumping and turbulent motions. We constrained the clumping factor, Mach number, and non-thermal pressure support in the different regions of interest of our clusters. We observe good agreement with clumping data from other works in the central parts, but not in the external parts, where it is $\sim 10\%$ higher. On the other hand, the non-thermal pressure profile agrees with numerical simulations in the external parts and a previous determination at $R_{500}$ for our sample, but is underestimated in the central parts up to a factor of 10. Due to the presence of assembly artefacts in the central regions, we discuss the notion that the density fluctuations in the central regions are dominated by pure clumping and density fluctuation originating from residual substructures, while turbulent motions dominate the outer fluctuations. 
\end{itemize}

Considering the sample variance obviously increases the total error in the modelling and parametrisation of the power spectrum of fluctuations.  However, this component of the error budget contributes significantly at most scales. Neglecting it could lead to over or false physical interpretations. 
For physical processes that are universal, such as turbulence, this sample variance can be tackled by increasing the number of sources. It is then assumed that each cluster of a sample would present an individual realisation of the same stochastic physical process, possibly rescaled with its typical size. Thus, we ought to extend our work to a larger sample to allow for a further refinement of the constraints on the velocity dispersion, slope, and  injection scales, or even test a more evolved turbulence model including, such as stratification or multiple injection scales (from e.g. AGN, sloshing, dynamical assembly), while retaining sufficient statistics to study relaxed and disturbed subsets to further characterise different behaviours as a function of the source dynamical state. For instance, in a future
work, we will present the application of our method to the CHEX-MATE sample \citep{chex-mate_collaboration_cluster_2021}.

Furthermore, upcoming and future direct measurements of turbulence will be a key step in this work and will become available in the coming years, first with the Resolve instrument on board of the XRISM missions \citep{terada_detailed_2021} and later with the X-IFU instrument on board the Athena mission \citep{barret_athena_2020}.

\begin{acknowledgements}
We would like to thank Giulia Gianfagna for sharing the MUSIC cluster non-thermal pressure profile from \cite{gianfagna_exploring_2021}, Irina Zhuravleva for sharing the turbulent velocity power spectra from \cite{zhuravleva_gas_2018}, Matteo Angelinelli and Thomas Jones for sharing the clumping data and non-thermal pressure profile of the ITASCA cluster sample from \cite{angelinelli_turbulent_2020, angelinelli_proprieties_2021} and Rajsekhar Mohapatra for the insightful discussions about the impact of stratification. This work was granted access to the HPC resources of CALMIP supercomputing center under the allocation 2022-22052. F.V. acknowledges financial support from the Horizon 2020 program under the ERC Starting Grant \texttt{magcow}, no. 714196. This work used various open-source packages such as
\texttt{matplotlib} \citep{hunter_matplotlib_2007},
\texttt{astropy} \citep{robitaille_astropy_2013, the_astropy_collaboration_astropy_2018},
\texttt{ChainConsumer} \citep{hinton_chainconsumer_2016},
\texttt{cmasher} \citep{velden_cmasher_2020},
\texttt{sbi} \citep{tejero-cantero_sbi_2020},
\texttt{pyro} \citep{bingham_pyro_2019},
\texttt{jax} \citep{bradbury_jax_2018},
\texttt{haiku} \citep{hennigan_haiku_2020},
\texttt{numpyro} \citep{bingham_pyro_2019, phan_composable_2019}

\end{acknowledgements}

%
%

\bibliographystyle{aa} 
\bibliography{references.bib}

\begin{thebibliography}{99}
\expandafter\ifx\csname natexlab\endcsname\relax\def\natexlab#1{#1}\fi

\bibitem[{Anders \& Grevesse(1989)}]{anders_abundances_1989}
Anders, E. \& Grevesse, N. 1989, Geochimica et Cosmochimica Acta, 53, 197

\bibitem[{Angelinelli {et~al.}(2021)Angelinelli, Ettori, Vazza, \&
  Jones}]{angelinelli_proprieties_2021}
Angelinelli, M., Ettori, S., Vazza, F., \& Jones, T.~W. 2021, Astronomy \&
  Astrophysics, 653, A171

\bibitem[{Angelinelli {et~al.}(2020)Angelinelli, Vazza, Giocoli, Ettori, Jones,
  Brunetti, Brüggen, \& Eckert}]{angelinelli_turbulent_2020}
Angelinelli, M., Vazza, F., Giocoli, C., {et~al.} 2020, Monthly Notices of the
  Royal Astronomical Society, 495, 864

\bibitem[{Arévalo {et~al.}(2012)Arévalo, Churazov, Zhuravleva,
  Hernández-Monteagudo, \& Revnivtsev}]{arevalo_mexican_2012}
Arévalo, P., Churazov, E., Zhuravleva, I., Hernández-Monteagudo, C., \&
  Revnivtsev, M. 2012, Monthly Notices of the Royal Astronomical Society, 426,
  1793, publisher: Oxford Academic

\bibitem[{Barret {et~al.}(2020)Barret, Decourchelle, Fabian, Guainazzi, Nandra,
  Smith, \& den Herder}]{barret_athena_2020}
Barret, D., Decourchelle, A., Fabian, A., {et~al.} 2020, Astronomische
  Nachrichten, 341, 224

\bibitem[{Bennett \& Sijacki(2021)}]{bennett_disturbing_2021}
Bennett, J.~S. \& Sijacki, D. 2021, arXiv:2110.07326 [astro-ph], arXiv:
  2110.07326

\bibitem[{Biffi {et~al.}(2016)Biffi, Borgani, Murante, Rasia, Planelles,
  Granato, Ragone-Figueroa, Beck, Gaspari, \& Dolag}]{biffi_nature_2016}
Biffi, V., Borgani, S., Murante, G., {et~al.} 2016, The Astrophysical Journal,
  827, 112, publisher: The American Astronomical Society

\bibitem[{Bingham {et~al.}(2019)Bingham, Chen, Jankowiak, Obermeyer, Pradhan,
  Karaletsos, Singh, Szerlip, Horsfall, \& Goodman}]{bingham_pyro_2019}
Bingham, E., Chen, J.~P., Jankowiak, M., {et~al.} 2019, Journal of Machine
  Learning Research, 20, 1

\bibitem[{Bolgiano~Jr.(1962)}]{bolgiano_jr_structure_1962}
Bolgiano~Jr., R. 1962, Journal of Geophysical Research (1896-1977), 67, 3015

\bibitem[{Botteon {et~al.}(2022)Botteon, van Weeren, Brunetti, Vazza, Shimwell,
  Brüggen, Röttgering, de~Gasperin, Akamatsu, Bonafede, Cassano, Cuciti,
  Dallacasa, Di~Gennaro, \& Gastaldello}]{botteon_magnetic_2022}
Botteon, A., van Weeren, R.~J., Brunetti, G., {et~al.} 2022, Science Advances,
  8, eabq7623, publisher: American Association for the Advancement of Science

\bibitem[{Bradbury {et~al.}(2018)Bradbury, Frostig, Hawkins, Johnson, Leary,
  Maclaurin, Necula, Paszke, VanderPlas, Wanderman-Milne, \&
  Zhang}]{bradbury_jax_2018}
Bradbury, J., Frostig, R., Hawkins, P., {et~al.} 2018, {JAX}: composable
  transformations of {Python}+{NumPy} programs

\bibitem[{Brüggen \& Vazza(2015)}]{lazarian_turbulence_2015}
Brüggen, M. \& Vazza, F. 2015, in Magnetic {Fields} in {Diffuse} {Media}, ed.
  A.~Lazarian, E.~M. de~Gouveia Dal~Pino, \& C.~Melioli, Vol. 407 (Berlin,
  Heidelberg: Springer Berlin Heidelberg), 599--614, series Title: Astrophysics
  and Space Science Library

\bibitem[{Campitiello {et~al.}(2022)Campitiello, Ettori, Lovisari, Bartalucci,
  Eckert, Rasia, Rossetti, Gastaldello, Pratt, Maughan, Pointecouteau, Sereno,
  Biffi, Borgani, De~Luca, De~Petris, Gaspari, Ghizzardi, Mazzotta, \&
  Molendi}]{campitiello_chex-mate_2022}
Campitiello, M.~G., Ettori, S., Lovisari, L., {et~al.} 2022, Astronomy \&
  Astrophysics, 665, A117

\bibitem[{Capalbo {et~al.}(2021)Capalbo, De Petris, De Luca, Cui, Yepes,
  Knebe, \& Rasia}]{capalbo_three_2021}
Capalbo, V., De Petris, M., De Luca, F., {et~al.} 2021, Monthly Notices of
  the Royal Astronomical Society, 503, 6155

\bibitem[{Cappellari \& Copin(2003)}]{cappellari_adaptive_2003}
Cappellari, M. \& Copin, Y. 2003, Monthly Notices of the Royal Astronomical
  Society, 342, 345

\bibitem[{Cavaliere \& Fusco-Femiano(1976)}]{cavaliere_reprint_1976}
Cavaliere, A. \& Fusco-Femiano, R. 1976, Astronomy and Astrophysics, 500, 95

\bibitem[{{Chex-Mate Collaboration} {et~al.}(2021){Chex-Mate Collaboration},
  Arnaud, Ettori, Pratt, Rossetti, Eckert, Gastaldello, Gavazzi, Kay, Lovisari,
  Maughan, Pointecouteau, Sereno, Bartalucci, Bonafede, Bourdin, Cassano,
  Duffy, Iqbal, Maurogordato, Rasia, Sayers, Andrade-Santos, Aussel, Barnes,
  Barrena, Borgani, Burkutean, Clerc, Corasaniti, Cuillandre, De~Grandi,
  De~Petris, Dolag, Donahue, Ferragamo, Gaspari, Ghizzardi, Gitti, Haines,
  Jauzac, Johnston-Hollitt, Jones, Kéruzoré, Le~Brun, Mayet, Mazzotta, Melin,
  Molendi, Nonino, Okabe, Paltani, Perotto, Pires, Radovich, Rubino-Martin,
  Salvati, Saro, Sartoris, Schellenberger, Streblyanska, Tarrío, Tozzi,
  Umetsu, van~der Burg, Vazza, Venturi, Yepes, \&
  Zarattini}]{chex-mate_collaboration_cluster_2021}
{Chex-Mate Collaboration}, Arnaud, M., Ettori, S., {et~al.} 2021, Astronomy and
  Astrophysics, 650, A104

\bibitem[{Choudhury \& Reynolds(2022)}]{choudhury_acoustic_2022}
Choudhury, P.~P. \& Reynolds, C.~S. 2022, Monthly Notices of the Royal
  Astronomical Society, 514, 3765

\bibitem[{Churazov {et~al.}(2012)Churazov, Vikhlinin, Zhuravleva, Schekochihin,
  Parrish, Sunyaev, Forman, Böhringer, \& Randall}]{churazov_x-ray_2012}
Churazov, E., Vikhlinin, A., Zhuravleva, I., {et~al.} 2012, Monthly Notices of
  the Royal Astronomical Society, 421, 1123

\bibitem[{Clerc {et~al.}(2019)Clerc, Cucchetti, Pointecouteau, \&
  Peille}]{clerc_towards_2019}
Clerc, N., Cucchetti, E., Pointecouteau, E., \& Peille, P. 2019, Astronomy \&
  Astrophysics, 629, A143

\bibitem[{Cucchetti {et~al.}(2019)Cucchetti, Clerc, Pointecouteau, Peille, \&
  Pajot}]{cucchetti_towards_2019}
Cucchetti, E., Clerc, N., Pointecouteau, E., Peille, P., \& Pajot, F. 2019,
  Astronomy \& Astrophysics, 629, A144

\bibitem[{Cuciti {et~al.}(2022)Cuciti, de~Gasperin, Brüggen, Vazza, Brunetti,
  Shimwell, Edler, van Weeren, Botteon, Cassano, Di~Gennaro, Gastaldello,
  Drabent, Röttgering, \& Tasse}]{cuciti_galaxy_2022}
Cuciti, V., de~Gasperin, F., Brüggen, M., {et~al.} 2022, Nature, 609, 911,
  number: 7929 Publisher: Nature Publishing Group

\bibitem[{de Vries {et~al.}(2023)de Vries, Mantz, Allen, Morris, Zhuravleva,
  Canning, Ehlert, Ogorzałek, Simionescu, \& Werner}]{devries_chandra_2023}
de Vries, M., Mantz, A.~B., Allen, S.~W., {et~al.} 2023, Monthly Notices of
  the Royal Astronomical Society, 518, 2954

\bibitem[{Eckert {et~al.}(2017)Eckert, Ettori, Pointecouteau, Molendi, Paltani,
  Tchernin, \& Collaboration}]{eckert_xmm_2017}
Eckert, D., Ettori, S., Pointecouteau, E., {et~al.} 2017, Astronomische
  Nachrichten, 338, 293, \_eprint:
  https://onlinelibrary.wiley.com/doi/pdf/10.1002/asna.201713345

\bibitem[{Eckert {et~al.}(2022)Eckert, Ettori, Robertson, Massey,
  Pointecouteau, Harvey, \& McCarthy}]{eckert_constraints_2022}
Eckert, D., Ettori, S., Robertson, A., {et~al.} 2022, Astronomy \&
  Astrophysics, 666, A41, publisher: EDP Sciences

\bibitem[{Eckert {et~al.}(2019)Eckert, Ghirardini, Ettori, Rasia, Biffi,
  Pointecouteau, Rossetti, Molendi, Vazza, Gastaldello, Gaspari, Grandi,
  Ghizzardi, Bourdin, Tchernin, \& Roncarelli}]{eckert_non-thermal_2019}
Eckert, D., Ghirardini, V., Ettori, S., {et~al.} 2019, Astronomy \&
  Astrophysics, 621, A40, publisher: EDP Sciences

\bibitem[{Eckert {et~al.}(2015)Eckert, Roncarelli, Ettori, Molendi, Vazza,
  Gastaldello, \& Rossetti}]{eckert_gas_2015}
Eckert, D., Roncarelli, M., Ettori, S., {et~al.} 2015, Monthly Notices of the
  Royal Astronomical Society, 447, 2198

\bibitem[{Ettori \& Eckert(2022)}]{ettori_tracing_2022}
Ettori, S. \& Eckert, D. 2022, Astronomy \& Astrophysics, 657, L1, publisher:
  EDP Sciences

\bibitem[{Ettori {et~al.}(2019)Ettori, Ghirardini, Eckert, Pointecouteau,
  Gastaldello, Sereno, Gaspari, Ghizzardi, Roncarelli, \&
  Rossetti}]{ettori_hydrostatic_2019}
Ettori, S., Ghirardini, V., Eckert, D., {et~al.} 2019, Astronomy \&
  Astrophysics, 621, A39

\bibitem[{Filippis {et~al.}(2005)Filippis, Sereno, Bautz, \&
  Longo}]{filippis_measuring_2005}
Filippis, E.~D., Sereno, M., Bautz, M.~W., \& Longo, G. 2005, The Astrophysical
  Journal, 625, 108

\bibitem[{Gaspari {et~al.}(2014)Gaspari, Churazov, Nagai, Lau, \&
  Zhuravleva}]{gaspari_relation_2014}
Gaspari, M., Churazov, E., Nagai, D., Lau, E.~T., \& Zhuravleva, I. 2014,
  Astronomy and Astrophysics, 569, A67

\bibitem[{Gatuzz {et~al.}(2022{\natexlab{a}})Gatuzz, Sanders, Canning, Dennerl,
  Fabian, Pinto, Russell, Tamura, Walker, \& ZuHone}]{gatuzz_velocity_2022}
Gatuzz, E., Sanders, J.~S., Canning, R., {et~al.} 2022{\natexlab{a}}, Monthly
  Notices of the Royal Astronomical Society, 513, 1932, aDS Bibcode:
  2022MNRAS.513.1932G

\bibitem[{Gatuzz {et~al.}(2022{\natexlab{b}})Gatuzz, Sanders, Dennerl, Pinto,
  Fabian, Tamura, Walker, \& ZuHone}]{gatuzz_measuring_2022}
Gatuzz, E., Sanders, J.~S., Dennerl, K., {et~al.} 2022{\natexlab{b}}, Monthly
  Notices of the Royal Astronomical Society, 511, 4511, aDS Bibcode:
  2022MNRAS.511.4511G

\bibitem[{Germain {et~al.}(2015)Germain, Gregor, Murray, \&
  Larochelle}]{germain_made_2015}
Germain, M., Gregor, K., Murray, I., \& Larochelle, H. 2015, in Proceedings of
  the 32nd {International} {Conference} on {Machine} {Learning} (PMLR),
  881--889, iSSN: 1938-7228

\bibitem[{Ghirardini {et~al.}(2019)Ghirardini, Eckert, Ettori, Pointecouteau,
  Molendi, Gaspari, Rossetti, De~Grandi, Roncarelli, Bourdin, Mazzotta, Rasia,
  \& Vazza}]{ghirardini_universal_2019}
Ghirardini, V., Eckert, D., Ettori, S., {et~al.} 2019, Astronomy \&
  Astrophysics, 621, A41

\bibitem[{Ghirardini {et~al.}(2018)Ghirardini, Ettori, Eckert, Molendi,
  Gastaldello, Pointecouteau, Hurier, \& Bourdin}]{ghirardini_xmm_2018}
Ghirardini, V., Ettori, S., Eckert, D., {et~al.} 2018, Astronomy \&
  Astrophysics, 614, A7, publisher: EDP Sciences

\bibitem[{Gianfagna {et~al.}(2021)Gianfagna, De Petris, Yepes, De Luca,
  Sembolini, Cui, Biffi, Kéruzoré, Macías-Pérez, Mayet, Perotto, Rasia, \&
  Ruppin}]{gianfagna_exploring_2021}
Gianfagna, G., De Petris, M., Yepes, G., {et~al.} 2021, Monthly Notices of the
  Royal Astronomical Society, 502, 5115

\bibitem[{Hennigan {et~al.}(2020)Hennigan, Cai, Norman, \&
  Babuschkin}]{hennigan_haiku_2020}
Hennigan, T., Cai, T., Norman, T., \& Babuschkin, I. 2020, Haiku: {Sonnet} for
  {JAX}

\bibitem[{{HI4PI Collaboration}(2016)}]{hi4pi_collaboration_hi4pi_2016}
{HI4PI Collaboration}. 2016, Astronomy and Astrophysics, 594, A116

\bibitem[{Hinton(2016)}]{hinton_chainconsumer_2016}
Hinton, S. 2016, Journal of Open Source Software, 1, 45

\bibitem[{Hoffman \& Gelman(2014)}]{hoffman_no-u-turn_2014}
Hoffman, M.~D. \& Gelman, A. 2014, Journal of Machine Learning Research, 15,
  1593

\bibitem[{Hofmann {et~al.}(2016)Hofmann, Sanders, Nandra, Clerc, \&
  Gaspari}]{hofmann_thermodynamic_2016}
Hofmann, F., Sanders, J.~S., Nandra, K., Clerc, N., \& Gaspari, M. 2016,
  Astronomy \& Astrophysics, 585, A130

\bibitem[{Hunter(2007)}]{hunter_matplotlib_2007}
Hunter, J.~D. 2007, Computing in Science \& Engineering, 9, 90, conference
  Name: Computing in Science \& Engineering

\bibitem[{Kawahara {et~al.}(2007)Kawahara, Suto, Kitayama, Sasaki, Shimizu,
  Rasia, \& Dolag}]{kawahara_radial_2007}
Kawahara, H., Suto, Y., Kitayama, T., {et~al.} 2007, The Astrophysical Journal,
  659, 257, publisher: IOP Publishing

\bibitem[{Khatri \& Gaspari(2016)}]{khatri_thermal_2016}
Khatri, R. \& Gaspari, M. 2016, Monthly Notices of the Royal Astronomical
  Society, 463, 655

\bibitem[{Lau {et~al.}(2021)Lau, Hearin, Nagai, \&
  Cappelluti}]{lau_correlations_2021}
Lau, E.~T., Hearin, A.~P., Nagai, D., \& Cappelluti, N. 2021, Monthly Notices
  of the Royal Astronomical Society, 500, 1029

\bibitem[{Lau {et~al.}(2009)Lau, Kravtsov, \& Nagai}]{lau_residual_2009}
Lau, E.~T., Kravtsov, A.~V., \& Nagai, D. 2009, The Astrophysical Journal, 705,
  1129, publisher: The American Astronomical Society

\bibitem[{Lovisari {et~al.}(2017)Lovisari, Forman, Jones, Ettori,
  Andrade-Santos, Arnaud, Démoclès, Pratt, Randall, \&
  Kraft}]{lovisari_x-ray_2017}
Lovisari, L., Forman, W.~R., Jones, C., {et~al.} 2017, The Astrophysical
  Journal, 846, 51, publisher: The American Astronomical Society

\bibitem[{McNamara \& Nulsen(2012)}]{mcnamara_mechanical_2012}
McNamara, B.~R. \& Nulsen, P. E.~J. 2012, New Journal of Physics, 14, 055023,
  publisher: IOP Publishing

\bibitem[{Meneghetti {et~al.}(2010)Meneghetti, Rasia, Merten, Bellagamba,
  Ettori, Mazzotta, Dolag, \& Marri}]{meneghetti_weighing_2010}
Meneghetti, M., Rasia, E., Merten, J., {et~al.} 2010, Astronomy and
  Astrophysics, 514, A93

\bibitem[{Mohapatra {et~al.}(2020)Mohapatra, Federrath, \&
  Sharma}]{mohapatra_turbulence_2020}
Mohapatra, R., Federrath, C., \& Sharma, P. 2020, Monthly Notices of the Royal
  Astronomical Society, 493, 5838

\bibitem[{Mohapatra {et~al.}(2021)Mohapatra, Federrath, \&
  Sharma}]{mohapatra_turbulent_2021}
Mohapatra, R., Federrath, C., \& Sharma, P. 2021, Monthly Notices of the Royal
  Astronomical Society, 500, 5072

\bibitem[{Mohapatra \& Sharma(2019)}]{mohapatra_turbulence_2019}
Mohapatra, R. \& Sharma, P. 2019, Monthly Notices of the Royal Astronomical
  Society, 484, 4881

\bibitem[{Mori \& Sugihara(2001)}]{mori_double-exponential_2001}
Mori, M. \& Sugihara, M. 2001, Journal of Computational and Applied
  Mathematics, 127, 287

\bibitem[{Nelson {et~al.}(2014)Nelson, Lau, Nagai, Rudd, \&
  Yu}]{nelson_weighing_2014}
Nelson, K., Lau, E.~T., Nagai, D., Rudd, D.~H., \& Yu, L. 2014, The
  Astrophysical Journal, 782, 107, publisher: The American Astronomical Society

\bibitem[{Nelson {et~al.}(2012)Nelson, Rudd, Shaw, \&
  Nagai}]{nelson_evolution_2012}
Nelson, K., Rudd, D.~H., Shaw, L., \& Nagai, D. 2012, The Astrophysical
  Journal, 751, 121, aDS Bibcode: 2012ApJ...751..121N

\bibitem[{Ota {et~al.}(2018)Ota, Nagai, \& Lau}]{ota_constraining_2018}
Ota, N., Nagai, D., \& Lau, E.~T. 2018, Publications of the Astronomical
  Society of Japan, 70, 51

\bibitem[{Papamakarios {et~al.}(2017)Papamakarios, Pavlakou, \&
  Murray}]{papamakarios_masked_2017}
Papamakarios, G., Pavlakou, T., \& Murray, I. 2017, in Advances in {Neural}
  {Information} {Processing} {Systems}, Vol.~30 (Curran Associates, Inc.)

\bibitem[{Papamakarios {et~al.}(2019)Papamakarios, Sterratt, \&
  Murray}]{papamakarios_sequential_2019}
Papamakarios, G., Sterratt, D., \& Murray, I. 2019, in Proceedings of the
  {Twenty}-{Second} {International} {Conference} on {Artificial} {Intelligence}
  and {Statistics} (PMLR), 837--848, iSSN: 2640-3498

\bibitem[{Perrone \& Latter(2022)}]{perrone_magneto-thermal_2022}
Perrone, L.~M. \& Latter, H. 2022, Monthly Notices of the Royal Astronomical
  Society, 513, 4605

\bibitem[{Phan {et~al.}(2019)Phan, Pradhan, \&
  Jankowiak}]{phan_composable_2019}
Phan, D., Pradhan, N., \& Jankowiak, M. 2019, Composable {Effects} for
  {Flexible} and {Accelerated} {Probabilistic} {Programming} in {NumPyro},
  number: arXiv:1912.11554 arXiv:1912.11554 [cs, stat]

\bibitem[{Piffaretti \& Valdarnini(2008)}]{piffaretti_total_2008}
Piffaretti, R. \& Valdarnini, R. 2008, Astronomy \& Astrophysics, 491, 71,
  number: 1 Publisher: EDP Sciences

\bibitem[{Pinto {et~al.}(2015)Pinto, Sanders, Werner, de~Plaa, Fabian, Zhang,
  Kaastra, Finoguenov, \& Ahoranta}]{pinto_chemical_2015}
Pinto, C., Sanders, J.~S., Werner, N., {et~al.} 2015, Astronomy \&
  Astrophysics, 575, A38

\bibitem[{Pratt {et~al.}(2019)Pratt, Arnaud, Biviano, Eckert, Ettori, Nagai,
  Okabe, \& Reiprich}]{pratt_galaxy_2019}
Pratt, G.~W., Arnaud, M., Biviano, A., {et~al.} 2019, Space Science Reviews,
  215, 25

\bibitem[{Robitaille {et~al.}(2013)Robitaille, Tollerud, Greenfield,
  Droettboom, Bray, Aldcroft, Davis, Ginsburg, Price-Whelan, Kerzendorf,
  Conley, Crighton, Barbary, Muna, Ferguson, Grollier, Parikh, Nair, Günther,
  Deil, Woillez, Conseil, Kramer, Turner, Singer, Fox, Weaver, Zabalza,
  Edwards, Bostroem, Burke, Casey, Crawford, Dencheva, Ely, Jenness, Labrie,
  Lim, Pierfederici, Pontzen, Ptak, Refsdal, Servillat, \&
  Streicher}]{robitaille_astropy_2013}
Robitaille, T.~P., Tollerud, E.~J., Greenfield, P., {et~al.} 2013, Astronomy \&
  Astrophysics, 558, A33, publisher: EDP Sciences

\bibitem[{Roncarelli {et~al.}(2018)Roncarelli, Gaspari, Ettori, Biffi,
  Brighenti, Bulbul, Clerc, Cucchetti, Pointecouteau, \&
  Rasia}]{roncarelli_measuring_2018}
Roncarelli, M., Gaspari, M., Ettori, S., {et~al.} 2018, Astronomy \&
  Astrophysics, 618, A39, publisher: EDP Sciences

\bibitem[{Rossetti {et~al.}(2013)Rossetti, Eckert, Grandi, Gastaldello,
  Ghizzardi, Roediger, \& Molendi}]{rossetti_abell_2013}
Rossetti, M., Eckert, D., Grandi, S.~D., {et~al.} 2013, Astronomy \&
  Astrophysics, 556, A44, publisher: EDP Sciences

\bibitem[{Ruszkowski {et~al.}(2017)Ruszkowski, Yang, \&
  Reynolds}]{ruszkowski_cosmic-ray_2017}
Ruszkowski, M., Yang, H.-Y.~K., \& Reynolds, C.~S. 2017, The Astrophysical
  Journal, 844, 13, publisher: The American Astronomical Society

\bibitem[{Sanders {et~al.}(2020)Sanders, Dennerl, Russell, Eckert, Pinto,
  Fabian, Walker, Tamura, ZuHone, \& Hofmann}]{sanders_measuring_2020}
Sanders, J.~S., Dennerl, K., Russell, H.~R., {et~al.} 2020, Astronomy \&
  Astrophysics, 633, A42

\bibitem[{Sanders {et~al.}(2011)Sanders, Fabian, \&
  Smith}]{sanders_constraints_2011}
Sanders, J.~S., Fabian, A.~C., \& Smith, R.~K. 2011, Monthly Notices of the
  Royal Astronomical Society, 410, 1797

\bibitem[{Sayers {et~al.}(2021)Sayers, Sereno, Ettori, Rasia, Cui, Golwala,
  Umetsu, \& Yepes}]{sayers_clump-3d_2021}
Sayers, J., Sereno, M., Ettori, S., {et~al.} 2021, Monthly Notices of the Royal
  Astronomical Society, 505, 4338

\bibitem[{Schuecker {et~al.}(2004)Schuecker, Finoguenov, Miniati, Böhringer,
  \& Briel}]{schuecker_probing_2004}
Schuecker, P., Finoguenov, A., Miniati, F., Böhringer, H., \& Briel, U.~G.
  2004, Astronomy \& Astrophysics, 426, 387, number: 2 Publisher: EDP Sciences

\bibitem[{Sereno {et~al.}(2017)Sereno, Ettori, Meneghetti, Sayers, Umetsu,
  Merten, Chiu, \& Zitrin}]{sereno_clump-3d_2017}
Sereno, M., Ettori, S., Meneghetti, M., {et~al.} 2017, Monthly Notices of the
  Royal Astronomical Society, 467, 3801, aDS Bibcode: 2017MNRAS.467.3801S

\bibitem[{Sereno {et~al.}(2006)Sereno, Filippis, Longo, \&
  Bautz}]{sereno_measuring_2006}
Sereno, M., Filippis, E.~D., Longo, G., \& Bautz, M.~W. 2006, The Astrophysical
  Journal, 645, 170

\bibitem[{Shi {et~al.}(2016)Shi, Komatsu, Nagai, \& Lau}]{shi_analytical_2016}
Shi, X., Komatsu, E., Nagai, D., \& Lau, E.~T. 2016, Monthly Notices of the
  Royal Astronomical Society, 455, 2936

\bibitem[{Simonte {et~al.}(2022)Simonte, Vazza, Brighenti, Brüggen, Jones, \&
  Angelinelli}]{simonte_exploring_2022}
Simonte, M., Vazza, F., Brighenti, F., {et~al.} 2022, Astronomy \&
  Astrophysics, 658, A149, publisher: EDP Sciences

\bibitem[{Takahasi \& Mori(1973)}]{takahasi_double_1973}
Takahasi, H. \& Mori, M. 1973, Publications of the Research Institute for
  Mathematical Sciences, 9, 721

\bibitem[{Tejero-Cantero {et~al.}(2020)Tejero-Cantero, Boelts, Deistler,
  Lueckmann, Durkan, Gonçalves, Greenberg, \& Macke}]{tejero-cantero_sbi_2020}
Tejero-Cantero, A., Boelts, J., Deistler, M., {et~al.} 2020, Journal of Open
  Source Software, 5, 2505

\bibitem[{Terada {et~al.}(2021)Terada, Holland, Loewenstein, Tashiro,
  Takahashi, Nobukawa, Mizuno, Tamura, Uno, Watanabe, Baluta, Burns, Ebisawa,
  Eguchi, Fukazawa, Hayashi, Iizuka, Katsuda, Kitaguchi, Kubota, Miller, Mukai,
  Nakashima, Nakazawa, Odaka, Ohno, Ota, Sato, Sawada, Sugawara, Shidatsu,
  Tamba, Tanimoto, Terashima, Tsuboi, Uchida, Uchiyama, Yamauchi, \&
  Yaqoob}]{terada_detailed_2021}
Terada, Y., Holland, M., Loewenstein, M., {et~al.} 2021, Journal of
  Astronomical Telescopes, Instruments, and Systems, 7, 037001, publisher: SPIE

\bibitem[{{The Astropy Collaboration} {et~al.}(2018){The Astropy
  Collaboration}, Price-Whelan, Sipőcz, Günther, Lim, Crawford, Conseil,
  Shupe, Craig, Dencheva, Ginsburg, VanderPlas, Bradley, Pérez-Suárez,
  Val-Borro, Contributors), Aldcroft, Cruz, Robitaille, Tollerud, Committee),
  Ardelean, Babej, Bach, Bachetti, Bakanov, Bamford, Barentsen, Barmby,
  Baumbach, Berry, Biscani, Boquien, Bostroem, Bouma, Brammer, Bray,
  Breytenbach, Buddelmeijer, Burke, Calderone, Rodríguez, Cara, Cardoso,
  Cheedella, Copin, Corrales, Crichton, D’Avella, Deil, Depagne, Dietrich,
  Donath, Droettboom, Earl, Erben, Fabbro, Ferreira, Finethy, Fox, Garrison,
  Gibbons, Goldstein, Gommers, Greco, Greenfield, Groener, Grollier, Hagen,
  Hirst, Homeier, Horton, Hosseinzadeh, Hu, Hunkeler, Ivezić, Jain, Jenness,
  Kanarek, Kendrew, Kern, Kerzendorf, Khvalko, King, Kirkby, Kulkarni, Kumar,
  Lee, Lenz, Littlefair, Ma, Macleod, Mastropietro, McCully, Montagnac, Morris,
  Mueller, Mumford, Muna, Murphy, Nelson, Nguyen, Ninan, Nöthe, Ogaz, Oh,
  Parejko, Parley, Pascual, Patil, Patil, Plunkett, Prochaska, Rastogi, Janga,
  Sabater, Sakurikar, Seifert, Sherbert, Sherwood-Taylor, Shih, Sick, Silbiger,
  Singanamalla, Singer, Sladen, Sooley, Sornarajah, Streicher, Teuben, Thomas,
  Tremblay, Turner, Terrón, Kerkwijk, Vega, Watkins, Weaver, Whitmore,
  Woillez, Zabalza, \& Contributors)}]{the_astropy_collaboration_astropy_2018}
{The Astropy Collaboration}, Price-Whelan, A.~M., Sipőcz, B.~M., {et~al.}
  2018, The Astronomical Journal, 156, 123, publisher: The American
  Astronomical Society

\bibitem[{{The Hitomi
  Collaboration}(2016)}]{the_hitomi_collaboration_quiescent_2016}
{The Hitomi Collaboration}. 2016, Nature, 535, 117

\bibitem[{{The Planck
  Collaboration}(2014)}]{the_planck_collaboration_planck_2014}
{The Planck Collaboration}. 2014, Astronomy \& Astrophysics, 571, A29,
  publisher: EDP Sciences

\bibitem[{Vazza {et~al.}(2018)Vazza, Angelinelli, Jones, Eckert, Brüggen,
  Brunetti, \& Gheller}]{vazza_turbulent_2018}
Vazza, F., Angelinelli, M., Jones, T.~W., {et~al.} 2018, Monthly Notices of the
  Royal Astronomical Society: Letters, 481, L120

\bibitem[{Vazza {et~al.}(2012)Vazza, Roediger, \&
  Brüggen}]{vazza_turbulence_2012}
Vazza, F., Roediger, E., \& Brüggen, M. 2012, Astronomy \& Astrophysics, 544,
  A103

\bibitem[{Vazza {et~al.}(2016)Vazza, Wittor, Brüggen, \&
  Gheller}]{vazza_non-thermal_2016}
Vazza, F., Wittor, D., Brüggen, M., \& Gheller, C. 2016, Galaxies, 4, 60,
  number: 4 Publisher: Multidisciplinary Digital Publishing Institute

\bibitem[{Vehtari {et~al.}(2021)Vehtari, Gelman, Simpson, Carpenter, \&
  Bürkner}]{vehtari_rank-normalization_2021}
Vehtari, A., Gelman, A., Simpson, D., Carpenter, B., \& Bürkner, P.-C. 2021,
  Bayesian Analysis, 16, 667, publisher: International Society for Bayesian
  Analysis

\bibitem[{Velden(2020)}]{velden_cmasher_2020}
Velden, E. v.~d. 2020, Journal of Open Source Software, 5, 2004

\bibitem[{Vikhlinin {et~al.}(2006)Vikhlinin, Kravtsov, Forman, Jones,
  Markevitch, Murray, \& Speybroeck}]{vikhlinin_chandra_2006}
Vikhlinin, A., Kravtsov, A., Forman, W., {et~al.} 2006, The Astrophysical
  Journal, 640, 691, publisher: IOP Publishing

\bibitem[{Voit {et~al.}(2017)Voit, Meece, Li, O'Shea, Bryan, \&
  Donahue}]{voit_global_2017}
Voit, G.~M., Meece, G., Li, Y., {et~al.} 2017, The Astrophysical Journal, 845,
  80, publisher: American Astronomical Society

\bibitem[{Weisstein(1995)}]{weisstein_fourier_1995}
Weisstein, E.~W. 1995, Fourier {Transform}, publisher: Wolfram Research, Inc.

\bibitem[{{XRISM Science Team}(2020)}]{xrism_science_team_science_2020}
{XRISM Science Team}. 2020, arXiv:2003.04962 [astro-ph], arXiv: 2003.04962

\bibitem[{Zhang {et~al.}(2022{\natexlab{a}})Zhang, Zhuravleva,
  Gendron-Marsolais, Churazov, Schekochihin, \&
  Forman}]{zhang_bubble-driven_2022}
Zhang, C., Zhuravleva, I., Gendron-Marsolais, M.-L., {et~al.}
  2022{\natexlab{a}}, Monthly Notices of the Royal Astronomical Society, 517,
  616

\bibitem[{Zhang {et~al.}(2022{\natexlab{b}})Zhang, Simionescu, Gastaldello,
  Eckert, Camillini, Natale, Rossetti, Brunetti, Akamatsu, Botteon, Cassano,
  Cuciti, Bruno, Shimwell, Jones, Kaastra, Ettori, Brüggen, de~Gasperin,
  Drabent, van Weeren, \& Röttgering}]{zhang_planck_2022}
Zhang, X., Simionescu, A., Gastaldello, F., {et~al.} 2022{\natexlab{b}}, The
  {Planck} clusters in the {LOFAR} sky. {III}. {LoTSS}-{DR2}: {Dynamic} states
  and density fluctuations of the intracluster medium, arXiv:2210.07284
  [astro-ph]

\bibitem[{Zhuravleva {et~al.}(2018)Zhuravleva, Allen, Mantz, \&
  Werner}]{zhuravleva_gas_2018}
Zhuravleva, I., Allen, S.~W., Mantz, A., \& Werner, N. 2018, The Astrophysical
  Journal, 865, 53, publisher: IOP Publishing

\bibitem[{Zhuravleva {et~al.}(2022)Zhuravleva, Chen, Churazov, Schekochihin,
  Zhang, \& Nagai}]{zhuravleva_indirect_2022}
Zhuravleva, I., Chen, M.~C., Churazov, E., {et~al.} 2022, Indirect
  {Measurements} of {Gas} {Velocities} in {Galaxy} {Clusters}: {Effects} of
  {Ellipticity} and {Cluster} {Dynamic} {State}, arXiv:2210.11544 [astro-ph]

\bibitem[{Zhuravleva {et~al.}(2015)Zhuravleva, Churazov, Arévalo,
  Schekochihin, Allen, Fabian, Forman, Sanders, Simionescu, Sunyaev, Vikhlinin,
  \& Werner}]{zhuravleva_gas_2015}
Zhuravleva, I., Churazov, E., Arévalo, P., {et~al.} 2015, Monthly Notices of
  the Royal Astronomical Society, 450, 4184

\bibitem[{Zhuravleva {et~al.}(2014)Zhuravleva, Churazov, Schekochihin, Lau,
  Nagai, Gaspari, Allen, Nelson, \& Parrish}]{zhuravleva_relation_2014}
Zhuravleva, I., Churazov, E.~M., Schekochihin, A.~A., {et~al.} 2014, The
  Astrophysical Journal, 788, L13, publisher: IOP Publishing

\bibitem[{ZuHone {et~al.}(2012)ZuHone, Markevitch, Brunetti, \&
  Giacintucci}]{zuhone_turbulence_2012}
ZuHone, J.~A., Markevitch, M., Brunetti, G., \& Giacintucci, S. 2012, The
  Astrophysical Journal, 762, 78, publisher: IOP Publishing

\bibitem[{ZuHone {et~al.}(2016)ZuHone, Markevitch, \&
  Zhuravleva}]{zuhone_mapping_2016}
ZuHone, J.~A., Markevitch, M., \& Zhuravleva, I. 2016, The Astrophysical
  Journal, 817, 110, publisher: The American Astronomical Society

\end{thebibliography}
\paragraph{List of objects :}
\object{ACO 1644}
\object{ACO 1795}
\object{ACO 2029}
\object{ACO 2142}
\object{ACO 2255}
\object{ACO 2319}
\object{ACO 3158}
\object{ACO 3266}
\object{ACO 644}
\object{ACO 85}
\object{RXC J1825.3+3026}
\object{ZwCl 1215}

\begin{appendix} 

\section{Additional table}

\begin{sidewaystable*}
\centering
\resizebox{\textheight}{!}{\begin{tabular}{|r|rrrrrrrrrrrrrr|}
\hline
\hline
Name & Prior & A1644 & A1795 & A2029 & A2142 & A2255 & A2319 & A3158 & A3266 & A644 & A85 & RXC1825 & ZW1215 & Joint\\[2mm]
\hline
$\sigma_{\delta}$ (I) & $10^{\mathcal{U}\left(-2, 1\right)}$ & $0.54^{+0.27}_{-0.18}$ & $0.2^{+0.08}_{-0.05}$ & $0.17^{+0.09}_{-0.06}$ & $0.12^{+0.05}_{-0.03}$ & $0.06^{+0.16}_{-0.04}$ & $0.19^{+0.1}_{-0.05}$ & $0.28^{+0.28}_{-0.14}$ & $0.42^{+0.28}_{-0.13}$ & $0.25^{+0.2}_{-0.11}$ & $0.09^{+0.04}_{-0.02}$ & $0.28^{+0.24}_{-0.1}$ & $0.05^{+0.24}_{-0.04}$ & $0.16^{+0.02}_{-0.01}$\\[2mm]
$\sigma_{\delta}$ (II) & $10^{\mathcal{U}\left(-2, 1\right)}$ & $0.31^{+0.23}_{-0.09}$ & $0.18^{+0.1}_{-0.04}$ & $0.09^{+0.04}_{-0.03}$ & $0.24^{+0.12}_{-0.06}$ & $0.05^{+0.13}_{-0.04}$ & $0.27^{+0.08}_{-0.05}$ & $0.11^{+0.09}_{-0.04}$ & $0.36^{+0.09}_{-0.05}$ & $0.06^{+0.03}_{-0.02}$ & $0.15^{+0.05}_{-0.03}$ & $0.31^{+0.17}_{-0.1}$ & $0.1^{+0.07}_{-0.04}$ & $0.23^{+0.03}_{-0.02}$\\[2mm]
$\sigma_{\delta}$ (III) & $10^{\mathcal{U}\left(-2, 1\right)}$ & $0.18^{+0.09}_{-0.05}$ & $0.22^{+0.21}_{-0.09}$ & $0.13^{+0.13}_{-0.05}$ & $0.18^{+0.07}_{-0.04}$ & $0.09^{+0.49}_{-0.07}$ & $0.27^{+0.09}_{-0.06}$ & $0.21^{+0.1}_{-0.05}$ & $0.23^{+0.06}_{-0.04}$ & $0.32^{+0.28}_{-0.11}$ & $0.28^{+0.12}_{-0.08}$ & $0.28^{+0.43}_{-0.15}$ & $0.18^{+0.07}_{-0.04}$ & $0.23^{+0.03}_{-0.02}$\\[2mm]
$\sigma_{\delta}$ (IV) & $10^{\mathcal{U}\left(-2, 1\right)}$ & $0.74^{+0.78}_{-0.35}$ & $0.42^{+0.03}_{-0.03}$ & $0.33^{+0.1}_{-0.07}$ & $0.57^{+0.25}_{-0.21}$ & $0.23^{+0.84}_{-0.19}$ & $0.45^{+0.21}_{-0.16}$ & $0.45^{+0.21}_{-0.14}$ & $0.77^{+0.21}_{-0.15}$ & $0.48^{+0.1}_{-0.08}$ & $0.41^{+0.38}_{-0.13}$ & $0.39^{+0.26}_{-0.11}$ & $0.25^{+0.17}_{-0.09}$ & $0.59^{+0.08}_{-0.07}$\\[2mm]
$\sigma_{\delta}$ in $R_{500}$ & $10^{\mathcal{U}\left(-2, 1\right)}$ & $0.37^{+0.11}_{-0.09}$ & $0.19^{+0.05}_{-0.04}$ & $0.14^{+0.07}_{-0.04}$ & $0.19^{+0.05}_{-0.03}$ & $0.17^{+0.18}_{-0.06}$ & $0.44^{+0.13}_{-0.08}$ & $0.28^{+0.19}_{-0.12}$ & $0.32^{+0.04}_{-0.04}$ & $0.14^{+0.08}_{-0.03}$ & $0.14^{+0.05}_{-0.04}$ & $0.41^{+0.16}_{-0.14}$ & $0.06^{+0.03}_{-0.02}$ & $0.18^{+0.01}_{-0.01}$\\[2mm]
$\ell_{\text{inj}} [R_{500}]$ (I) & $10^{\mathcal{U}\left(-2, 1\right)}$ & $0.46^{+1.92}_{-0.33}$ & $0.44^{+1.2}_{-0.28}$ & $1.41^{+3.49}_{-1.05}$ & $0.15^{+0.43}_{-0.05}$ & $0.06^{+0.38}_{-0.04}$ & $0.19^{+0.55}_{-0.08}$ & $1.77^{+4.76}_{-1.49}$ & $0.34^{+0.5}_{-0.15}$ & $1.37^{+2.68}_{-0.8}$ & $0.81^{+3.07}_{-0.65}$ & $0.49^{+2.16}_{-0.31}$ & $0.12^{+1.69}_{-0.1}$ & $0.26^{+0.12}_{-0.07}$\\[2mm]
$\ell_{\text{inj}} [R_{500}]$ (II) & $10^{\mathcal{U}\left(-2, 1\right)}$ & $0.6^{+2.02}_{-0.35}$ & $0.49^{+0.95}_{-0.22}$ & $3.34^{+3.72}_{-2.34}$ & $0.47^{+0.64}_{-0.2}$ & $0.53^{+3.01}_{-0.47}$ & $0.34^{+0.51}_{-0.12}$ & $0.78^{+2.07}_{-0.43}$ & $0.29^{+0.27}_{-0.09}$ & $1.99^{+4.74}_{-1.55}$ & $0.77^{+1.67}_{-0.42}$ & $1.29^{+2.84}_{-0.87}$ & $1.38^{+3.03}_{-0.9}$ & $2.55^{+1.09}_{-0.76}$\\[2mm]
$\ell_{\text{inj}} [R_{500}]$ (III) & $10^{\mathcal{U}\left(-2, 1\right)}$ & $0.47^{+2.1}_{-0.22}$ & $0.61^{+1.05}_{-0.31}$ & $1.16^{+2.52}_{-0.62}$ & $0.91^{+1.65}_{-0.44}$ & $0.1^{+1.32}_{-0.08}$ & $2.75^{+3.06}_{-1.55}$ & $0.54^{+0.96}_{-0.24}$ & $0.55^{+0.69}_{-0.21}$ & $1.82^{+4.6}_{-1.14}$ & $2.87^{+3.4}_{-1.63}$ & $1.14^{+4.14}_{-0.71}$ & $1.08^{+1.48}_{-0.52}$ & $1.65^{+0.55}_{-0.4}$\\[2mm]
$\ell_{\text{inj}} [R_{500}]$ (IV) & $10^{\mathcal{U}\left(-2, 1\right)}$ & $2.48^{+3.91}_{-2.36}$ & $0.12^{+0.03}_{-0.02}$ & $1.22^{+1.37}_{-0.45}$ & $5.88^{+2.63}_{-2.88}$ & $0.35^{+4.45}_{-0.33}$ & $4.73^{+3.34}_{-2.51}$ & $1.71^{+3.41}_{-1.18}$ & $1.19^{+2.15}_{-0.63}$ & $0.32^{+0.43}_{-0.14}$ & $2.27^{+4.33}_{-1.64}$ & $2.59^{+3.23}_{-1.6}$ & $1.15^{+3.12}_{-0.71}$ & $6.53^{+1.45}_{-1.21}$\\[2mm]
$\ell_{\text{inj}} [R_{500}]$ in $R_{500}$ & $10^{\mathcal{U}\left(-2, 1\right)}$ & $0.32^{+0.19}_{-0.12}$ & $1.31^{+1.58}_{-0.61}$ & $0.32^{+0.39}_{-0.13}$ & $0.52^{+0.38}_{-0.16}$ & $0.57^{+0.74}_{-0.26}$ & $5.6^{+2.35}_{-1.82}$ & $2.17^{+3.65}_{-1.52}$ & $0.35^{+0.12}_{-0.08}$ & $0.5^{+1.15}_{-0.32}$ & $0.73^{+2.25}_{-0.44}$ & $2.87^{+4.98}_{-2.4}$ & $0.73^{+0.99}_{-0.28}$ & $0.48^{+0.06}_{-0.05}$\\[2mm]
$\alpha$ (I) & $\mathcal{U}\left(0, 7\right)$ & $4.0^{+1.24}_{-0.85}$ & $2.83^{+0.61}_{-0.47}$ & $3.18^{+0.61}_{-0.41}$ & $5.17^{+1.19}_{-1.31}$ & $4.02^{+2.12}_{-2.73}$ & $4.89^{+1.22}_{-1.07}$ & $2.9^{+1.37}_{-0.73}$ & $5.21^{+1.09}_{-0.96}$ & $5.86^{+0.76}_{-0.82}$ & $3.06^{+0.98}_{-0.63}$ & $4.97^{+1.41}_{-1.33}$ & $1.81^{+2.2}_{-1.22}$ & $4.08^{+0.35}_{-0.28}$\\[2mm]
$\alpha$ (II) & $\mathcal{U}\left(0, 7\right)$ & $3.81^{+1.39}_{-0.73}$ & $5.94^{+0.73}_{-0.91}$ & $3.79^{+1.31}_{-0.7}$ & $6.01^{+0.7}_{-1.09}$ & $4.73^{+1.47}_{-2.56}$ & $4.81^{+1.35}_{-0.88}$ & $3.69^{+1.5}_{-0.95}$ & $4.94^{+0.88}_{-0.65}$ & $4.97^{+1.35}_{-1.44}$ & $4.1^{+1.11}_{-0.61}$ & $3.37^{+0.92}_{-0.51}$ & $4.53^{+1.65}_{-1.34}$ & $3.68^{+0.18}_{-0.14}$\\[2mm]
$\alpha$ (III) & $\mathcal{U}\left(0, 7\right)$ & $4.85^{+1.52}_{-1.54}$ & $2.96^{+1.03}_{-0.52}$ & $3.4^{+1.13}_{-0.64}$ & $5.13^{+1.15}_{-1.01}$ & $2.27^{+2.45}_{-1.59}$ & $3.54^{+0.59}_{-0.37}$ & $4.3^{+1.63}_{-1.18}$ & $3.97^{+1.15}_{-0.61}$ & $4.59^{+1.59}_{-1.16}$ & $2.96^{+0.37}_{-0.31}$ & $2.85^{+1.47}_{-0.58}$ & $4.77^{+1.22}_{-1.1}$ & $3.15^{+0.14}_{-0.12}$\\[2mm]
$\alpha$ (IV) & $\mathcal{U}\left(0, 7\right)$ & $3.72^{+2.12}_{-1.35}$ & $5.69^{+0.74}_{-0.81}$ & $4.85^{+1.49}_{-1.1}$ & $3.31^{+0.89}_{-0.37}$ & $3.54^{+1.99}_{-2.01}$ & $4.01^{+1.44}_{-0.6}$ & $3.05^{+0.99}_{-0.37}$ & $2.72^{+0.26}_{-0.21}$ & $4.48^{+1.51}_{-1.25}$ & $4.1^{+2.32}_{-1.15}$ & $3.93^{+0.81}_{-0.58}$ & $4.68^{+1.68}_{-1.56}$ & $2.9^{+0.1}_{-0.09}$\\[2mm]
$\alpha$ in $R_{500}$ & $\mathcal{U}\left(0, 7\right)$ & $3.4^{+1.01}_{-0.5}$ & $3.07^{+0.38}_{-0.33}$ & $2.9^{+0.66}_{-0.6}$ & $4.05^{+0.52}_{-0.42}$ & $3.5^{+1.22}_{-0.87}$ & $3.2^{+0.22}_{-0.22}$ & $2.83^{+0.83}_{-0.44}$ & $4.26^{+0.45}_{-0.38}$ & $3.32^{+2.45}_{-0.72}$ & $3.88^{+1.66}_{-0.86}$ & $2.81^{+0.37}_{-0.26}$ & $4.89^{+1.29}_{-1.24}$ & $3.67^{+0.12}_{-0.12}$\\[2mm]
\hline
$\sqrt{C}$ (I) &  & $1.14^{+0.15}_{-0.07}$ & $1.02^{+0.02}_{-0.01}$ & $1.02^{+0.02}_{-0.01}$ & $1.01^{+0.01}_{-0.0}$ & $1.0^{+0.02}_{-0.0}$ & $1.02^{+0.02}_{-0.01}$ & $1.04^{+0.11}_{-0.03}$ & $1.09^{+0.14}_{-0.04}$ & $1.03^{+0.07}_{-0.02}$ & $1.0^{+0.0}_{-0.0}$ & $1.04^{+0.09}_{-0.02}$ & $1.0^{+0.04}_{-0.0}$ & $1.01^{+0.0}_{-0.0}$\\[2mm]
$\sqrt{C}$ (II) &  & $1.05^{+0.09}_{-0.02}$ & $1.02^{+0.02}_{-0.01}$ & $1.0^{+0.0}_{-0.0}$ & $1.03^{+0.03}_{-0.01}$ & $1.0^{+0.02}_{-0.0}$ & $1.04^{+0.02}_{-0.01}$ & $1.01^{+0.01}_{-0.0}$ & $1.06^{+0.03}_{-0.02}$ & $1.0^{+0.0}_{-0.0}$ & $1.01^{+0.01}_{-0.0}$ & $1.05^{+0.06}_{-0.03}$ & $1.01^{+0.01}_{-0.0}$ & $1.03^{+0.01}_{-0.01}$\\[2mm]
$\sqrt{C}$ (III) &  & $1.02^{+0.02}_{-0.01}$ & $1.02^{+0.06}_{-0.02}$ & $1.01^{+0.03}_{-0.0}$ & $1.02^{+0.02}_{-0.01}$ & $1.0^{+0.15}_{-0.0}$ & $1.04^{+0.03}_{-0.01}$ & $1.02^{+0.03}_{-0.01}$ & $1.03^{+0.01}_{-0.01}$ & $1.05^{+0.12}_{-0.03}$ & $1.04^{+0.04}_{-0.02}$ & $1.04^{+0.19}_{-0.03}$ & $1.02^{+0.01}_{-0.01}$ & $1.03^{+0.01}_{-0.0}$\\[2mm]
$\sqrt{C}$ (IV) &  & $1.24^{+0.58}_{-0.17}$ & $1.09^{+0.01}_{-0.01}$ & $1.05^{+0.03}_{-0.02}$ & $1.15^{+0.14}_{-0.09}$ & $1.03^{+0.44}_{-0.02}$ & $1.1^{+0.1}_{-0.05}$ & $1.09^{+0.1}_{-0.05}$ & $1.26^{+0.14}_{-0.09}$ & $1.11^{+0.05}_{-0.03}$ & $1.08^{+0.19}_{-0.04}$ & $1.07^{+0.12}_{-0.04}$ & $1.03^{+0.06}_{-0.02}$ & $1.16^{+0.05}_{-0.04}$\\[2mm]
$\sqrt{C}$ in $R_{500}$ &  & $1.07^{+0.04}_{-0.03}$ & $1.02^{+0.01}_{-0.01}$ & $1.01^{+0.01}_{-0.0}$ & $1.02^{+0.01}_{-0.01}$ & $1.01^{+0.05}_{-0.01}$ & $1.09^{+0.06}_{-0.03}$ & $1.04^{+0.07}_{-0.03}$ & $1.05^{+0.01}_{-0.01}$ & $1.01^{+0.02}_{-0.0}$ & $1.01^{+0.01}_{-0.0}$ & $1.08^{+0.07}_{-0.05}$ & $1.0^{+0.0}_{-0.0}$ & $1.02^{+0.0}_{-0.0}$\\[2mm]
$\mathcal{M}$ (I) &  & $0.34^{+0.18}_{-0.11}$ & $0.13^{+0.05}_{-0.03}$ & $0.11^{+0.06}_{-0.04}$ & $0.08^{+0.03}_{-0.02}$ & $0.04^{+0.1}_{-0.02}$ & $0.12^{+0.06}_{-0.03}$ & $0.17^{+0.18}_{-0.09}$ & $0.27^{+0.18}_{-0.08}$ & $0.16^{+0.13}_{-0.07}$ & $0.06^{+0.02}_{-0.01}$ & $0.18^{+0.15}_{-0.07}$ & $0.03^{+0.15}_{-0.02}$ & $0.1^{+0.01}_{-0.01}$\\[2mm]
$\mathcal{M}$ (II) &  & $0.19^{+0.15}_{-0.06}$ & $0.11^{+0.06}_{-0.03}$ & $0.06^{+0.03}_{-0.02}$ & $0.15^{+0.08}_{-0.04}$ & $0.03^{+0.09}_{-0.02}$ & $0.17^{+0.05}_{-0.03}$ & $0.07^{+0.06}_{-0.02}$ & $0.23^{+0.06}_{-0.04}$ & $0.04^{+0.02}_{-0.01}$ & $0.1^{+0.03}_{-0.02}$ & $0.19^{+0.11}_{-0.06}$ & $0.06^{+0.05}_{-0.02}$ & $0.15^{+0.02}_{-0.02}$\\[2mm]
$\mathcal{M}$ (III) &  & $0.11^{+0.06}_{-0.03}$ & $0.14^{+0.13}_{-0.06}$ & $0.08^{+0.08}_{-0.03}$ & $0.11^{+0.05}_{-0.03}$ & $0.05^{+0.3}_{-0.04}$ & $0.17^{+0.06}_{-0.04}$ & $0.13^{+0.06}_{-0.03}$ & $0.15^{+0.04}_{-0.02}$ & $0.2^{+0.18}_{-0.07}$ & $0.18^{+0.07}_{-0.05}$ & $0.18^{+0.27}_{-0.09}$ & $0.11^{+0.05}_{-0.02}$ & $0.15^{+0.02}_{-0.02}$\\[2mm]
$\mathcal{M}$ (IV) &  & $0.47^{+0.5}_{-0.23}$ & $0.27^{+0.03}_{-0.03}$ & $0.21^{+0.06}_{-0.04}$ & $0.36^{+0.16}_{-0.13}$ & $0.14^{+0.53}_{-0.12}$ & $0.28^{+0.13}_{-0.1}$ & $0.28^{+0.13}_{-0.09}$ & $0.48^{+0.14}_{-0.1}$ & $0.3^{+0.07}_{-0.05}$ & $0.26^{+0.24}_{-0.08}$ & $0.25^{+0.16}_{-0.07}$ & $0.16^{+0.11}_{-0.06}$ & $0.37^{+0.06}_{-0.05}$\\[2mm]
$\mathcal{M}$ in $R_{500}$ &  & $0.23^{+0.07}_{-0.06}$ & $0.12^{+0.03}_{-0.02}$ & $0.09^{+0.05}_{-0.02}$ & $0.12^{+0.03}_{-0.02}$ & $0.11^{+0.11}_{-0.04}$ & $0.27^{+0.08}_{-0.05}$ & $0.18^{+0.12}_{-0.07}$ & $0.2^{+0.03}_{-0.03}$ & $0.09^{+0.05}_{-0.02}$ & $0.09^{+0.03}_{-0.02}$ & $0.26^{+0.1}_{-0.09}$ & $0.04^{+0.02}_{-0.01}$ & $0.11^{+0.01}_{-0.01}$\\[2mm]
$P_{\text{NTH}}/P_{\text{TOT}}$ (I) &  & $5.99^{+6.83}_{-3.18}$ & $0.87^{+0.83}_{-0.38}$ & $0.67^{+0.87}_{-0.41}$ & $0.32^{+0.34}_{-0.13}$ & $0.07^{+0.93}_{-0.06}$ & $0.83^{+1.04}_{-0.36}$ & $1.65^{+4.8}_{-1.26}$ & $3.79^{+6.12}_{-1.93}$ & $1.39^{+2.92}_{-0.92}$ & $0.19^{+0.2}_{-0.09}$ & $1.73^{+3.98}_{-1.04}$ & $0.06^{+1.78}_{-0.06}$ & $0.54^{+0.14}_{-0.11}$\\[2mm]
$P_{\text{NTH}}/P_{\text{TOT}}$ (II) &  & $2.06^{+3.98}_{-1.08}$ & $0.68^{+1.0}_{-0.3}$ & $0.19^{+0.22}_{-0.1}$ & $1.23^{+1.59}_{-0.54}$ & $0.06^{+0.7}_{-0.05}$ & $1.59^{+1.1}_{-0.52}$ & $0.24^{+0.58}_{-0.14}$ & $2.81^{+1.53}_{-0.77}$ & $0.08^{+0.1}_{-0.04}$ & $0.51^{+0.4}_{-0.19}$ & $2.05^{+2.71}_{-1.12}$ & $0.23^{+0.44}_{-0.14}$ & $1.2^{+0.34}_{-0.26}$\\[2mm]
$P_{\text{NTH}}/P_{\text{TOT}}$ (III) &  & $0.71^{+0.87}_{-0.32}$ & $1.03^{+2.84}_{-0.7}$ & $0.38^{+1.12}_{-0.22}$ & $0.72^{+0.7}_{-0.28}$ & $0.17^{+6.57}_{-0.16}$ & $1.59^{+1.27}_{-0.61}$ & $0.97^{+1.09}_{-0.4}$ & $1.16^{+0.64}_{-0.35}$ & $2.22^{+5.17}_{-1.26}$ & $1.71^{+1.67}_{-0.82}$ & $1.75^{+8.31}_{-1.35}$ & $0.68^{+0.65}_{-0.26}$ & $1.19^{+0.33}_{-0.25}$\\[2mm]
$P_{\text{NTH}}/P_{\text{TOT}}$ (IV) &  & $10.75^{+23.14}_{-7.57}$ & $3.79^{+0.76}_{-0.66}$ & $2.36^{+1.58}_{-0.85}$ & $6.53^{+6.4}_{-3.8}$ & $1.11^{+19.13}_{-1.08}$ & $4.19^{+4.51}_{-2.38}$ & $4.18^{+4.4}_{-2.14}$ & $11.58^{+6.14}_{-3.89}$ & $4.81^{+2.21}_{-1.52}$ & $3.53^{+8.66}_{-1.82}$ & $3.28^{+5.22}_{-1.61}$ & $1.39^{+2.42}_{-0.83}$ & $7.2^{+2.17}_{-1.74}$\\[2mm]
$P_{\text{NTH}}/P_{\text{TOT}}$ in $R_{500}$ &  & $2.91^{+1.96}_{-1.26}$ & $0.81^{+0.52}_{-0.28}$ & $0.45^{+0.56}_{-0.2}$ & $0.79^{+0.5}_{-0.26}$ & $0.66^{+2.04}_{-0.36}$ & $4.02^{+2.63}_{-1.32}$ & $1.74^{+2.96}_{-1.12}$ & $2.14^{+0.68}_{-0.51}$ & $0.45^{+0.66}_{-0.19}$ & $0.44^{+0.37}_{-0.2}$ & $3.57^{+3.14}_{-2.02}$ & $0.09^{+0.11}_{-0.04}$ & $0.7^{+0.12}_{-0.1}$\\[2mm]
\hline
\end{tabular}}
\caption{Prior distribution, inferred median and 16\textsuperscript{th}-84\textsuperscript{th} percentiles for the density fluctuation parameters, and deduced parameters, for each cluster in the X-COP sample and for a joint fit over the whole sample. The first section represents the density fluctuation parameters defined in Sec.~\ref{sec:density-fluctuations-as-grf}. The second section corresponds to the parameters deduced from the density fluctuations, namely, the clumping factor $\sqrt{C}$ (Sect. \ref{sec:clumping}), Mach number $\mathcal{M}$ (Sect.~\ref{sec:mach}) and ratio of non-thermal pressure to total pressure $P_{\text{NTH}}/P_{\text{TOT}}$ (Sect.~\ref{sec:nth-pressure}). The parameters of the density fluctuations as well as the inferred parameters are calculated in the four regions defined in Table~(\ref{tab:region_definition}) and within $R_{500}$. We ensured the proper convergence of the MCMC sampling by assessing that $\hat{R} < 1.01$ for each parameter.}
\label{tab:all_parameters}
\end{sidewaystable*}

\section{Fourier transform convention}
\label{app:fourier_convention}

In this paper, we define the Fourier transform with the classical signal processing convention, namely $(0, -2\pi)$, see \cite{weisstein_fourier_1995}. This pairs results in the forward transform highlighted in Eq. \ref{eq:tf2D} and \ref{eq:tf3D}. We use $\hat{f}$ and $\tilde{f}$ to refer, respectively, to the 2D and 3D Fourier transform of a function, $f$.

\begin{equation}
\label{eq:tf2D}
    \mathcal{FT}_{2D}\left\{f\right\} \equiv \int \diff^2 \Vec{\rho} \, f(\Vec{\rho}) e^{ -2i\pi\Vec{k_\rho}.\Vec{\rho}} =  \hat{f}(\vec{k}_\rho),
\end{equation}

\begin{equation}
\label{eq:tf3D}
    \mathcal{FT}_{3D}\left\{f\right\} \equiv \int \diff^3\Vec{r} \, f(\vec{r}) e^{ -2i\pi\Vec{k_r}.\Vec{r}} =  \tilde{f}(\vec{k}_r).
\end{equation}

\section{Functional approximation for  $\Psi$ }
\label{app:cooling_function}

The cooling function as seen by XMM-Newton in the [0.7 - 1.2] keV energy band can be phenomenologically modelled with a smooth-broken power law for the temperature dependency and an exponential absorption for the column density:  

\begin{equation}
    \Psi(N_H, T) \simeq \Lambda_0 e^{- N_H \sigma} \left( \frac{T}{T_{\text{break}}}\right)^{-\alpha_{1}}\left(\frac{1}{2} + \frac{1}{2}\left(\frac{T}{T_{\text{break}}}\right)^{1/\Delta}\right)^{(\alpha_1 - \alpha_2)\Delta}
.\end{equation}

To determine the best set of parameters for this analytical approximation, we minimised the least-square of the residuals between the functional given in the previous equation and the cooling function estimated with \texttt{XSPEC} for each cluster of the sample. The true cooling rate is determined by computing the  count rate for XMM-Newton with an \texttt{PhABS*APEC} model. The temperature at a given radius is set using the universal profile derived from the X-COP sample by \cite{ghirardini_universal_2019}. The exposure time is arbitrarily set to 1 Ms and split among PN, MOS-1 and MOS-2 with a 60\%, 20\%, 20\% ratio. The abundance is fixed to $0.3 \, Z_\odot$. The count-rate in an element of volume is computed in the [0.7-1.2] keV band, assuming that $n_e = 1.17n_H$ for a fully ionised ICM \citep{anders_abundances_1989}, and with the following normalisation:  

    $$\mathcal{N} = \frac{10^{-14}}{4\pi [d_A(1+z)]^2} \int \diff^3V n_e n_H,  $$ 
    
where $d_A$ is the angular diameter distance at the redshift $z$ of the cluster. The \texttt{XSPEC} cooling function $\Bar{\Lambda}$ is computed on a $10 \times 10$ grid with $k_B T \in [1, 10] \text{ keV}$ and $N_H \in [1\times 10^{19}, 2\times 10^{21}] \text{ cm}^{-2}$ to cover the ranges expected in the X-COP clusters. 

\section{Binning approximation}
\label{app:binning_approximation}

\begin{figure}
\centering
\includegraphics[width=0.47\textwidth]{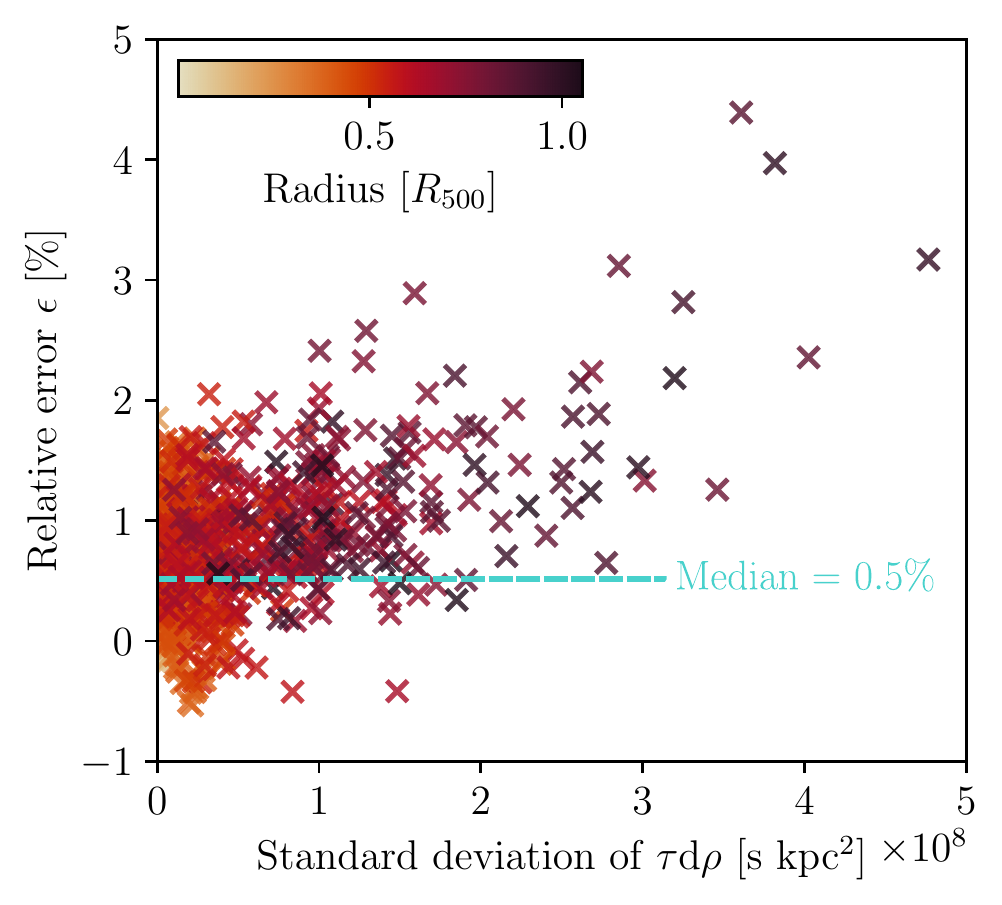}
\caption{Evaluating the uncertainties induced by our approximation of a flat surface brightness in 2D spatial bins. The relative error, $\epsilon,$ is compared to the standard deviation of effective exposure in each bin and the color of each point indicate the distance of the bin to the centre of the cluster image.}
\label{fig:mean_model_approximation}
\end{figure}

In Sect.~\ref{sec:mean-profile}, we make the approximation that the model counts in each bin is given by the product of the surface brightness, $S_X$, estimated in the centre times the effective exposure $\int_{\text{Bin}} \tau \diff \vec{\rho}$, instead of the integral over the bin geometry:

\begin{equation}
    \text{Model Counts} = \int_{\text{Bin}} S_{X}(\vec{\rho}) ~ \tau \diff \vec{\rho} \simeq S_{X}(\vec{\rho}_0) \times \int_{\text{Bin}} \tau \diff \vec{\rho}
\label{eq:model_counts}
,\end{equation}

where $\tau$ is the exposure time in each pixel and $\diff \vec{\rho}$ is a surface element of the bin area. To validate this approximation, we used a A3266 image, exposure map, and binning, and we define the relative error $\epsilon$ between the true integral and the approximation as follows : 

$$\epsilon = \frac{\int_{\text{Bin}} S_{X}(\vec{\rho}) ~ \tau \diff \vec{\rho} - S_{X}(\vec{\rho}_0) \times \int_{\text{Bin}} \tau \diff \vec{\rho}}{\int_{\text{Bin}} S_{X}(\vec{\rho}) ~ \tau \diff \vec{\rho}}.$$
We plot in Figure (\ref{fig:mean_model_approximation}),  $\epsilon$ compared to the standard deviation of $\tau \diff \vec{\rho}$ within the bin. We observe that $\epsilon$ is contained around $0.5\%$ for most of the bins. It rises to $4\%$ for some bins located in areas of the image away from the centre, where the mosaic exposure varies greatly. These regions correspond to the places where the \XMM mosaics overlap, and produce strong variations of exposure within the same bin. In any case, the bins with the coarsest approximation are those that contribute the least to the likelihood of our average model.

\section{Mean model parameters}
\label{app:mean_model_parameters}
Here, we provide the parameters obtained for the average surface gloss profile, as presented in Sect.~\ref{sec:mean-profile}. The notation $\mathcal{U}$ represent a uniform prior distribution, while $\mathcal{N}$ stands for a normal distribution. The density parameters are displayed in Table~\ref{tab:mean-model-density} and the spatial parameters are displayed in Table~\ref{tab:mean-model-morpho}
\begin{table*}
\centering
\begin{tabular}{l|cccccc}
  \hline
    \hline
 & $\log_{10} n_{e,0}^2$ & $\log_{10} r_c$ & $\log_{10} r_s$ & $\beta$ & $\epsilon$ & $\log_{10} B$ \\
 \hline

 Prior & $\mathcal{U}(-8,-3)$& $\mathcal{U}(-2,0)$& $\mathcal{U}(-1,1)$& $\mathcal{U}(0,5)$& $\mathcal{U}(0,5)$& $\mathcal{U}(-10,-4)$\\
 \hline
A1644 & $-4.057\pm0.084$ & $-1.934\pm0.049$ & $-0.091\pm0.103$ & $0.341\pm0.004$ & $3.254\pm1.221$ & $-7.162\pm0.438$ \\
A1795 & $-3.574\pm0.005$ & $-1.475\pm0.004$ & $-0.254\pm0.034$ & $0.502\pm0.002$ & $2.075\pm0.321$ & $-7.34\pm0.039$ \\
A2029 & $-3.241\pm0.014$ & $-1.912\pm0.016$ & $-0.988\pm0.007$ & $0.31\pm0.004$ & $1.647\pm0.022$ & $-7.334\pm0.012$ \\
A2142 & $-4.267\pm0.006$ & $-1.274\pm0.004$ & $-0.08\pm0.007$ & $0.476\pm0.002$ & $4.891\pm0.104$ & $-7.44\pm0.01$ \\
A2255 & $-6.09\pm0.034$ & $-0.6\pm0.098$ & $0.021\pm0.431$ & $0.526\pm0.122$ & $2.643\pm1.49$ & $-7.925\pm0.65$ \\
A2319 & $-4.978\pm0.007$ & $-1.033\pm0.007$ & $-0.001\pm0.018$ & $0.459\pm0.004$ & $4.56\pm0.365$ & $-6.907\pm0.019$ \\
A3158 & $-5.146\pm0.026$ & $-1.234\pm0.051$ & $-0.527\pm0.025$ & $0.27\pm0.021$ & $2.522\pm0.115$ & $-7.458\pm0.055$ \\
A3266 & $-5.606\pm0.008$ & $-0.992\pm0.012$ & $-0.202\pm0.006$ & $0.327\pm0.006$ & $4.976\pm0.025$ & $-7.229\pm0.011$ \\
A644 & $-4.678\pm0.009$ & $-1.049\pm0.01$ & $-0.112\pm0.048$ & $0.565\pm0.009$ & $3.914\pm0.77$ & $-7.486\pm0.029$ \\
A85 & $-3.224\pm0.008$ & $-1.987\pm0.006$ & $-0.639\pm0.008$ & $0.357\pm0.001$ & $1.381\pm0.019$ & $-8.785\pm0.548$ \\
RXC1825 & $-5.527\pm0.072$ & $-1.528\pm0.145$ & $-0.56\pm0.022$ & $0.149\pm0.016$ & $2.995\pm0.098$ & $-7.178\pm0.023$ \\
ZW1215 & $-5.447\pm0.019$ & $-0.997\pm0.037$ & $-0.298\pm0.059$ & $0.412\pm0.03$ & $3.554\pm0.54$ & $-7.545\pm0.035$ \\
 \hline
\end{tabular}
\caption{Mean and standard deviation of the density model parameters from Eq.~\ref{eq:vikhlinin_density_profile}. The prior distributions are displayed in the first line.}
\label{tab:mean-model-density}
\end{table*}

\begin{table*}
\centering
\begin{tabular}{l|cccc}
\hline
\hline
 & $\theta$ & $e$ & Right ascension & Declination \\
 \hline
 Prior & $\mathcal{U}(-\pi/2,+\pi/2)$ & $\mathcal{U}(0,0.99)$ & $\text{RA}_{Planck}\times\mathcal{N}(1,0.5)$ & $\text{DEC}_{Planck}\times\mathcal{N}(1,0.5)$ \\
 \hline
A1644 & $1.148\pm0.048$ & $0.41\pm0.02$ & $12h57m10.923s\pm0.07s$ & $-17^{\circ}24'41.001"\pm0.75"$ \\
A1795 & $-0.197\pm0.005$ & $0.578\pm0.002$ & $13h48m52.778s\pm0.007s$ & $26^{\circ}35'33.404"\pm0.126"$ \\
A2029 & $-0.347\pm0.004$ & $0.612\pm0.002$ & $15h10m56.236s\pm0.005s$ & $5^{\circ}44'42.049"\pm0.087"$ \\
A2142 & $0.854\pm0.003$ & $0.753\pm0.001$ & $15h58m20.173s\pm0.013s$ & $27^{\circ}13'55.763"\pm0.17"$ \\
A2255 & $1.481\pm0.036$ & $0.553\pm0.016$ & $17h12m51.890s\pm0.397s$ & $64^{\circ}03'48.037"\pm2.149"$ \\
A2319 & $0.445\pm0.005$ & $0.651\pm0.003$ & $19h21m10.511s\pm0.038s$ & $43^{\circ}57'21.618"\pm0.527"$ \\
A3158 & $1.418\pm0.01$ & $0.639\pm0.005$ & $3h42m52.761s\pm0.09s$ & $-53^{\circ}37'44.784"\pm0.624"$ \\
A3266 & $-1.136\pm0.006$ & $0.58\pm0.003$ & $4h31m25.116s\pm0.079s$ & $-61^{\circ}25'33.411"\pm0.548"$ \\
A644 & $-0.205\pm0.01$ & $0.621\pm0.005$ & $8h17m25.447s\pm0.025s$ & $-7^{\circ}31'02.495"\pm0.459"$ \\
A85 & $0.355\pm0.005$ & $0.509\pm0.002$ & $0h41m50.414s\pm0.005s$ & $-9^{\circ}18'11.713"\pm0.074"$ \\
RXC1825 & $1.57\pm0.0$ & $0.624\pm0.006$ & $18h25m21.609s\pm0.069s$ & $30^{\circ}26'22.100"\pm0.679"$ \\
ZW1215 & $-1.094\pm0.011$ & $0.637\pm0.005$ & $12h17m41.309s\pm0.051s$ & $3^{\circ}39'30.396"\pm0.617"$ \\
\end{tabular}
\caption{Mean and standard deviation of the spatial parametrisation from Eq.~\ref{eq:ellipse}. The prior distributions are displayed in the first line.}
\label{tab:mean-model-morpho}
\end{table*}

\section{Mexican hat filtering}
\label{app:mexican_hat_analytic}

\begin{figure*}[t]
\centering
\subfigure{
\label{fig:subfig:mexican_hat_img} 
\includegraphics[width=0.5\textwidth]{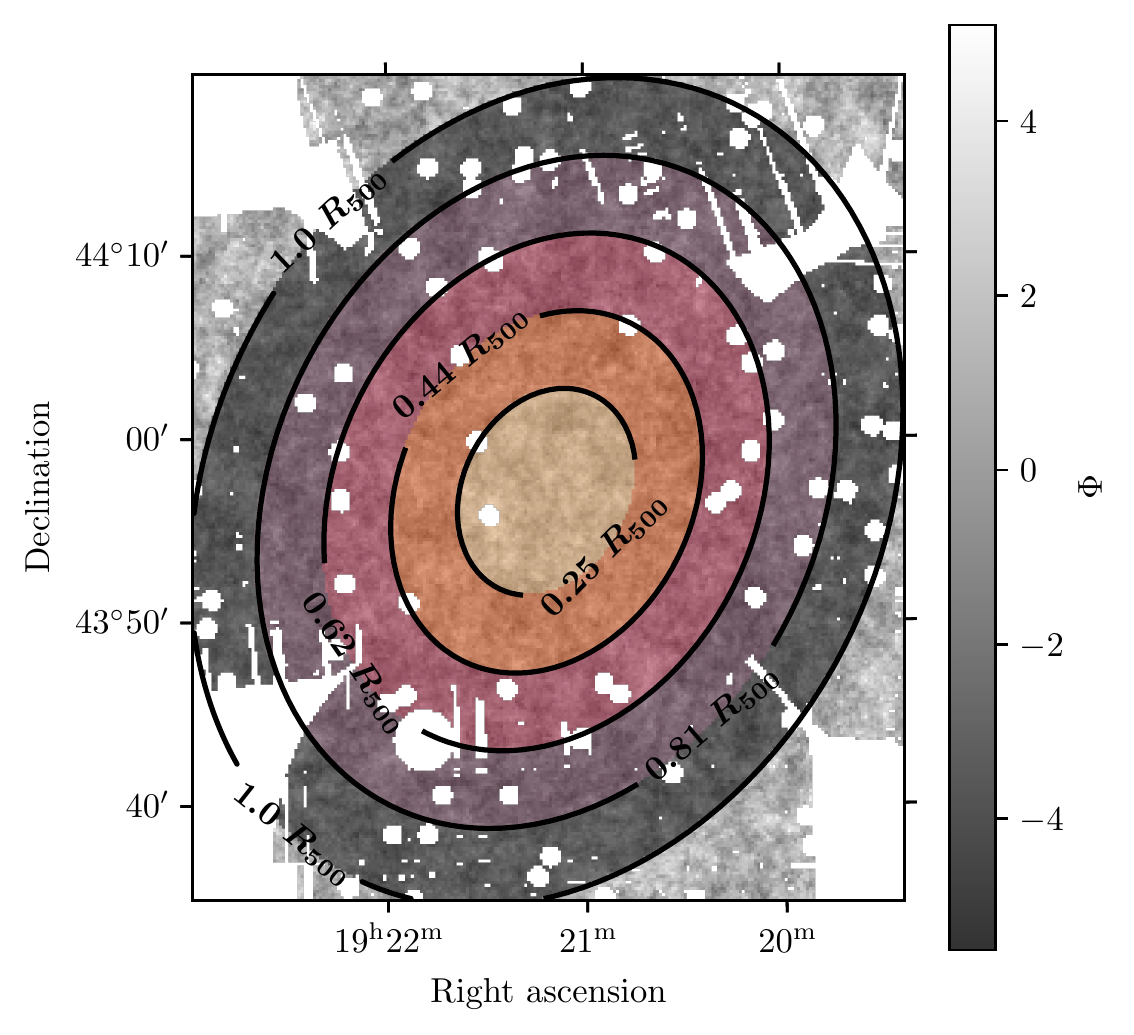}}
\hspace{0.2in}
\subfigure{
\label{fig:subfig:mexican_hat_calibration} 
\includegraphics[width=0.45\textwidth]{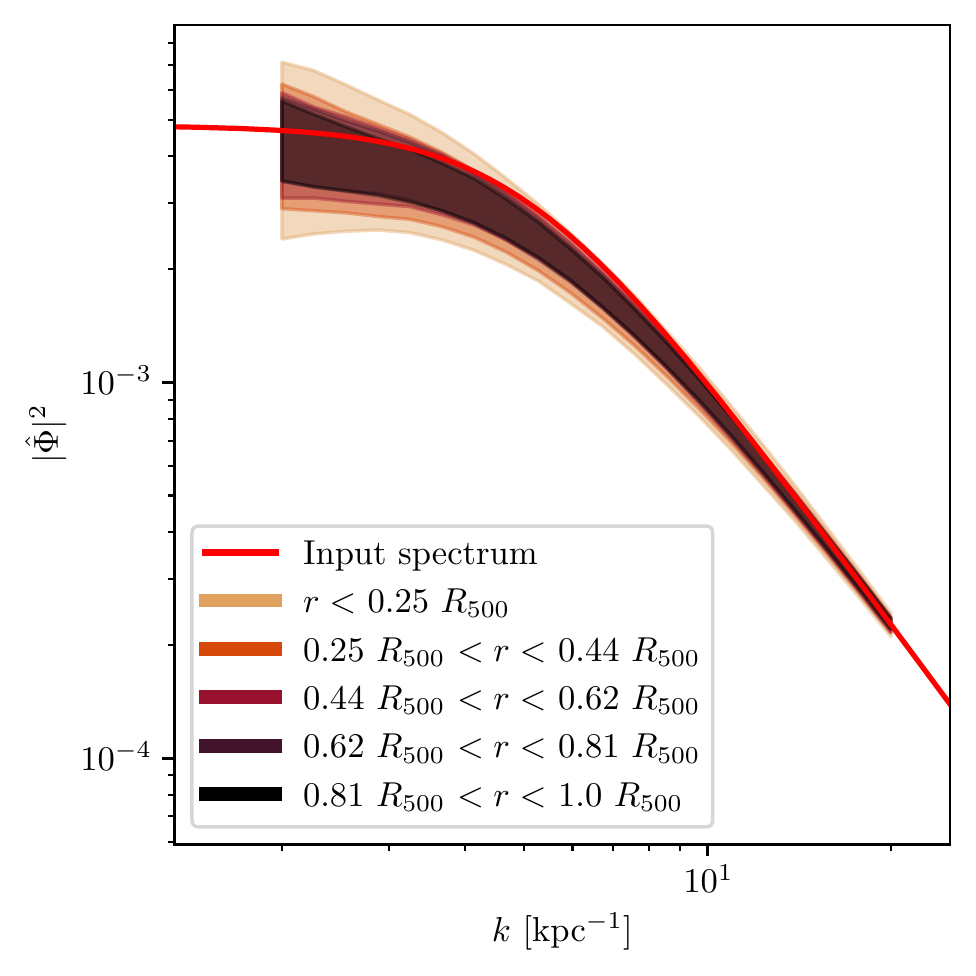}}
\caption{Assessing the efficiency of Mexican hats to compute the power spectrum of a field with arbitrary masks by computing the power spectra of 1000 realisations of a random field with known power spectrum in various ring regions. Left panel : regions of interest overplotted on a single realisation of the known random field with the A2319 exposure map. Right panel : recovered power spectra in the various regions of interest for 1000 realisations of the known random field, and comparison with the input power spectrum}
\label{fig:mexican_hat_calibration}
\end{figure*}

We redirect  the readers to the original article by \cite{arevalo_mexican_2012} for additional details on the Mexican hat filtering method. Here, we extend this formalism to arbitrary conventions for Fourier transforms $(a,b)$, see \cite{weisstein_fourier_1995}. To do so, we first define the $n$-dimensional Fourier transform for a radially symmetric function $f$ as the following Hankel transform:

\begin{equation}
    \mathcal{FT}_{nD}\left\{f\right\} = \left(\frac{|b|}{(2\pi)^{1-a}}\right)^{n/2}
\frac{(2\pi)^{n/2}}{(bk)^{n/2-1}}
\int_0^\infty r^{n/2} f(r) J_{n/2-1}(bkr) \diff r
.\end{equation}

In the following, we designate $\tilde{f}$ as the N-dimension Fourier transform of $f$. We define the $n$-dimensional Gaussian kernel of $\sigma$ standard deviation and its Fourier transform as follows :

\begin{equation}
    G_{\sigma, nD}(r) = \frac{1}{\left(2\pi\sigma^2 \right)^{n/2}} \exp \left(-\frac{r^2}{2\sigma^2}\right) 
,\end{equation}

\begin{equation}
\Tilde{G}_{\sigma, nD}= 
     \left(2\pi\right)^{\frac{n \left(a - 1\right)}{2}} \left|{b}\right|^{\frac{n}{2}} \exp \left(- \frac{b^{2} k^{2} \sigma^{2}}{2}\right)
.\end{equation}

The kernel of the Mexican hat is formally defined as the difference of two Gaussian kernels, whose standard deviations are, respectively, $\sigma_1 = \sigma/\sqrt{1+\epsilon}$ and $\sigma_2 = \sigma\times\sqrt{1+\epsilon}$ with $\epsilon = 10^{-3}$. Computing the Taylor expansion of this kernel gives:

\begin{equation}
     \Tilde{F}(k) = (\Tilde{G}_{\sigma_1, nD} - \Tilde{G}_{\sigma_2, nD})(k) \simeq \left|{b}\right|^{\frac{n}{2}+2} \epsilon k^{2} \sigma^{2} \left(2\pi\right)^{\frac{n \left(a - 1\right)}{2}} e^{- \frac{b^{2} k^{2} \sigma^{2}}{2}} 
    \label{eq:mexican_hat_taylor}
.\end{equation}

The relation between the filtered frequency $k_{\max}$, which corresponds to the maximum value of the previous filter, and the standard deviation $\sigma$ is given by $k_{\max} = \sqrt{2}/(\left|{b}\right|\sigma)$. We then assume that we set $\sigma$ in Eq.~\ref{eq:mexican_hat_taylor} so that $k_r$ is the frequency of maximum value. We convolve the true image $I$ with this kernel : 

\begin{equation}
    I_{c}(\Vec{r}) = \left(G_{\sigma_1}- G_{\sigma_2}\right)*I(\Vec{r})
\end{equation}

The variance of the convolved image $I_c$ is directly related to the power spectrum of $I$ evaluated at $k_r$, with a simple proportionality relation : $\operatorname{Var} I_c \propto |\Tilde{I}(k_r)|^2$. It can be shown using Plancherel theorem : 

\begin{equation}
    \int \diff^n\Vec{r}~ |I_c(\Vec{r})|^2 = \int \diff^n\Vec{k}~ |\Tilde{I}_c(\Vec{k})|^2 = \int \diff^n\Vec{k}~|\Tilde{I}(\Vec{k})|^2|\Tilde{F}(\Vec{k})|^2
    \label{eq:plancherel}
\end{equation}

The kernel $|\Tilde{F}(\Vec{k})|^2$ is supposed to be thin enough to act like a Dirac function, and be close to zero when away from $k_r$. We can then rewrite Eq.~\ref{eq:plancherel} as follows:

\begin{equation}
\begin{split}
\int \diff^n\Vec{r}~ |I_c(\Vec{r})|^2 \simeq& |\Tilde{I}(k_r)|^2 \int \diff^n\Vec{k}~|\Tilde{F}(\Vec{k})|^2\\
=&|\Tilde{I}(k_r)|^2 \epsilon^2 \Upsilon(n) k_r^n
\end{split}
\label{eq:relationship_convolved_ps}
,\end{equation}

where $\Upsilon(n) = 2^{n \left(a - \frac{1}{2}\right) -1} \pi^{n \left(a + \frac{1}{2}\right)}  n \left(\frac{n}{2} + 1\right)$ is a constant which depends on the dimension $n$. The idea now is to incorporate an arbitrary mask, $M,$ to remove unwanted pixels on the image. We defined the new convolved image as follows :

\begin{equation}
    I_{c}(\Vec{r})  =\ \left( \ \frac{G_{\sigma_1}*I}{G_{\sigma_1}*M} \ -\ \frac{G_{\sigma_2}\ *\ I}{G_{\sigma_2}*M}\right) \times M
.\end{equation}

Then, including these changes in Eq.~\ref{eq:relationship_convolved_ps}, we can estimate the power spectrum of the image by computing:

\begin{equation}
    |\hat{I}(k_r)|^2 \simeq \frac{1}{\epsilon^2 \Upsilon(n) k_r^n} \frac{\int \diff^n\Vec{r}~ I_c(\Vec{r})^2}{\int \diff^n\Vec{r}~ M(\Vec{r})} 
    \label{eq:mexican_hat_formula}
.\end{equation}

The integrals in Eq.~\ref{eq:mexican_hat_formula}  are simply expressed as the sum over all cells (or pixels in 2D). This methodology intrinsically handles border effects as well as point source exclusions in the data. We test the effectiveness of this approach for an image generated via a GRF $\Phi$, with the exposure masks of A3266. In Fig.~\ref{fig:mexican_hat_calibration}, we show a single realisation of this GRF along with the regions used to compute the power spectra, along with the distribution of their estimations for 1000 realisations of the GRF. We see a good agreement between the power spectrum used to generate the GRF and the reconstructed ones, with a slight bias, which is investigated in detail by \cite{arevalo_mexican_2012}.

\begin{figure}[ht]
\includegraphics[width=\linewidth]{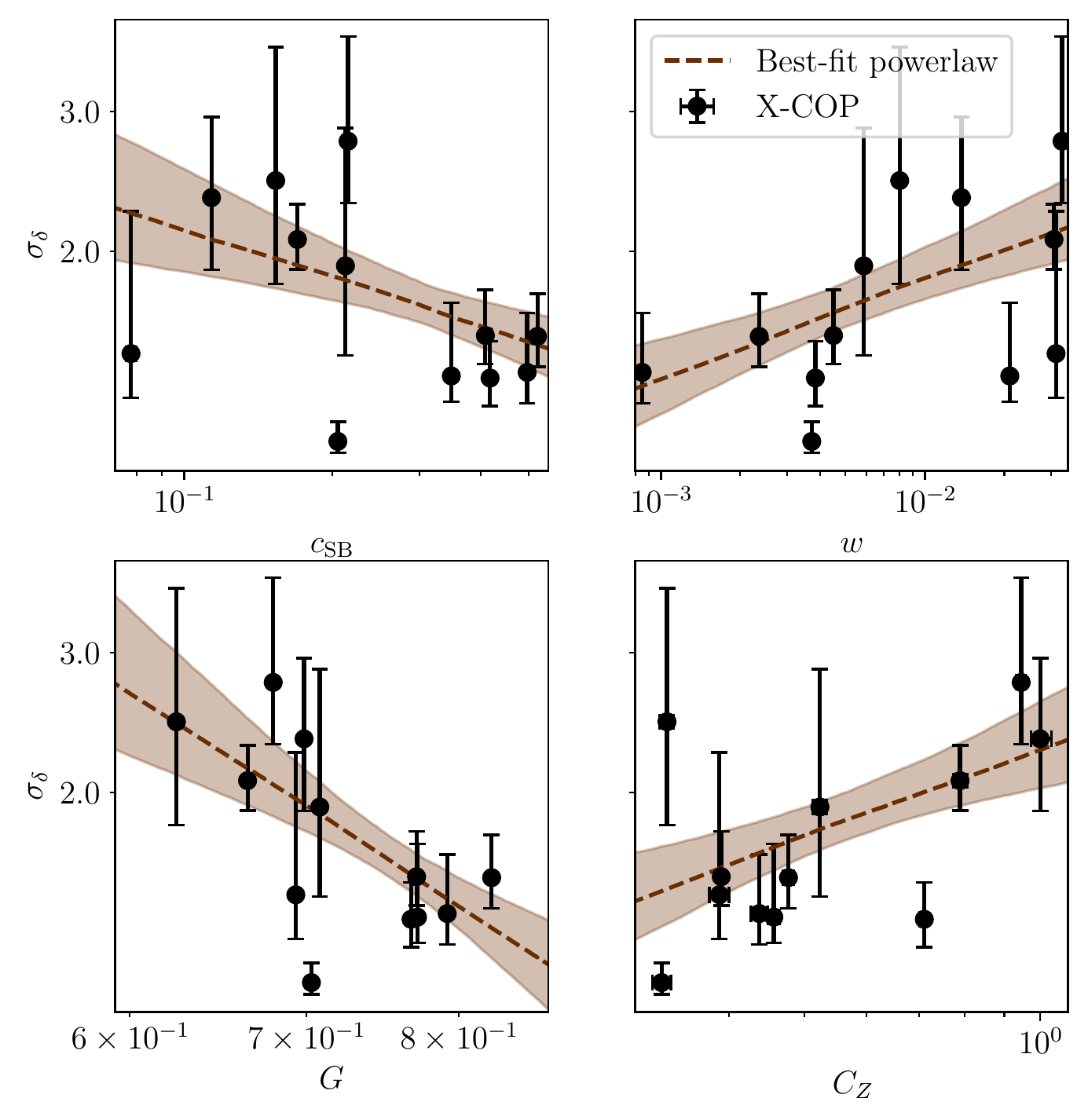}
\caption{Correlation between $\sigma_\delta$ evaluated in $R_{500}$ for each cluster and the morphological indicators defined in Sect.~\ref{sec:morpho_indicators}. The plain line and envelop represent the median and 16\textsuperscript{th}-84\textsuperscript{th} percentiles of the best fit of a power law scaling.}
\label{fig:correlation_sigma_indicators}
\end{figure}

\section{Correlation between $\sigma_\delta$ and the morphological indicators}
\label{app:correlation_morpho}

Here, we show  the individual correlations between the normalisation of density fluctuation estimated inside $R_{500}$ and the various morphological indicators introduced in Sect.~\ref{sec:morpho_indicators} for each cluster in the X-COP sample in Fig.~\ref{fig:correlation_sigma_indicators}. We plot the best-fit distribution of a power law scaling for each correlation. This scaling is defined as $\log_{10} y = a_1 \log_{10} x + a_2$. The best-fit values of $a_1$ and $a_2$ for each correlation are displayed in Table~\ref{tab:correlation_fit}.

\begin{table}
    \centering
    \begin{tabular}{c|c|c}
    \hline
    \hline
    & $a_1$ & $a_2$ \\
    \hline
    
$\sigma_\delta = f(c_{\text{SB}})$ &$-0.198^{+0.097}_{-0.134}$ &$0.129^{+0.06}_{-0.069}$ \\[2mm]
$\sigma_\delta = f(w)$ &$0.123^{+0.057}_{-0.045}$ &$0.507^{+0.141}_{-0.1}$ \\[2mm]
$\sigma_\delta = f(G)$ &$-2.15^{+0.771}_{-0.988}$ &$-0.05^{+0.101}_{-0.118}$ \\[2mm]
$\sigma_\delta = f(C_Z)$ &$0.37^{+0.174}_{-0.179}$ &$0.355^{+0.061}_{-0.048}$
    \end{tabular}
    \caption{Median values of $a_1$ and $a_2$  and difference with the 16\textsuperscript{th}-84\textsuperscript{th} percentiles of the best fit for each scaling}
    \label{tab:correlation_fit}
\end{table}
\end{appendix}
\end{document}